\documentclass[useAMS,usenatbib]{mnras}
\usepackage{graphicx}
\usepackage{latexsym}
\usepackage{psfig}
\usepackage{amssymb,amsmath}
\usepackage{subfig}
\usepackage{caption}
\usepackage{times}
\usepackage{color}
\title[Tidally--induced star formation in CMZ clouds]{The dynamical evolution of molecular clouds near the Galactic Centre -- III. Tidally--induced star formation in protocluster clouds}

\author[]{James E. Dale$^{1}$, J. M. Diederik Kruijssen$^{2}$ and S.~N.~ Longmore$^{3}$ \vspace*{0.1cm}\\
$^{1}$Centre for Astrophysics Research, University of Hertfordshire, Hatfield, AL10 9AB, UK\\
$^{2}$Astronomisches Rechen-Institut, Zentrum f\"{u}r Astronomie der Universit\"{a}t Heidelberg, M\"{o}nchhofstra\ss e 12-14, 69120 Heidelberg, Germany\\$^{3}$Astrophysics Research Institute, Liverpool John Moores University, 146 Brownlow Hill, Liverpool L3 5RF, UK}

\begin{document}
                             
\pagerange{\pageref{firstpage}--\pageref{lastpage}} \pubyear{2019}

\maketitle

\label{firstpage}

\def\mnras{MNRAS}
\def\apj{ApJ}
\def\aj{AJ}
\def\aap{A\&A}
\def\apjl{ApJL}
\def\apjs{ApJS}
\def\araa{ARA\&A}
\def\pasp{PASP}

\begin{abstract}
As part of a series of papers aimed at understanding the evolution of the Milky Way's Central Molecular Zone (CMZ), we present hydrodynamical simulations of turbulent molecular clouds orbiting in an accurate model of the gravitational potential extant there. We consider two sets of model clouds differing in the energy content of their velocity fields. In the first, \textit{self--virialised} set, the turbulent kinetic energies are chosen to be close in magnitude to the clouds' \textit{self--gravitational potential energies}. Comparison with isolated clouds evolving without an external potential shows that the self--virialised clouds are unable to withstand the compressive tidal field of the CMZ and rapidly collapse, forming stars much faster and reaching gas exhaustion after a small fraction of a Galactocentric orbit. In the second, \textit{tidally--virialised}, set of simulations, the clouds' turbulent kinetic energies are in equilibrium with the \textit{external tidal field}. These models are better supported against the field and the stronger turbulence suppresses star formation. Our results strongly support the inference that anomalously low star formation rates in the CMZ are due primarily to high velocity dispersions in the molecular gas. The clouds follow open, eccentric orbits oscillating in all three spatial coordinates. We examine the consequences of the orbital dynamics, particularly pericentre passage, by performing companion simulations of clouds on circular orbits. The increased tidal forces at pericentre produce transient accelerations in star formation rates of at most a factor of 2.7. Our results demonstrate that modelling star formation in galactic centres \textit{requires} the inclusion of tidal forces.\\
\end{abstract}


\section{Introduction}
\begin{figure}
\includegraphics[width=0.48\textwidth]{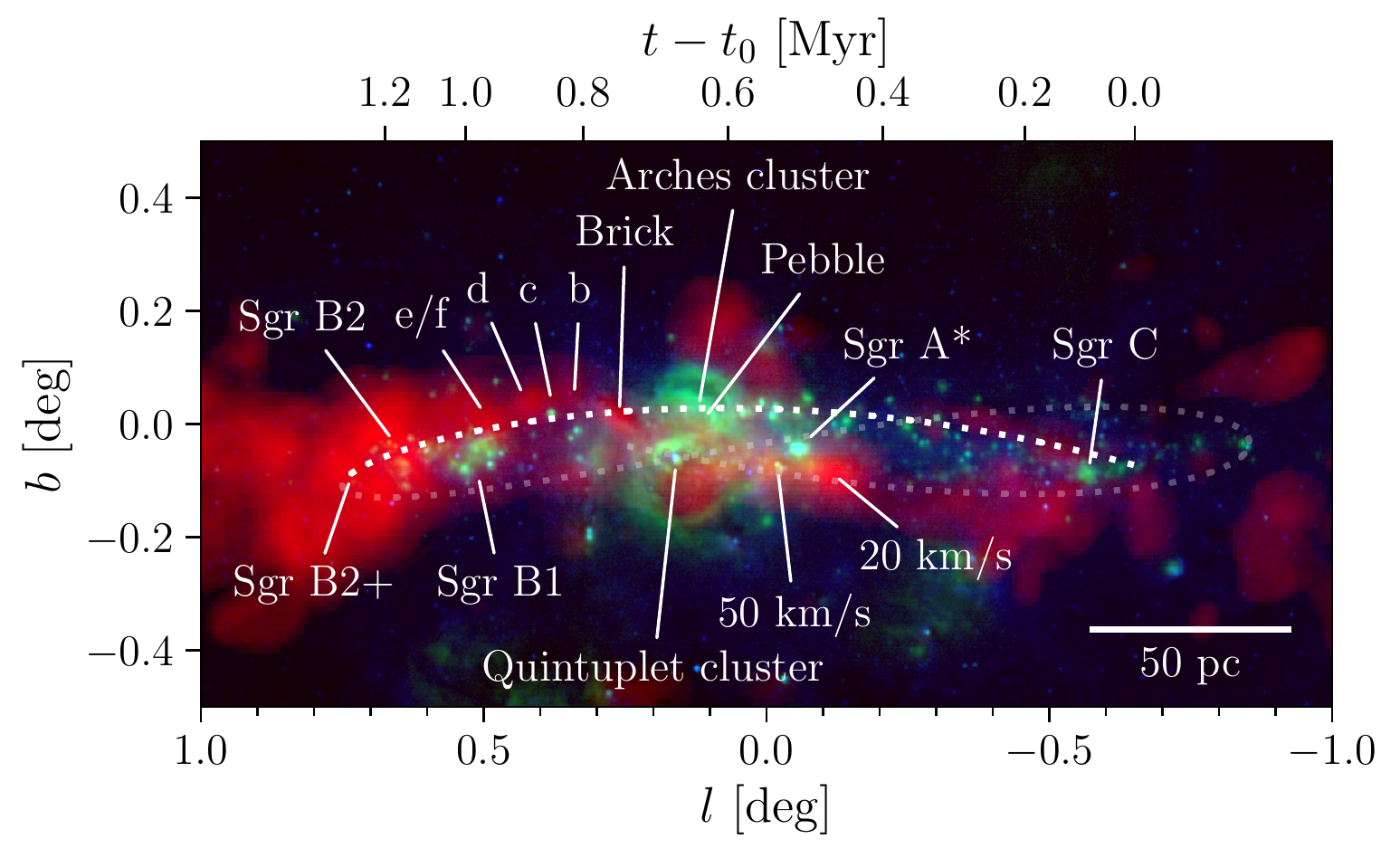}
\caption{Three--colour composite image of the CMZ. Red shows an integrated-intensity map of the HOPS NH$_{3}$(1,1) emission (revealing gas with a number density $n>$several$\times10^{3}$\,cm$^{-3}$, green shows the MSX 21.3\,$\mu$m image \citep[][]{1998ApJ...494L.199E,2001AJ....121.2819P}, and blue shows the MSX 8.28\,$\mu$m image. The fitted orbit from Kruijssen et al (2015) is shown as a white dotted line and the positions of several well--known objects are indicated.}
\label{fig:cmz_mol}
\end{figure}

\indent Star formation is one of the most important processes in astrophysics and the Central Molecular Zone (CMZ) of the Milky Way presents us with an invaluable opportunity to test our understanding of it in an extreme environment. The CMZ (usually defined as the central 500\,pc of the Milky Way) contains a reservoir of $M\approx 5\times10^{7}$\,M$_{\odot}$ of molecular gas \citep[e.g.][]{2007A&A...467..611F}. The thermal and dynamical state of this material is very different from that of typical molecular interstellar medium (ISM) gas elsewhere in the Galaxy. Molecular gas in the CMZ is two orders of magnitude denser ($n_{\rm CMZ}\sim10^{4}$\,cm$^{-3}$, \cite{doi:10.1093/mnras/sts376}), up to more than one order of magnitude warmer ($T_{\rm CMZ}\sim50-400$\,K, \cite{2013ApJ...772..105M,2013A&A...550A.135A,2016A&A...586A..50G,2017ApJ...850...77K}) and has turbulent Mach numbers several times higher ($\mathcal{M}_{\rm CMZ}\approx30$, \cite{2001ApJ...562..348O,2014MNRAS.440.3370K,2016MNRAS.457.2675H}) than in typical Galactic disk GMCs. These properties are much more reminiscent of gas in $z=2$--3 galaxies \citep[][]{2013MNRAS.435.2598K}, when the cosmic star formation rate peaked \citep[][]{2014ARA&A..52..415M}, such that the CMZ may provide key insight into the extreme conditions under which many of the stars in the Milky Way and other galaxies formed.\\
\indent It has been known for some time that the star formation rates in the Galactic Centre clouds are very low given their high volume and surface densities, placing them far off well--known correlations between these quantities, such as the Kennicutt--Schmidt relation \citep[e.g.][]{1998ApJ...498..541K}. One well--studied cloud in particular, G0.253 + 0.016, often known as `the Brick', has few detectable signs of star formation \citep[e.g.][]{kauffmann17}, despite having a mass of $\sim10^{5}$\,M$_{\odot}$ and a freefall time of $\sim10^{5}$\,yr \citep[][]{2014ApJ...786..140R}. \cite{2016ApJ...832..143F} interpret the low star formation rate of the Brick as the imprint of solenoidally driven turbulence, which is likely an imprint of high levels of shear experienced by the cloud \citep{kruijssen18}.\\
\indent In Figure \ref{fig:cmz_mol}, we show a three--colour composite image of the CMZ. Red shows an integrated-intensity map of the HOPS NH$_{3}$(1,1) emission (revealing gas with a number density $n$$>$several$\times10^{3}$\,cm$^{-3}$, green shows the MSX 21.3\,$\mu$m image \citep[][]{1998ApJ...494L.199E,2001AJ....121.2819P}, and blue shows the MSX 8.28\,$\mu$m image. The locations of several well--known objects are indicated.\\
\indent As well as having unusual intrinsic properties, the molecular clouds in the CMZ are distributed in a configuration that is unique within the Milky Way. A large fraction of the molecular gas is contained in an approximately ring--shaped structure orbiting about the Galactic Centre at a radius of $r\approx 100$\,pc \citep[][]{1995PASJ...47..527S,2011ApJ...735L..33M}. The dense gas (red in Figure \ref{fig:cmz_mol}) and young stars (blue) are distributed asymmetrically as a function of longitude. This likely results from instabilities as the gas flows in along the bar, resulting in interruptions in the gas morphology, causing the instantaneous spatial distribution in the central $\sim100$~pc to change due to the regular orbital motion of the gas \citep{2018MNRAS.475.2383S}. In \cite{2015MNRAS.447.1059K}, we presented a detailed analysis of the dynamics of the gas stream and used it to construct the most accurate model to date of the orbit on which the clouds lie, and the potential required to produce it 
\citep[derived from][]{2002A&A...384..112L}. This model represents the gas as an eccentric, vertically-oscillatory, and open ended stream orbiting the Galactic Centre, and is supported by the detailed kinematic analysis presented in \cite{2016MNRAS.457.2675H}. The fitted orbit from \cite{2015MNRAS.447.1059K} is shown in Figure \ref{fig:cmz_mol} as a white dotted line.\\
\indent \cite{2015MNRAS.453..739K} and \cite{2017MNRAS.466.1213K} developed a dynamical model for the formation and evolution of the molecular stream. Gas is taken to be fed at a steady rate into a disk of radius $\sim500$\,pc by the Galactic bar \citep[][]{1999ApJ...520..592M}, which also drives acoustic instabilities in the disk. The instabilities maintain a high velocity dispersion which prevents the disk from becoming gravitationally unstable, and the radial shear transports angular momentum outwards, allowing matter to flow inwards. At a radius of $\approx100$\,pc, the Galactic rotation curve morphs from flat to solid--body, leading to a local radial minimum in shear. Angular momentum transport therefore fails and gas accumulates at this radius. There, the turbulent energy dissipates and the gas becomes gravitationally unstable, leading to a burst of star formation.\\
\indent As already remarked, the CMZ molecular clouds do indeed have unusually high velocity dispersions, leading \cite{doi:10.1093/mnras/sts376} and \cite{2014MNRAS.440.3370K} to suggest that this suppresses star formation. However, since the gas is on an eccentric orbit in the \cite{2015MNRAS.447.1059K} model, tidal forces associated with passage through pericentre may be sufficient to overcome turbulent support and induce the clouds to begin forming stars. \cite{2013MNRAS.433L..15L} proposed that an understanding of the 100--pc stream clouds in various stages of star formation may allow us to derive an absolute, as opposed to merely relative, timescale for the star formation process.\\
\indent Alternatively, once the gas detaches from the large--scale disk structure, acoustic instabilities cease driving turbulence within it. Dissipation of the turbulent motions by internal shocks may then sufficiently deprive the gas of support that it becomes locally gravitationally unstable, and star formation begins on the timescale on which the turbulence dies away. However, the tidal shear generated by the external potential is likely to be a source of turbulence, stabilising the gas and in particular setting a minimum value for the clouds's internal velocity dispersions, as discussed in \cite{kruijssen18}. This complicates any attempt at a simple derivation of this timescale.\\
\indent There are several alternatives to the \cite{2015MNRAS.447.1059K} model of an open--ended stream oscillating in all three spatial coordinates. \cite{2011ApJ...735L..33M} proposed that the gas follows a closed stream, and the work of \cite{2015MNRAS.447.1059K} was in part intended as an improvement on this idea aimed at a more consistent fit with available position--position and position--velocity observations.\\
\indent Alternatively, \citet{1995PASJ...47..527S} and \citet{2004MNRAS.349.1167S} modelled the CMZ streams as spiral arms. The feasibility of this geometry was recently demonstrated by \citet{ridley17} using hydrodynamical simulations. However, this model retains fundamental topological problems relative to the observed position-position-velocity structure, and 2D simulations provide no insight in the vertical structure. We therefore adopt the geometry and orbital model derived by\citep{2015MNRAS.447.1059K}, which best reproduces the observations \citep[see e.g.][]{2016MNRAS.457.2675H} and is similar to the transient structures found in the simulation by \citet{2015MNRAS.446.2468E}.\\
\indent In this series of papers, we will use the opportunity presented to us by the CMZ to compare targeted simulations directly with observations and answer two questions fundamental to the whole field of star formation: (i) what are the absolute timescales on which molecular clouds evolve from quiescent to star--forming to dispersal?; (ii) to what extent does the external environment, particularly the local tidal field, influence the evolution of GMCs?\\
\indent The first paper in the series, \cite{2015MNRAS.447.1059K}, laid the groundwork for this project by proposing an orbit which fits the observed position--position and position--velocity structure in the CMZ, and provides a physically and observationally self--consistent model for the gravitational potential required to generate such an orbit. In this paper and in \cite{kruijssen18}, we use the models for the gravitational potential and the orbit to begin addressing these questions directly using numerical simulations. This paper presents the simulations themselves and describes and explains their behaviour entirely from a theoretical perspective. \cite{kruijssen18} instead examines in detail the \textit{observable} properties of the model clouds and compares them explicitly with the Galactic Centre clouds.\\
\indent The layout of this paper is as follows: in Section 2, we briefly discuss why the simulations presented here are so different from most models of molecular clouds. Section 3 describes our numerical methods, in particular how the external potential used to represent the Milky Way's nuclear stellar mass distribution is modelled. In Section 4, we present the results of two classes of simulation. In the first, clouds are supported by turbulence only against their self--gravity. We compare the evolution of clouds on eccentric and circular orbits to examine the influence of the orbital dynamics and in particular the pericentre passage. We also evolve these clouds in the absence of the background potential in order to make explicit comparisons with more traditional models of isolated turbulent objects. In the second set of simulations, clouds are supported against the compressive tidal field in the CMZ by stronger turbulent velocity fields. We again model clouds following eccentric and circular orbits to extract the influence of pericentre passage on their evolution. In Section 5, we compare the two sets of simulations in the context of explaining the behaviour of the CMZ clouds. In Section 6, we present our conclusions.\\
\section{Molecular clouds evolving in strong tidal fields}
\indent The majority of numerical simulations of molecular cloud evolution performed to date have considered either isolated objects, with no boundary conditions of any kind \citep[e.g.][]{2014MNRAS.442..285B}, simple outflow boundary conditions \citep[e.g.][]{2016MNRAS.461.2953H}, or the contents of periodic turbulent boxes, often with an artificial driving mechanism to create or maintain the turbulence \citep[e.g.][]{2015MNRAS.450.4035F}. In all these cases, the progression of the simulations is in large part determined by twice the ratio of turbulent kinetic energy $T$ to the modulus of the self--generated gravitational potential energy $V$ on the largest meaningful scales, so that $\alpha_{\rm vir}=2T/|V|$.\\
\indent As will become clear, the evolution of the clouds in the CMZ is dominated by the \textit{external} tidal forces acting upon them, and the clouds' self--gravity is, at least initially, \textit{almost irrelevant to their evolution}. The dynamical states of the clouds are therefore not well described by the traditional virial ratio.\\
\indent Simulations of galactic disks aimed at \textit{forming} molecular clouds \citep[e.g.][]{2009ApJ...700..358T,2013MNRAS.432..653D} clearly do include the external tidal forces acting on the clouds. However, in most cases, these forces are not likely to dominate the clouds' behaviour \citep[e.g.][]{1995ApJ...451..167D} unless the clouds are very elongated \citep[][]{2009MNRAS.393.1563B}. This is mainly because the tidal forces exerted on a cloud due to its being embedded in an external potential only become dominant when the density of the mass distribution responsible for the potential exceeds that of the cloud itself. Molecular clouds have densities of a few to tens of solar masses per cubic parsec, two or three orders of magnitude greater than the average mass density of the stellar component of a typical disk galaxy. Clearly, the stellar mass density can be much larger in the vicinity of large stellar clusters, leading to complex tidal interactions between clouds and clusters. However, most stellar clusters are small compared to molecular clouds, and mutual encounters are brief.\\
\indent The only locations in a galaxy where there is an \textit{extended} region in which the ambient stellar mass density is comparable to that of typical molecular cloud densities is near galactic nuclei. The clouds which are the subject of this study orbit the centre of the Milky Way at a distance of $r\approx 100$\,pc, where the enclosed stellar mass is $M(r)\approx 7\times10^{8}$\,M$_{\odot}$ \citep[][]{2002A&A...384..112L}. Even if this mass were spherically distributed (which it is not; the results of \citealt{2002A&A...384..112L} show that the mass distribution is highly flattened), this would be equivalent to a mass density of $\sim100$\,$M_{\odot}$\,pc$^{-3}$, which coincides with the lower end of the range of densities seen in CMZ clouds \citep[e.g.][]{henshaw16b,henshaw17}. Hence, even though the clouds in the CMZ are substantially denser than typical disk molecular clouds, their evolution is strongly influenced -- if not dominated -- by the deep potential generated by the old stellar population there. Molecular cloud evolution and star formation in the centres of most spiral galaxies are likely to be influenced by the pre--existing stellar population in much the same fashion as in the CMZ of the Milky Way and this work therefore readily offers insights into the structure and kinematics of many systems other than the CMZ.\\
\section{Numerical Methods}
\indent We use the state--of--the--art Smoothed Particle Hydrodynamics code \textsc{gandalf} \citep[][]{2018MNRAS.473.1603H}. \textsc{gandalf} is a hybrid code in which the fluid equations are solved using the grad--h SPH formalism \citep[][]{2002MNRAS.333..649S,2004MNRAS.348..139P} and a leapfrog kick--drift--kick integrator, while point--mass sink particles are used to model star formation. We do not have sufficient resolution to model the formation of single stars and the sink particles created in our simulations instead can be taken to represent stellar subclusters. However, we are mainly interested in the properties of the gas in the simulations and the sink particles are used simply as a numerical device to avoid having to follow the evolution of very dense condensations of gas. Apart from using them as a measure of how much gas has crossed into the regime of local gravitational collapse and star formation, we do not discuss them at length.\\
\indent The code computes self--gravitational forces using an octal tree. Artificial viscosity forces are computed according to the \cite{1997JCoPh.136..298M} scheme, with $\alpha=1$, $\beta=2$. The gas thermodynamics is computed using a barotropic equation of state \citep[e.g.][]{2011A&A...529A..27H} with a critical density of 10$^{-16}$\,g\,cm$^{-3}$, although the threshold for sink particle formation is 10$^{-17}$\,g\,cm$^{-3}$, so the simulations are effectively isothermal, with a gas temperature of 65\,K. The gas is taken to have a mean molecular weight $\mu=2.35$.\\
\indent Our clouds are all modelled with $10^{6}$ SPH particles and all possess the same initially divergence--free turbulent velocity field, of which the power spectrum follows $P(k)\propto k^{-4}$ and is undriven. Particle masses, positions and velocities are scaled to model clouds with different intrinsic characteristics. The parameters we choose to vary are the mass $M$, initial radius $R_{0}$, the initial one--dimensional velocity dispersion $\sigma_{1D,0}$, the initial mass volume density $\rho_{0}$, and the initial mass surface density $\Sigma_{0}$. The clouds all have the same turbulent velocity field, scaled so that $v_{\rm RMS}\equiv\sigma_{\rm 3D,0}=\sqrt{3}\sigma_{\rm 1D,0}$.\\
\indent The mass, radius and mass volume density are related as usual by
\begin{eqnarray}
\rho_{0}=\frac{3 M}{4\pi R_{0}^{3}}
\end{eqnarray}
and the mass surface density is given by
\begin{eqnarray}
\Sigma_{0}=\frac{M}{\pi R_{0}^{2}}.
\end{eqnarray}
and the clouds' initial freefall times are given by
\begin{eqnarray}
t_{\rm ff}=\left(\frac{3\pi}{32G\rho_{0}}\right)^{\frac{1}{2}}
\end{eqnarray}
The virial ratios are calculated accurately and directly by computing the sum $T$ of the turbulent kinetic energies of all the SPH particles, and the gravitational potential, using the tree, of the gas and sink particles where appropriate (neglecting the external tidal field) $V$, and setting $\alpha_{\rm vir}=2T/|V|$.\\
\subsection{Background tidal field}
\indent The purpose of this paper is to study the evolution of a set of molecular clouds orbiting in the gravitational potential of the Galactic Centre region, which is dominated by pre--existing stellar mass. Observations by \cite{2002A&A...384..112L} derived the mass volume density and \cite{2015MNRAS.447.1059K} showed that the distribution could be modelled by a power law $\rho(r)\propto r^{-\gamma}$ in the galactocentric radius $r$ in the range of interest here, 1--300\,pc, with an exponent $\gamma=1.7-1.9$ (although note that we compute enclosed masses at a given value of $r$ directly by logarithmic interpolation of the enclosed mass inferred by \citealt{2002A&A...384..112L}).\\
\indent \cite{2002A&A...384..112L} show that the distribution of pre--existing stellar mass is strongly flattened. In order to reproduce the motions of the Galactic Centre clouds perpendicular to the Galactic plane, \cite{2015MNRAS.447.1059K} constructed a flattened axisymmetric analytic potential model by setting
\begin{eqnarray}
\Phi(r)\rightarrow\Phi(r',z),
\end{eqnarray} 
where $R$ is the spherical radial coordinate $(x^{2}+y^{2}+z^{2})^{\frac{1}{2}}$, $r'$ is the cylindrical radial coordinate $(x^{2}+y^{2})^{\frac{1}{2}}$ in the Galactic plane and $z$ becomes the axial cylindrical coordinate perpendicular to the Galactic plane. The coordinate transform
\begin{eqnarray}
r^{2}\equiv r'^{2}+\frac{z^{2}}{q_{\Phi}^{2}}
\label{eqn:R}
\end{eqnarray} 
with $q_{\Phi}<1$ produces a potential flattened in the $z$--direction, yielding forces that point towards the $z=0$ plane. The best--fit value of $q_{\Phi}=0.63$ was then extracted from a formal fit to the orbital parameters of the observed Galactic Centre clouds derived from the NH$_3$(1,1) emission from the  HOPS survey \citep[][]{2012MNRAS.426.1972P}, taking care that such a potential corresponds to a physically meaningful density distribution.\\
\indent For the purposes of these simulations, we impose external forces on all SPH and sink particles appropriate to such an external potential. We compute for every particle at every timestep the enclosed mass $M(r)$ due to the stellar mass distribution(which we do not model explicitly) at the particle's instantaneous galactocentric spherical radius $r$ by logarithmic interpolation of the enclosed mass inferred by \cite{2002A&A...384..112L}. Forces in the $x$--, $y$-- and $z$--directions are computed from\\
\begin{eqnarray}
F_{x}=-\frac{GM(r)}{r^{2}}\hat{\bf x},\\
F_{y}=-\frac{GM(r)}{r^{2}}\hat{\bf y},\\
F_{z}=-\frac{GM(r)}{r^{2}}\hat{\bf z},
\end{eqnarray}
with $r$ defined as in Equation \ref{eqn:R}, and the unit vectors 
\begin{eqnarray}
\hat{\bf x}\equiv\frac{x}{(x^{2}+y^{2}+z^{2})^{\frac{1}{2}}},\\
\hat{\bf y}\equiv\frac{y}{(x^{2}+y^{2}+z^{2})^{\frac{1}{2}}},\\
\hat{\bf z}\equiv\frac{z}{q_{\Phi}(x^{2}+y^{2}+z^{2})^{\frac{1}{2}}}.
\end{eqnarray}
The potential is centred on the location of Sgr A$^{*}$ at $(x_{\rm SgrA^{*}},y_{\rm SgrA^{*}},z_{\rm SgrA^{*}})=(8.08,0.00,-6.68)$\,pc. The assumption that the gravitational forces at a given radius are due only to mass internal to that radius is strictly correct only in spherical symmetry, but corrections due to the breaking of spherical symmetry are likely to be small owing to the central concentration of the mass distribution.\\
\begin{table*}
\begin{tabular}{|l|l|l|l|l|l|l|l|l|l|}
\hline
Run name & Short name &M(M$_{\odot}$)&$R_{0}$(pc)&$\sigma_{\rm 1D,0}$(km\,s$^{-1}$)&$\alpha_{\rm vir,0}$& $\rho_{0}$(M$_{\odot}$\,pc$^{-3}$) & $\Sigma_{0}$(M$_{\odot}$\,pc$^{-2}$) & t$_{\rm ff}$ (Myr)\\
\hline
\textit{Self--virialised clouds:} &&&&&&&&\\
Fiducial&Fid&773015&13.50&12.7&2.6&75.0&1350&0.93\\
Low density&LDens&1338901&23.38&12.7&2.6&25.0&779&1.62\\
High density&HDens&446300&7.79&12.7&2.6&225.0&2338&0.54\\
Low velocity dispersion&LVDis&96627&6.75&6.3&2.6&75.0&675&0.93\\
High velocity dispersion&HVDis&2608925&20.25&19.0&2.6&75.0&2025&0.93\\
Low virial ratio&LVir&4016704&23.38&12.7&0.87&75.0&2338&0.93\\
High virial ratio&HVir&148766&7.79&12.7&7.8&75.0&779&0.93\\
&&&&&&&&\\
\textit{Tidally--virialised clouds:} &&&&&&&\\
Fiducial&Fid&773015&13.50&24.1&9.4&75.0&1350&0.93\\
Low density&LDens&1338901&23.38&41.2&27.4&25.0&779&1.62\\
High density&HDens&446300&7.79&13.9&3.2&225.0&2338&0.54\\
Low velocity dispersion&LVDis&96627&6.75&12.1&9.4&75.0&675&0.93\\
High velocity dispersion&HVDis&2608925&20.25&36.2&9.4&75.0&2025&0.93\\
\end{tabular}
\caption{Initial properties of model clouds. Column 1: Run name; column 2: short run name; column 3; cloud mass in solar masses; column 4; cloud initial radius in pc; column 5: initial one--dimensional velocity dispersion in kilometres per second; column 6: initial virial parameter; column 7; initial density in solar masses per cubic parsec; column 8: initial surface density in solar masses per square parsec; column 9; initial freefall time in megayears.}
\label{tab:tidal_sims}
\end{table*}
\subsection{Self--virialised model clouds}
\indent We perform two very different sets of simulations. In the first set, which we refer to as `self--virialised', we construct traditional models of turbulent clouds in which the velocity dispersion is characterised by the ratio of the turbulent kinetic energy to the clouds' self--gravitational binding energies.\\
\indent Our Fiducial (Fid) cloud has a volume density of 75\,M$_{\odot}$\,pc$^{-3}$ ($\approx$1300\,cm$^{-3}$) and a surface density of 1350\,M$_{\odot}$\,pc$^{-2}$, representative of the CMZ clouds, and a virial parameter of 1.3. These specifications result in a mass of 7.73$\times10^{5}$\,M$_{\odot}$, a radius of 13.5\,pc and a one--dimensional velocity dispersion of 12.7\,km\,s$^{-1}$.\\
\indent In addition, we explore the parameter space of model clouds, varying their characteristics to determine which are best able to reproduce the observations of the CMZ. We construct clouds whose volume densities are respectively a factor of three lower and higher (LDens and HDens), clouds whose velocity dispersions are 50\% lower and higher (LVDis and HVDis), and clouds whose virial ratios are a factor of three lower and higher (LVir and HVir). These clouds also have attendant differences in mass and size, as detailed in the first section of Table \ref{tab:tidal_sims}. These are the natural characteristics to vary in models of self--gravitating clouds, and we explore the ability of the models to explain the properties of the CMZ clouds.\\
\subsection{Tidally--virialised model clouds}
\begin{figure}
\includegraphics[width=0.48\textwidth]{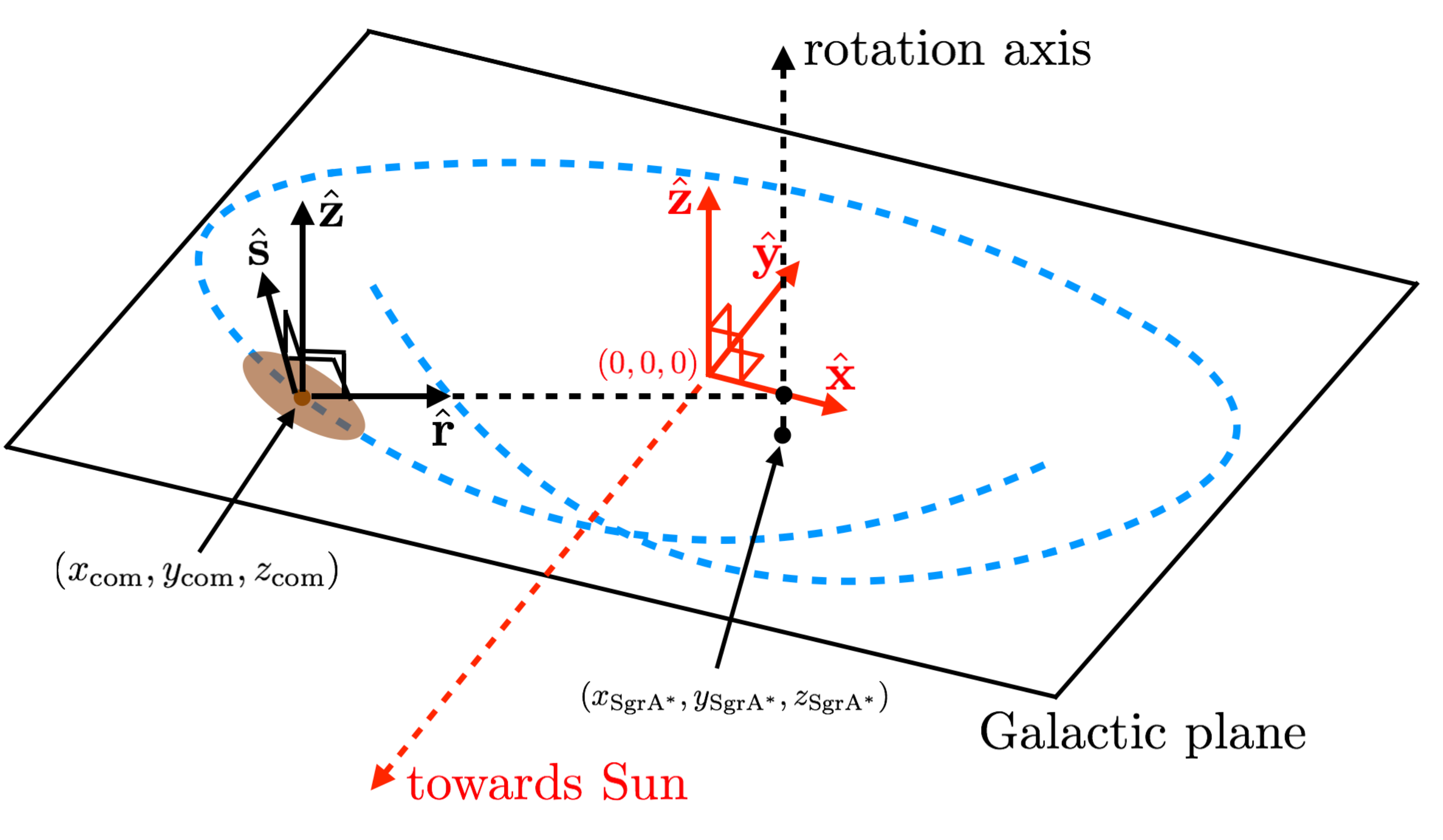}
\caption{Illustration of the two coordinate systems used in this paper. The $(\hat{\bf x},\hat{\bf y},\hat{\bf z})$ system, shown in red, is fixed and centred on the Galactic Centre. The $\hat{\bf x}$ direction is aligned with Galactic longitude, with increasing $x$ corresponding to decreasing longitude. The perpendicular $\hat{\bf y}$ vector points along the line--of--sight from the Sun to the Galactic Centre. The $\hat{\bf z}$ is mutually perpendicular to $\hat{\bf x}$ and $\hat{\bf y}$ and points towards the Galactic North Pole. Alternatively, the $(\hat{\bf r},\hat{\bf s},\hat{\bf z})$ coordinate system, shown in black, follows the clouds on their orbits, with $\hat{\bf r}$ pointing from the cloud centre of mass to the projection of the Sgr A$^{*}$'s position onto the Galactic plane. $\hat{\bf z}$ is common to both coordinate systems, and $\hat{\bf s}$ is mutually perpendicular to $\hat{\bf r}$ and $\hat{\bf z}$.}
\label{fig:coords}
\end{figure}

\indent In the second set of models, dubbed `tidally--virialised', we take a different approach. We abandon the traditional idea that the clouds' velocity dispersions are determined by an approximate equilibrium between the turbulent kinetic energy and the clouds' self--gravities. Instead, we contend that the strength of the turbulence is determined by processes external to the clouds, as postulated in the model of \cite{2015MNRAS.453..739K} and \cite{2017MNRAS.466.1213K}. We instead scale the clouds' internal velocity dispersions so the turbulent kinetic energy approximately balances the gravitational potential energy of the \textit{tidal field} within the volume of each cloud.\\
\indent The tidal forces experienced by the CMZ clouds can be examined in terms of the tidal tensor $T_{ij}$. For a gravitational potential $\phi({\bf r})$ as a function of a generic position vector ${\bf r}$ and generic coordinates $x_{i}$, $x_{j}$,
\begin{eqnarray}
T_{ij}({\bf r})=-\frac{\partial^{2}\phi({\bf r})}{\partial x_{i}\partial x_{j}}.
\end{eqnarray}
For the remainder of the paper, it becomes necessary to introduce two coordinate systems. The first Cartesian system is used to describe the properties of the clouds as seen from the Earth's perspective. The axes are fixed, with the origin at $(0, 0, 0)$ and, once more, Sgr A$^{*}$ located at $(8.08,0.00,-6.68)$\,pc. In this system, $\hat{\bf x}$ is a vector lying in the Galactic plane perpendicular to the line of sight from the Sun and in the opposite sense of Galactic longitude, $\hat{\bf y}$ points along the observational line of sight from the Sun to the Galactic Centre and $\hat{\bf z}$ is a mutually perpendicular vector pointing towards Galactic north. For brevity, we sometimes refer to $\hat{\bf z}$ as the `vertical' direction.\\
\indent We use the second coordinate system to describe the clouds in a more natural fashion accessible only via simulations. This system is cylindrical and local to each simulated cloud at each moment in time. The origin is the cloud's centre of mass. The vector $\hat{\bf r}$ joins the instantaneous cloud centre of mass $(x_{\rm com},y_{\rm com},z_{\rm com})$ perpendicularly to the rotation axis passing through the location of Sgr A$^{*}$, and is always parallel to the Galactic plane. $\hat{\bf z}$ is common with the fixed coordinate system, and $\hat{\bf s}$ is a vector perpendicular $\hat{\bf r}$ and $\hat{\bf z}$, parallel to the Galactic plane and pointing in the same sense of the cloud's orbital motion. Note that, for clouds following eccentric orbits in the background potential, $\hat{\bf s}$ is \textit{not} parallel to the instantaneous orbital velocity. We reiterate, however, that this coordinate system is defined only for convenience in describing the extents of the clouds. In the hope of avoiding confusion, Figure \ref{fig:coords} shows the relation of these two coordinate systems.\\
\indent The best known form of the tidal tensor, appropriate for a central point--mass is not relevant here. Firstly, our clouds reside in the potential generated by an extended mass distribution. We show in \cite{2015MNRAS.447.1059K} that the mass distribution inferred by \cite{2002A&A...384..112L} which we use to generate our potential is well--approximated in the Galactocentric radial range of interest here by $M(r)=Ar^{\alpha}$ with $A$ a constant and $\alpha=2.2$ (note however that the potential used in the simulations is generated using logarithmic interpolation of the raw \cite{2002A&A...384..112L} data and not from this approximation). In addition, the potential is flattened in the $z$--direction (see Equation \ref{eqn:R}). These considerations result in the non--zero components of the tidal tensor being
\begin{eqnarray}
T_{rr}=(2-\alpha)\frac{GM(r)}{r^{3}},\\
T_{ss}=-\frac{GM(r)}{r^{3}},\\
T_{zz}=-\frac{GM(r)}{q_{\phi}^{2}r^{3}}.
\end{eqnarray}
Note that, with $\alpha>2$, \textit{all} components of the tidal tensor are \textit{compressive}, unlike in the familiar Keplerian case.\\
\indent The initial tidal \textit{forces} can be computed from the tidal tensor by multiplying each component by the mass of the cloud and by its extension in each coordinate direction:
\begin{eqnarray}
\label{eqn:tforces1}
F^{\rm tidal}_{rr}=(2-\alpha)\frac{GM(r)}{r^{3}}\times2R_{0}M_{\rm cloud},\\
F^{\rm tidal}_{ss}=-\frac{GM(r)}{r^{3}}\times2R_{0}M_{\rm cloud},\\
\label{eqn:tforces2}
F^{\rm tidal}_{zz}=-\frac{GM(r)}{q_{\phi}^{2}r^{3}}\times2R_{0}M_{\rm cloud},
\label{eqn:tforces3}
\end{eqnarray}
where we have made use of the fact that the clouds are initially spherical so that their extents in the $r$, $s$ and $z$ coordinates are equal at the beginning of the simulation.\\
\indent We import the results of Equations \ref{eqn:tforces1}--\ref{eqn:tforces3} and multiply each tidal force component by the initial cloud radius $R_{0}$ to yield an equivalent energy. This energy is the work that would be done by the tidal field in compressing the clouds to zero thickness in each direction.  Stabilising the clouds in each direction therefore requires that the energy stored in the relevant component of the turbulent velocity field is at least as large. We therefore define the energies of the three turbulent velocity components required to achieve this by:
\begin{eqnarray}
E_{\rm tidal}\{r,s,z\}=\{0.2,1.0,2.6\}\frac{GM(r_{0})}{r_{0}^{3}}R_{0}^{2}M_{\rm cloud}\\
=\frac{1}{2}M_{\rm cloud}\sigma_{\rm 1D,0\{r,s,z\}}^{2},
\end{eqnarray}
where the enclosed mass $M(r_{0})$ is that of the old stellar population (which generates the background potential) enclosed by the location on the orbit where the simulations are started. We choose a location such that $r_{0}=90.3$\,pc, resulting in $M(r_{0})=5.01\times10^{8}$\,M$_{\odot}$. This generates three different one--dimensional velocity dispersions for each cloud, dependent on their mass and radius, with the initial radial component being the smallest, the vertical component the largest, and the tangential component intermediate between the other two in the ratio 0.2:1.0:2.6. These velocity dispersions are essentially what would be required to support the clouds against each component of the tidal field.\footnote{In \cite{kruijssen18} we discuss the evolution of the clouds in the fully--compressive external tidal field mainly by comparing the relative degree of compression felt by the clouds in different spatial directions at any given time. Here, we focus more on explaining the dynamics of the clouds, often using quantities such as the gravitational potential or the velocity dispersion perpendicular to the line of the sight, which are not observable. We therefore instead make use of the terms `over--supported' and `under--supported' to describe the dynamical states of the clouds, even when establishing observationally whether a cloud is in such a state may be difficult or impossible. We note, however, that being `under--supported' in a given direction means that the cloud is being tidally compressed in that direction.}\\
\indent It is not physically appropriate to set the initial radially-- and tangentially--pointing velocity components of the gas particles to be different (the computed values typically differ by a factor a little larger than two), since the clouds are not in a tidally--locked state. We choose instead to give the clouds an initially \textit{isotropic} velocity field whose one--dimensional velocity dispersion is the geometric average of the three computed tidal components. The clouds are therefore oversupported in the initially--radial direction, undersupported in the vertical direction, and roughly in equilibrium in the initially--tangential direction. As might be expected, the velocity dispersions and formal virial parameters of these models are substantially larger than are observed in typical molecular clouds, and are certainly larger than those required to support the model clouds only against their own self--gravity. We detail the properties of the tidally--virialised clouds in the second section of Table \ref{tab:tidal_sims}. Note that we do not repeat the LVir and HVir simulations in this fashion, since we are now asserting that the virial ratio is fixed by influences external to the cloud. For completeness, we also perform one calculation, described in the Appendix, with an anisotropic velocity field where the velocity dispersions are initially set to the values determined directly from the instantaneous tidal force components at the clouds' initial location.\\
\indent Note that the above process results in all clouds having the same turbulent crossing time, since\\
\begin{eqnarray}
\tau_{\rm cross}\sim\frac{R_{0}}{\sigma_{\rm 1D,0}}
\end{eqnarray}
and the velocity dispersions computed above are proportional to $R_{0}$. All tidally virialised clouds have turbulent crossing times of 0.55\,Myr. This has the obvious corollary that the linewidth--size relation for these clouds $\sigma=Ar^{b}$ has $b=1$, considerably steeper than the canonical Larson value of $b=0.5$ observed in typical Milky Way GMCs. \cite{2012MNRAS.425..720S} report that the size--linewidth relation in the CMZ is indeed steeper than is seen in the Galactic disk, deriving a best--fit value of $b=0.78$. This is evidently not as steep as the value we propose here, but it does demonstrate that the CMZ clouds are not in virial equilibrium. We aim to investigate this interesting aspect of the CMZ clouds in a future paper.\\
\subsection{Cloud angular momentum}
\indent Velocity shear is expected to impart on inspiralling gas in the Galactic Centre a spin angular velocity in the $\hat{\bf z}$ direction in the opposite sense to the orbital angular velocity. The initial spin angular velocity $\omega_z^{\rm shear,0}$ in question is related to the gradient of the orbital angular velocity $\Omega$ at the clouds' initial locations by $\omega_z^{\rm shear,0}=-r_{0}{\rm d}\Omega/{\rm d}r|_{r_{0}}=[(3-\alpha)/2]\Omega(r_{0})$. At the circular radius $r_{0}=90.3$\,pc from which our clouds are launched, we compute numerically $\omega_z^{\rm shear,0}=2.2\times10^{-14}$\,s$^{-1}=0.7$\,Myr$^{-1}$.\\
\indent Our model clouds are all initialised with a purely solenoidal (i.e. divergence--free) turbulent velocity field. The field has non--zero angular momentum about all three principal axes. The value of the angular momentum and consequent angular velocity $\omega_z^{\rm spin,0}$ is determined by the size of the cloud and the velocity field and is not trivial to control. We compute the spin angular momenta $S_{z,0}$ of the clouds' initial states in their centre--of--mass frames as
\begin{eqnarray}
S_{z,0}=\sum_{i}^{N_{\rm SPH}}m_{i}({\bf v_{xy,i}\times r_{xy,i}}),
\end{eqnarray}
the moment of inertia about the $\hat{\bf z}$ axis $I_{z,0}$ as
\begin{eqnarray}
I_{z,0}=\sum_{i}^{N_{\rm SPH}}m_{i}r_{xy,i}^{2},
\end{eqnarray}
yielding 
\begin{eqnarray}
\omega_z^{\rm spin,0}=\frac{S_{z,0}}{I_{z,0}}.
\end{eqnarray}
\indent The self--virialised clouds have values of $\omega_z^{\rm spin,0}$ ranging from $6.8\times10^{-15}-2.0\times10^{-14}$\,s$^{-1}$ ($0.21-0.63$\,Myr$^{-1}$), considerably lower than the angular velocities implied by the potential. In contrast, the tidally--virialised clouds, since they obey a linear size--linewidth relation, all have the same value of $\omega_z^{\rm spin,0}=3.1\times10^{-14}$\,s$^{-1}$ ($0.98$\,Myr$^{-1}$), close to the value computed from the velocity shear of the potential.  We have therefore selected tidally--virialised cloud models with spin angular velocities in close agreement with observed values.\\
\indent During the evolution of the clouds, tidal shear induces an exchange of orbital and spin angular momentum, which moves the clouds from a counter--rotating to a corotating state after about half an orbit (in other words, roughly by the time they reach the position of Sgr B2). A more detailed discussion of the angular momentum evolution of the clouds is given in \cite{kruijssen18}.\\
\subsection{Orbits around the Galactic Centre}
\indent For all self--virialised and tidally--virialised model clouds, two simulations are run in the external potential described above. `Eccentric' calculations refer to those where the cloud is placed on the accurately--determined Galactocentric orbit on which the CMZ clouds lie from \cite{2015MNRAS.447.1059K}. The initial location of the clouds is chosen to be halfway between the orbital apocentre at negative Galactic longitude and the pericentre. The initial positions and velocities of the of the clouds on the inferred orbit is given in the first row of Table \ref{tab:orbits}. The initial separation from the centre of the potential is $r_{0}=90.3$\,pc.\\
\indent `Circular' simulations place the cloud on a circular orbit in the same potential such that the initial coordinates of the cloud in the $xy$--plane $(x_{0},y_{0})$ are the same as those of the eccentric simulations, but with a $z$--coordinate of $z_{\rm SgrA^{*}}$, a $z$--velocity of zero and $x$-- and $y$--velocities such that the cloud executes a circular orbit about the location of Sgr A$^{*}$ parallel to (but, due to the vertical offset of Sgr A*, below) the Galactic plane (see the second row of Table \ref{tab:orbits}). The purpose of the circular runs is to explore the effect of the changing tidal forces the clouds experience as they follow the eccentric orbit.\\
\begin{table*}
\begin{tabular}{l|l|l|l|l|l|l|}
Orbit & x$_{0}$(pc) & y$_{0}$(pc) & z$_{0}$(pc) & v$_{x,0}$(km\,s$^{-1}$) & v$_{y,0}$(km\,s$^{-1}$) & v$_{z,0}$(km\,s$^{-1}$)\\
\hline
Eccentric & 96.63 & -17.32 & -10.63 & -115.62 & -113.98 & 32.42\\
Circular & 96.63 & -17.32 & -6.68 & -29.62 & -151.45 & 0.00\\
\end{tabular}
\caption{Initial positions and velocities of orbiting clouds. Note that the potential is centred on Sgr A* at $(x_{\rm SgrA^{*}},y_{\rm SgrA^{*}},z_{\rm SgrA^{*}})=(8.08,0.00,-6.68)$\,pc.}
\label{tab:orbits}
\end{table*}
\indent Finally, we run `Isolated' simulations of the self--virialised clouds, in the absence of the tidal field or any bulk cloud motion, to allow us to compare the models with more traditional molecular cloud simulations. Clearly, this is not a sensible set of simulations for the tidally-virialised clouds, which would undergo catastrophic expansion due to the absence of a compressive tidal field that balances their internal motion.\\
\indent Simulations by \cite{2015MNRAS.451.3679B} investigated the possibility of explaining the low star--formation rates/efficiencies in the CMZ by modelling isolated molecular clouds with a range of virial ratios, going up to $\alpha_{\rm vir}=16$. They concluded that increasing the virial ratio did indeed suppress star formation. However, since their models had no confining tidal field, their model clouds with higher virial parameters expanded to large sizes, inconsistent with the observed CMZ clouds. With the models presented here, the feasibility of suppressing star formation with high virial ratios can be reassessed thanks to the presence of the compressive tidal field.\\
\section{Results}
\subsection{Self--virialised clouds}
\subsubsection{Star formation efficiencies and rates}
\begin{figure*}
\captionsetup[subfigure]{labelformat=empty}
\centering
\subfloat[]{\includegraphics[width=0.32\textwidth]{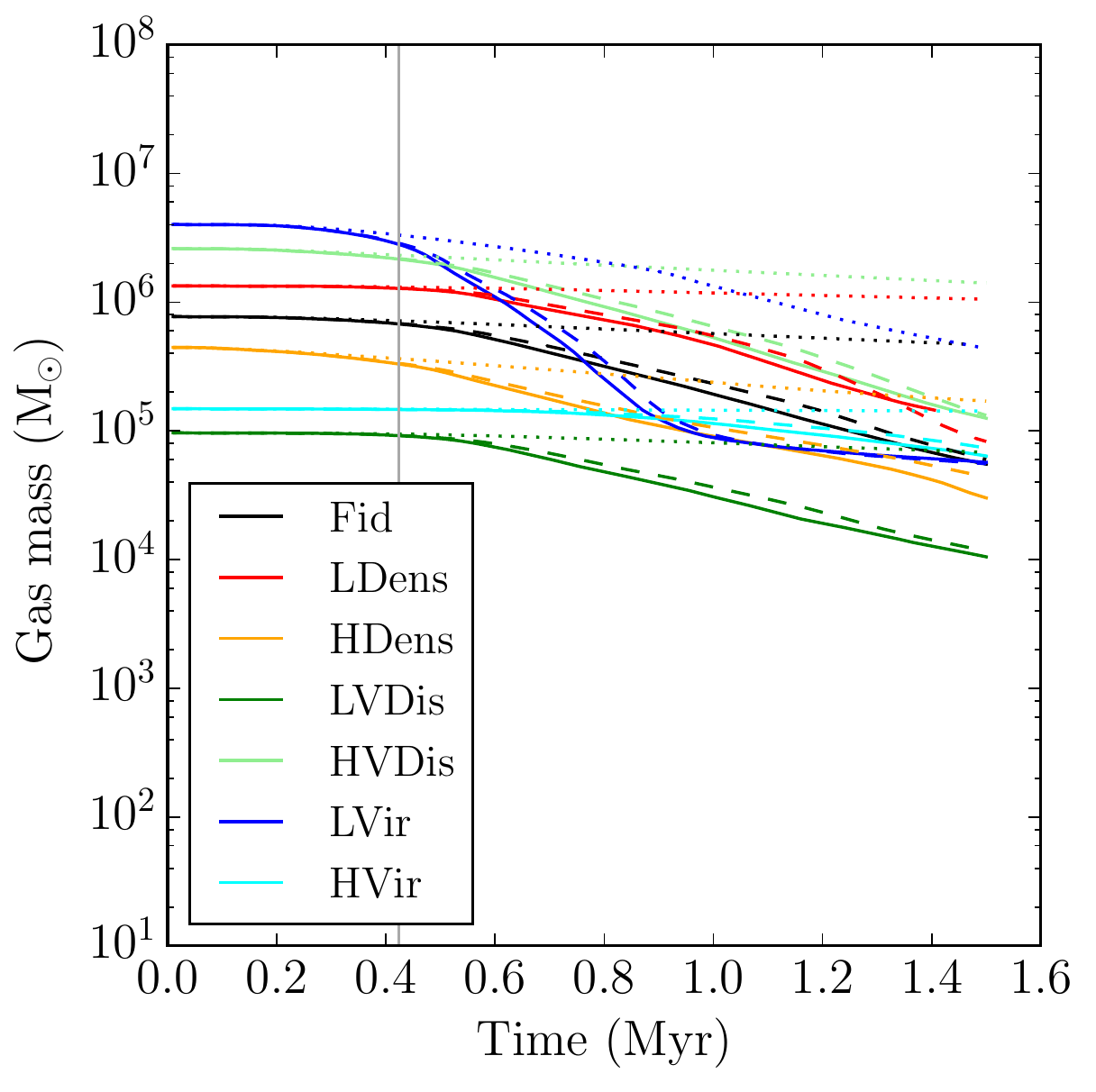}}     
     \hspace{.1in}
\subfloat[]{\includegraphics[width=0.32\textwidth]{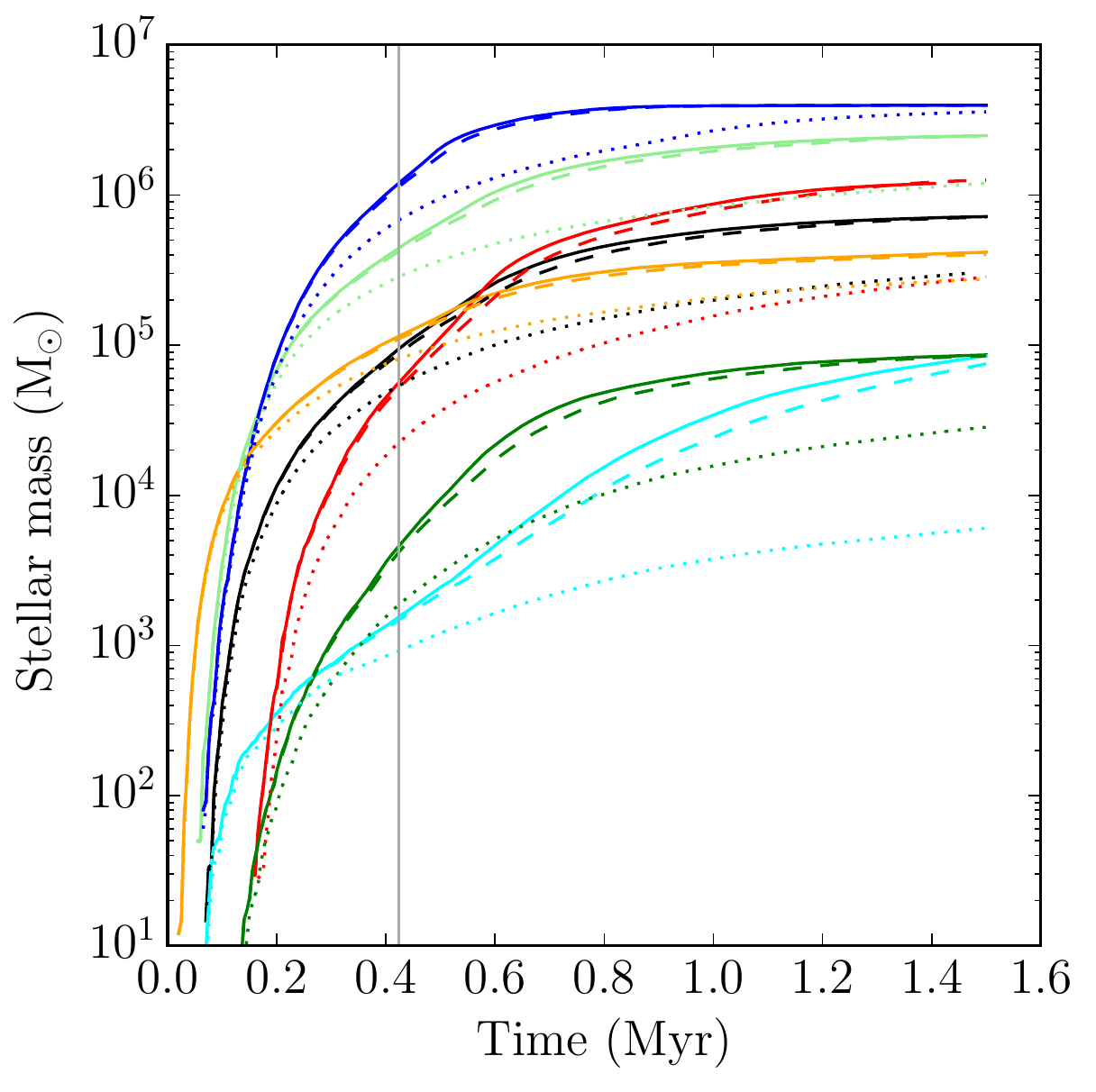}}
     \hspace{.1in}   
\subfloat[]{\includegraphics[width=0.32\textwidth]{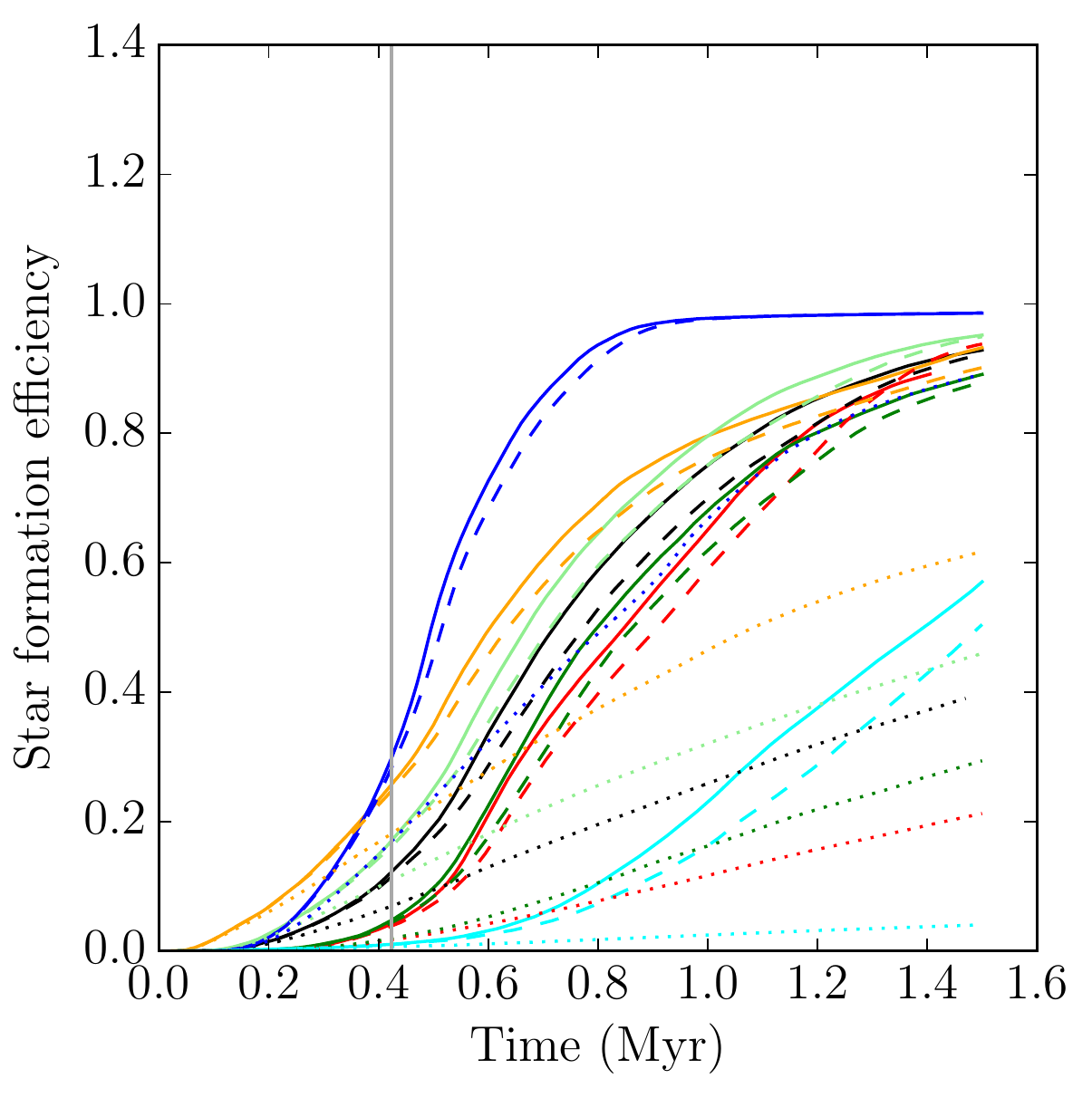}}
     \vspace{-.3in}   
\subfloat[]{\includegraphics[width=0.32\textwidth]{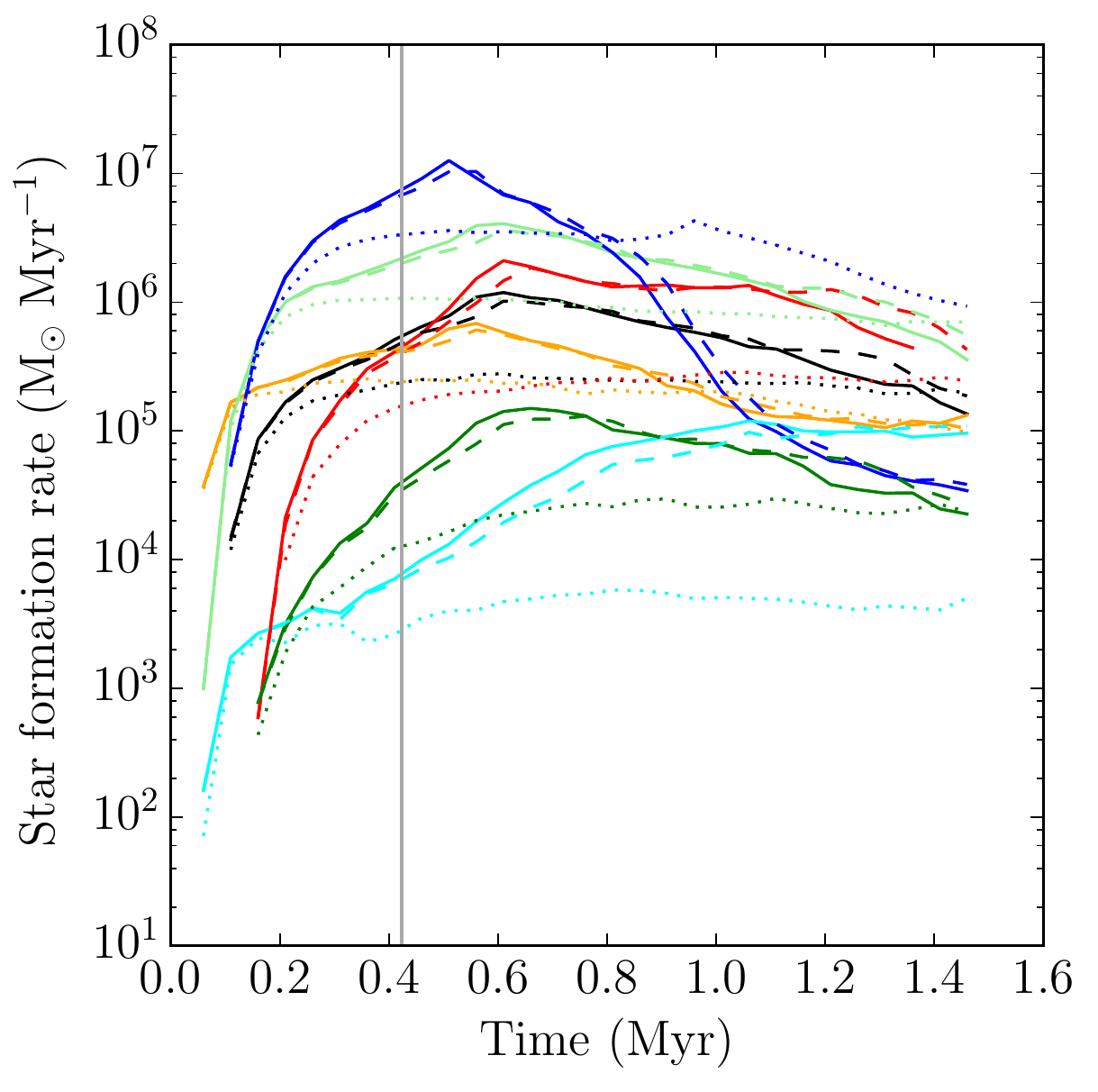}}     
     \hspace{.1in}
\subfloat[]{\includegraphics[width=0.32\textwidth]{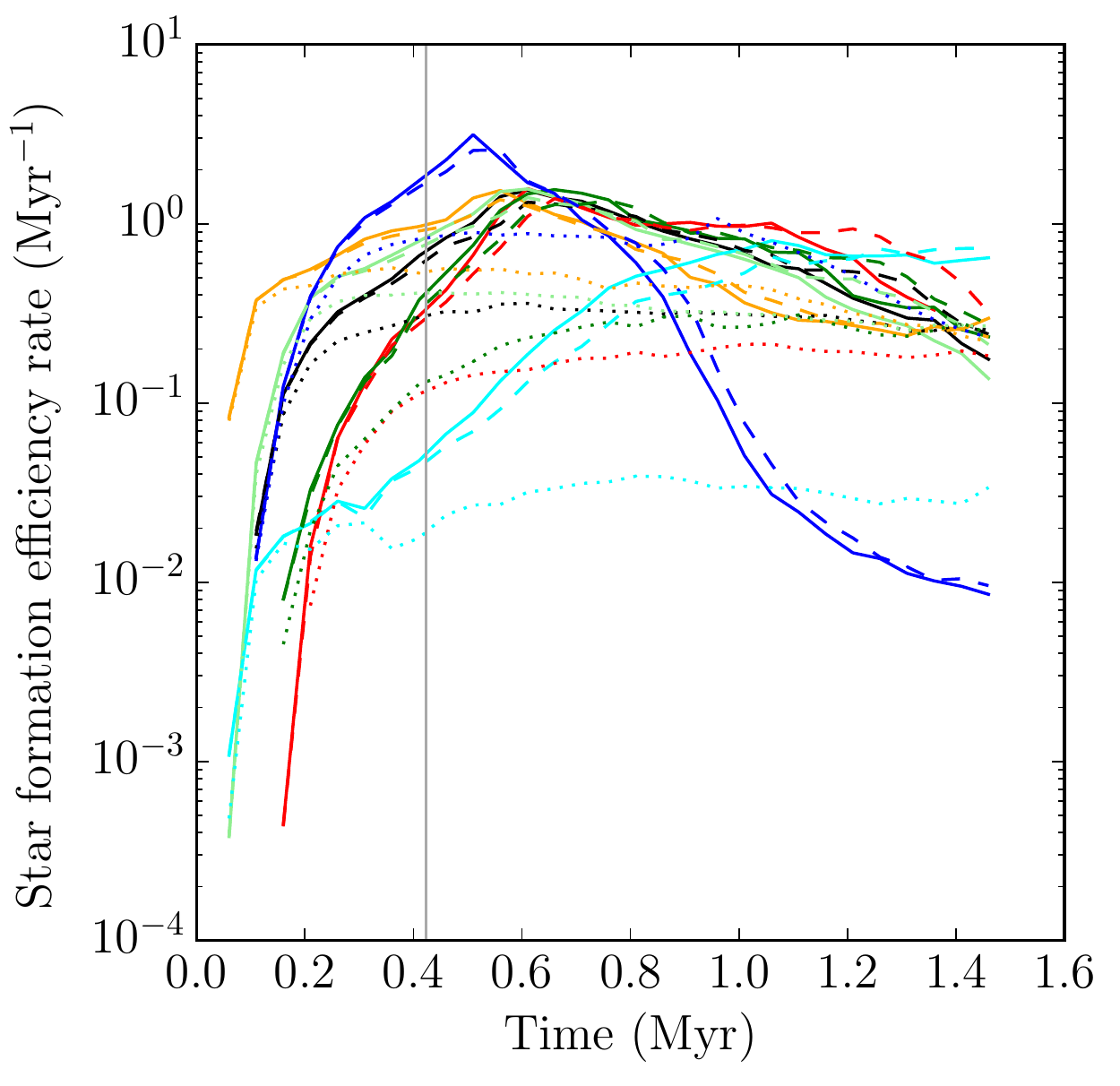}}
     \hspace{.1in}   
\subfloat[]{\includegraphics[width=0.32\textwidth]{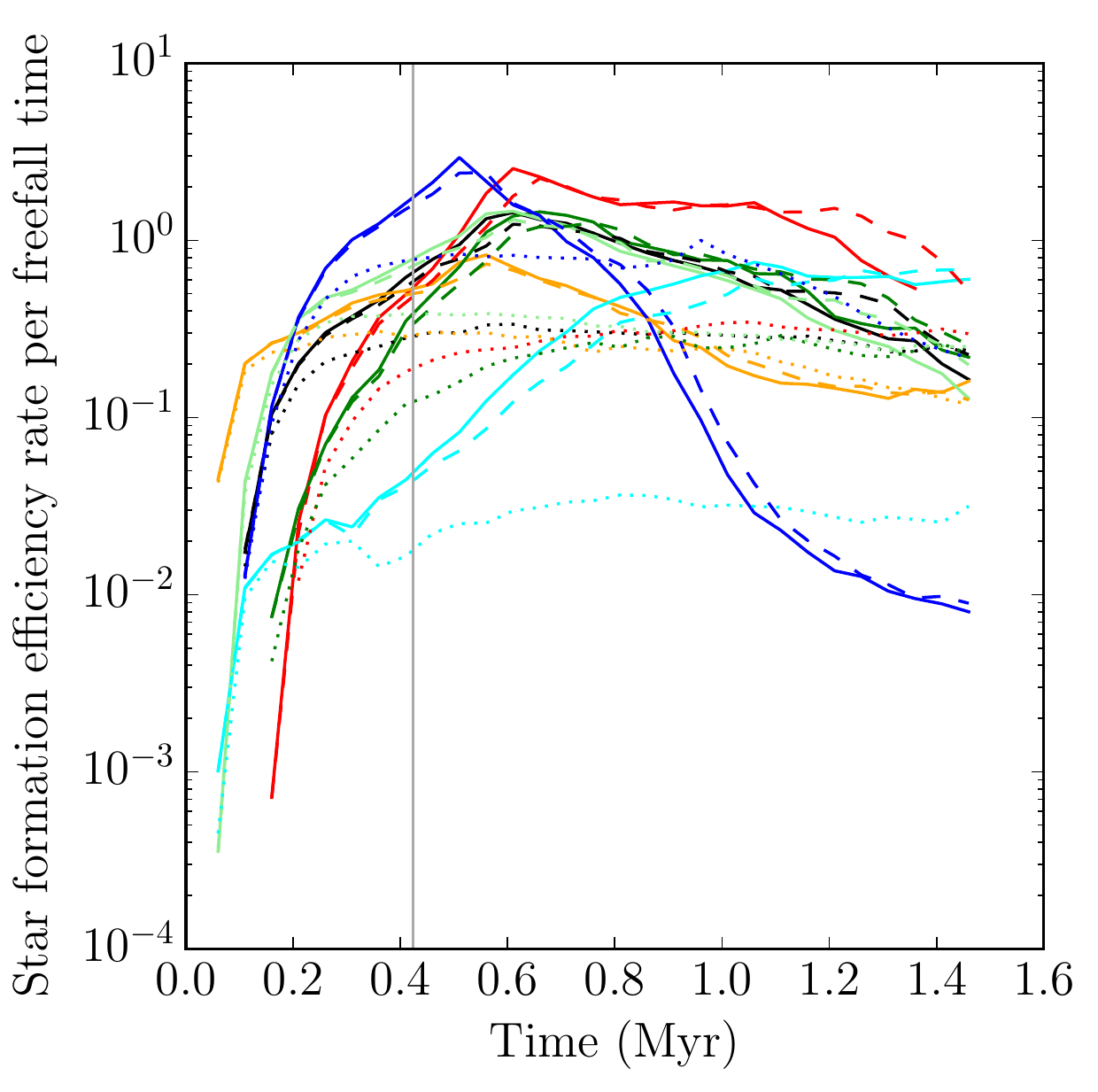}}
\caption{Time evolution of several quantities describing all of the self-virialised simulations. The panels show the gas mass (top left), stellar mass (top centre), star formation efficiency (top right), absolute star formation rate (bottom left), star formation efficiency rate (bottom centre), star formation efficiency rate per free-fall time (bottom right). In all panels, solid lines are eccentric simulations, dashed lines are circular simulations, dotted lines are isolated simulations, the colours correspond to the different model clouds (as indicated by the legend in the first panel), and the vertical grey line marks the time of pericentre passage at 0.41\,Myr. This figure demonstrates rapid star formation in the self--virialised clouds.}
\label{fig:traditional_sf_rates}
\end{figure*}
\begin{figure*}
\captionsetup[subfigure]{labelformat=empty}
\centering
\subfloat[]{\includegraphics[width=0.48\textwidth]{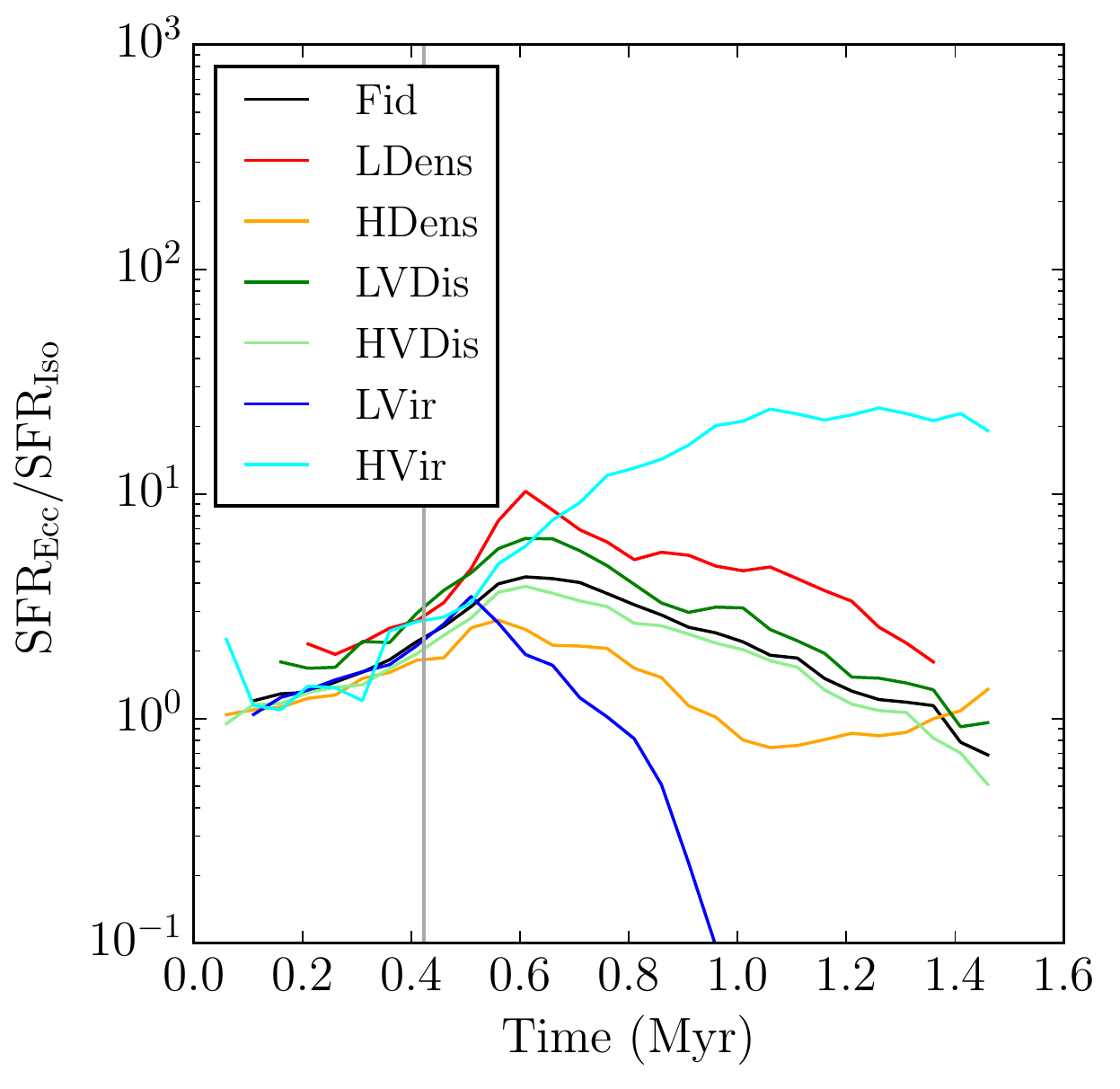}}
\hspace{-.1in}
\subfloat[]{\includegraphics[width=0.48\textwidth]{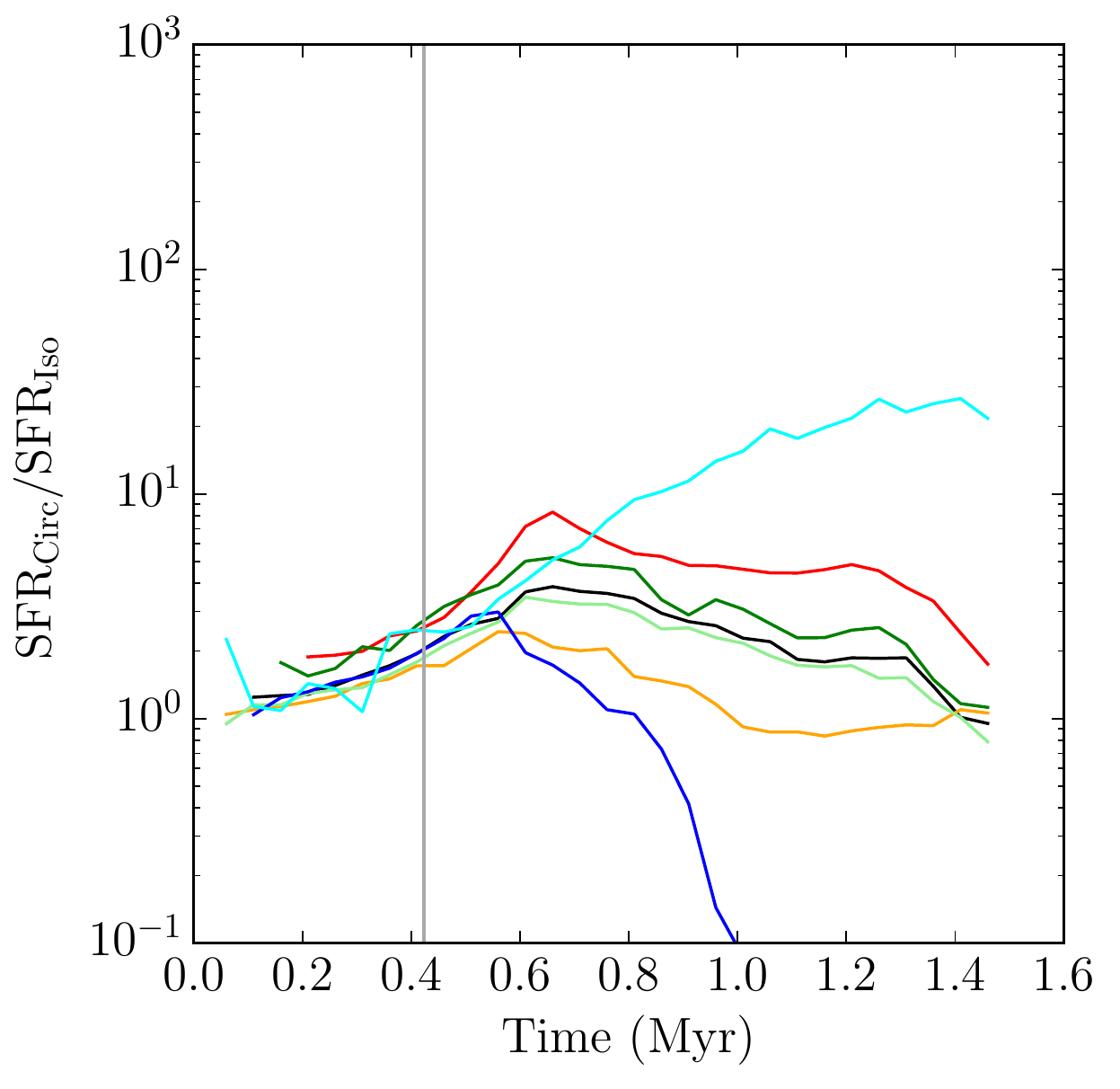}}
\caption{Ratio as a function of time of the star formation rates in the self--virialised eccentric and isolated (left panel) and circular and isolated (right panel) simulations, showing that star formation in both sets of tidally--influenced clouds is substantially faster than in the corresponding isolated clouds, until the tidally--influenced clouds become deprived of gas. The colours denote the different model clouds, as indicated by the legend in the left panel, and the vertical grey line represents pericentre passage in the eccentric orbit at 0.41\,Myr.}
\label{fig:ecc_over_iso}
\end{figure*}

\indent All simulations are permitted to form sink particles to replace locally--gravitationally unstable pockets of gas. We do not have sufficient resolution to track the formation of individual stars, but we use the sink particles to gauge an upper limit to the amount of gas likely to be involved in star formation at a given time in each model. We can thus compute upper limits to expected star formation rates and efficiencies as the clouds travel along their orbits.\\
\indent In Figure \ref{fig:traditional_sf_rates}, we show the stellar mass, gas mass, star formation efficiency, and three different measures of the star formation rate as functions of time in the self--virialised clouds. Since the clouds have a wide range of masses, comparison of the absolute star formation rates may be misleading, and we compute the star formation efficiency rate (SFER -- the rate of change of the star formation efficiency in absolute time units) and the star formation efficiency rate per initial freefall time (SFER$_{\rm ff}$). Note that in all plots in this paper plotting quantities against time, the time of pericentre is shown by a vertical dark grey line.\\
\indent Looking at the isolated clouds first (dotted lines), the HDens and LVir simulations form stars at the highest rates. This is not surprising, since they have respectively the shortest initial freefall time, and the least support against gravitational collapse. The LVir calculation in particular is close to reaching gas exhaustion after $1.5$\,Myr. The Fid simulation achieves a star--formation efficiency of 0.4 over the same time period, and is bracketed by the HVDis (0.45) and LVDis (0.31) clouds. All three of these clouds have the same virial ratio and initial freefall time, but the higher turbulent velocities assist in creating locally unstable gas by generating stronger shocks. The LDens cloud has an initial freefall time longer than the duration of the simulation, and reaches an SFE of 0.21. Finally, the HVir cloud, which is strongly unbound, is the slowest star former, achieving an efficiency of 0.04. Clouds of this kind have been studied before, for example by \cite{2005MNRAS.359..809C}. Readers are reminded that these simulations are the closest in mindset to most molecular cloud simulations performed in the past.\\
\indent Turning to the orbiting clouds, we see that in all cases, these clouds form stars substantially faster than their isolated counterparts. We quantify this statement in Figure \ref{fig:ecc_over_iso}, where we show as functions of time the ratio of the star formation rates in the eccentric and isolated clouds, and the circular and isolated clouds.\\
\indent In both sets of orbiting clouds, the star formation rates exceed those in the corresponding isolated clouds by factors of a few up to more than an order of magnitude. For most of the clouds, the relative acceleration in star formation rate tails off after $\approx0.6$\,Myr, but this is due mainly to the fact that most of the orbiting clouds come close to gas exhaustion (${\rm SFE}>0.85$) by this time. The exception is the HVir cloud. In isolation, this cloud has a persistently very low star formation rate. On both of the orbits, the HVir cloud is induced to form stars more than order of magnitude faster for the latter $\approx0.8$\,Myr of the simulations, but even this increase in star formation rate is not enough to exhaust the clouds' gas reserves after 1.5\,Myr.\\
\indent The external tidal field on both the eccentric and circular orbit drives all but the HVir cloud, which has the strongest internal turbulent support, to gas exhaustion by $1.5$\,Myr, about one half of a Galactocentric orbit. Most of the orbiting clouds achieve star--formation efficiency rates per freefall time greater than unity at some point in their evolution, and some of them do so for extended periods of time.  The isolated clouds, except for the strongly--bound and unstable LVir cloud, achieve maximum efficiency rates per freefall time of a few tens of percent. This demonstrates that \textit{self--gravity is not the main driver of star formation in the orbiting clouds and that collapse is strongly accelerated by the external tidal forces, regardless of the detailed shape of the orbit. It is the tidal forces which overwhelm the clouds' turbulent support and initiate collapse, until self--gravitational forces acting on small scales are able to drive the formation of stellar mass.}\\
\begin{figure}
\centering
\includegraphics[width=0.48\textwidth]{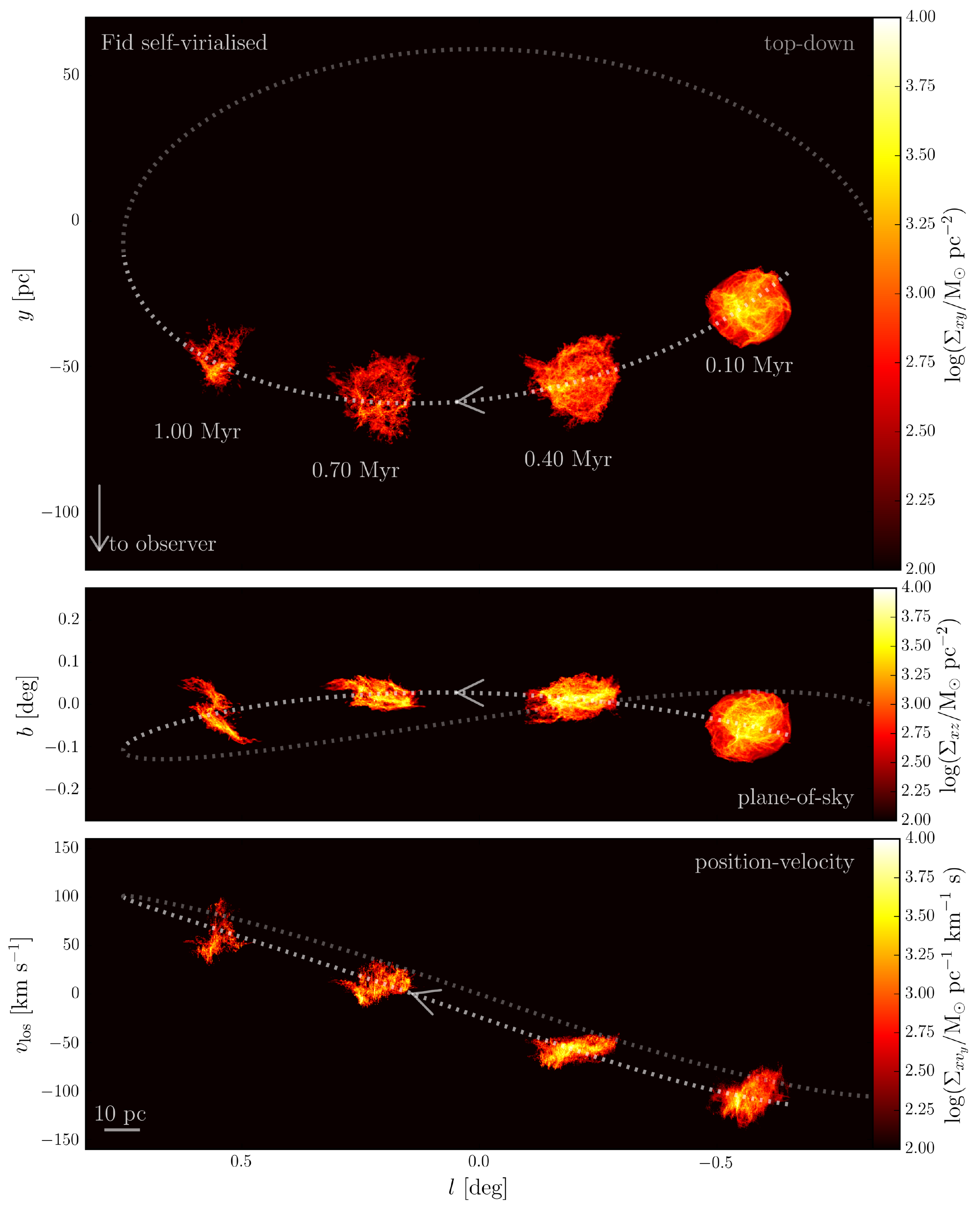}
\caption{Position--position and position--velocity renders of the self--virialised Fiducial simulation at four timesteps. All images are logarithmically--scaled. Dashed lines represent the orbit fitted by . Top panel: Top--down position--position view (along the $z$--axis) with an arrow indicating the direction of the Sun. Middle panel: Line--of--sight position--position view (along the $y$--axis). Bottom panel: Line--of--sight position--velocity view.}
\label{fig:sim_PP_PV_trad}
\end{figure}
\indent Most of the self-virialised clouds on an eccentric or circular orbit around the Galactic centre become gas--exhausted after about half an orbit. In Figure \ref{fig:sim_PP_PV_trad}, we show position--position images viewed along the $z$--axis and the $y$--axis (upper panels) and a position--velocity image viewed along the $y$--axis (lower panel) of the self--virialised Fiducial cloud on the eccentric orbit (indicated by the dashed line), shown at four different times and locations. For clarity, sink particles are omitted from these images, so that we may concentrate on the gas. The cloud clearly becomes very strongly flattened in the vertical direction and it also contracts to a lesser extent in the radial and tangential directions due to the fully--compressive nature of the tidal field. The cloud evidently begins to rotate clockwise about the $y$--axis, acquiring a very pronounced tilt between its projected long axis and the orbit by $1.0$\,Myr. The position--velocity plot initially becomes simpler as the turbulent velocity field with which the cloud is seeded is restructured by the tidal field, so that the form of the position--velocity distribution shows counter---rotation by $0.4$\,Myr. However, by later times, the position--velocity plot becomes dominated by features due to localised cloud collapse. Another striking change visible is the apparent fading of the cloud due to gas exhaustion.\\
\subsubsection{Response of the self--virialised clouds to the tidal field}
\begin{figure}
\includegraphics[width=0.48\textwidth]{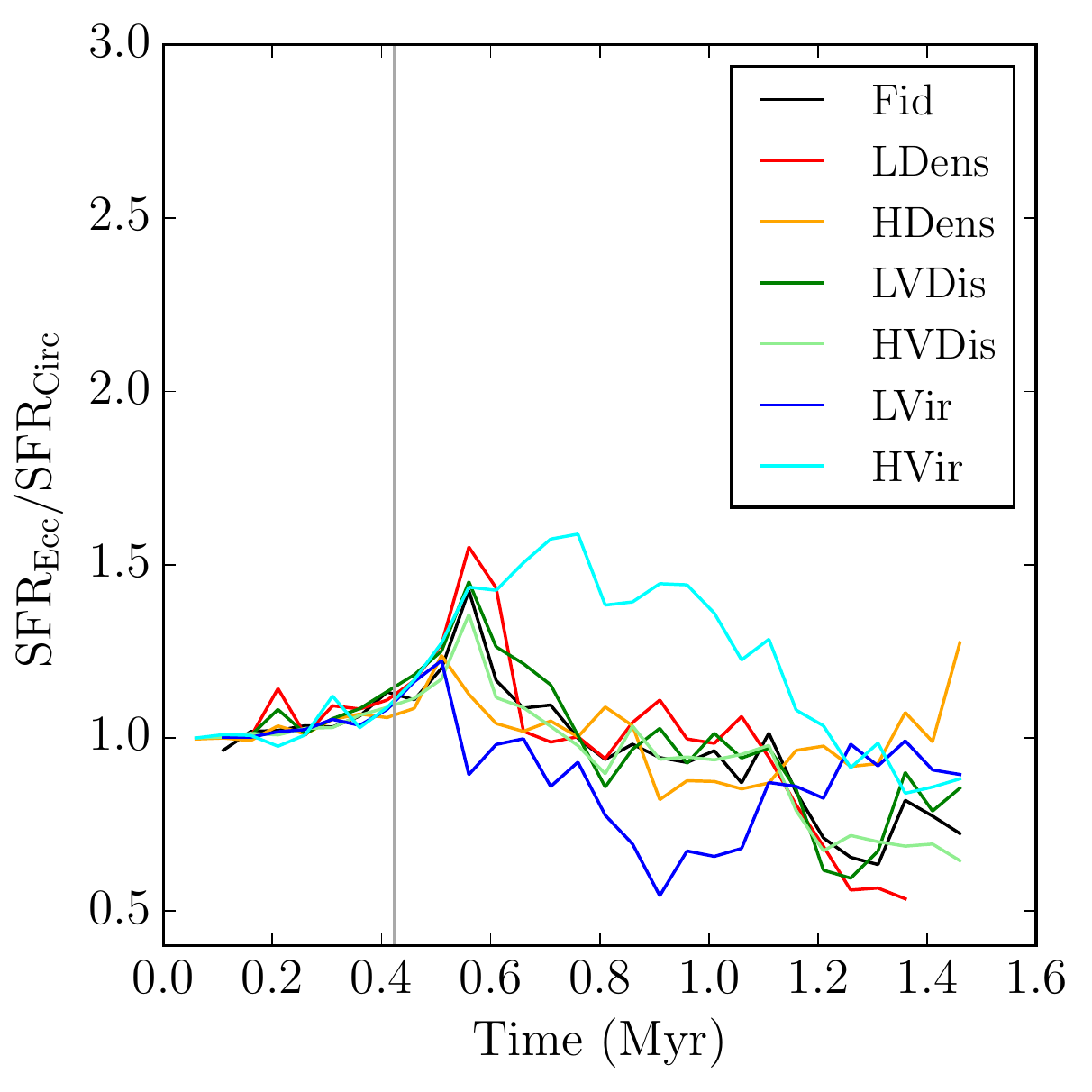}
\caption{Time evolution of the star formation rates of the self-virialised clouds on eccentric relative to those on circular orbits, demonstrating only modest differences. The colours denote the different model clouds, as indicated by the legend, and the vertical grey line represents pericentre passage in the eccentric orbit at 0.41\,Myr..}
\label{fig:sfr_ecc_over_circ}
\end{figure}

\begin{figure}
\includegraphics[width=0.48\textwidth]{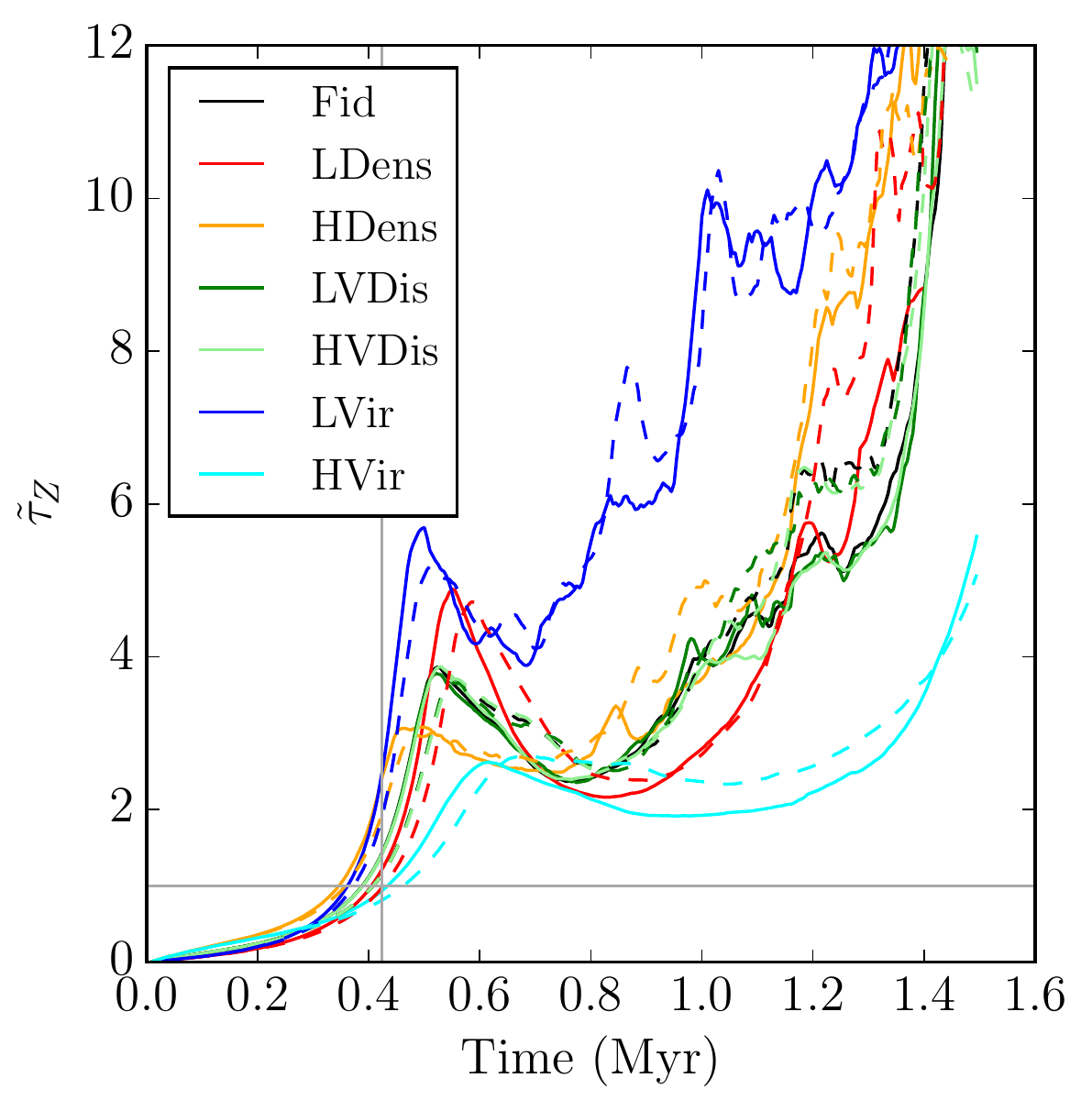}
\caption{Absolute time--evolution of the normalised vertical crossing time defined in the text in the self--virialised clouds. Clouds on eccentric orbits are represented by solid lines and those on circular orbits by dashed lines. The colours denote the different model clouds, as indicated by the legend in the left panel, and the vertical grey line represents pericentre passage in the eccentric orbit at 0.41\,Myr, while the horizontal grey line denotes a normalised crossing time of unity.}
\label{fig:delsigma}
\end{figure}

\begin{figure}
\includegraphics[width=0.48\textwidth]{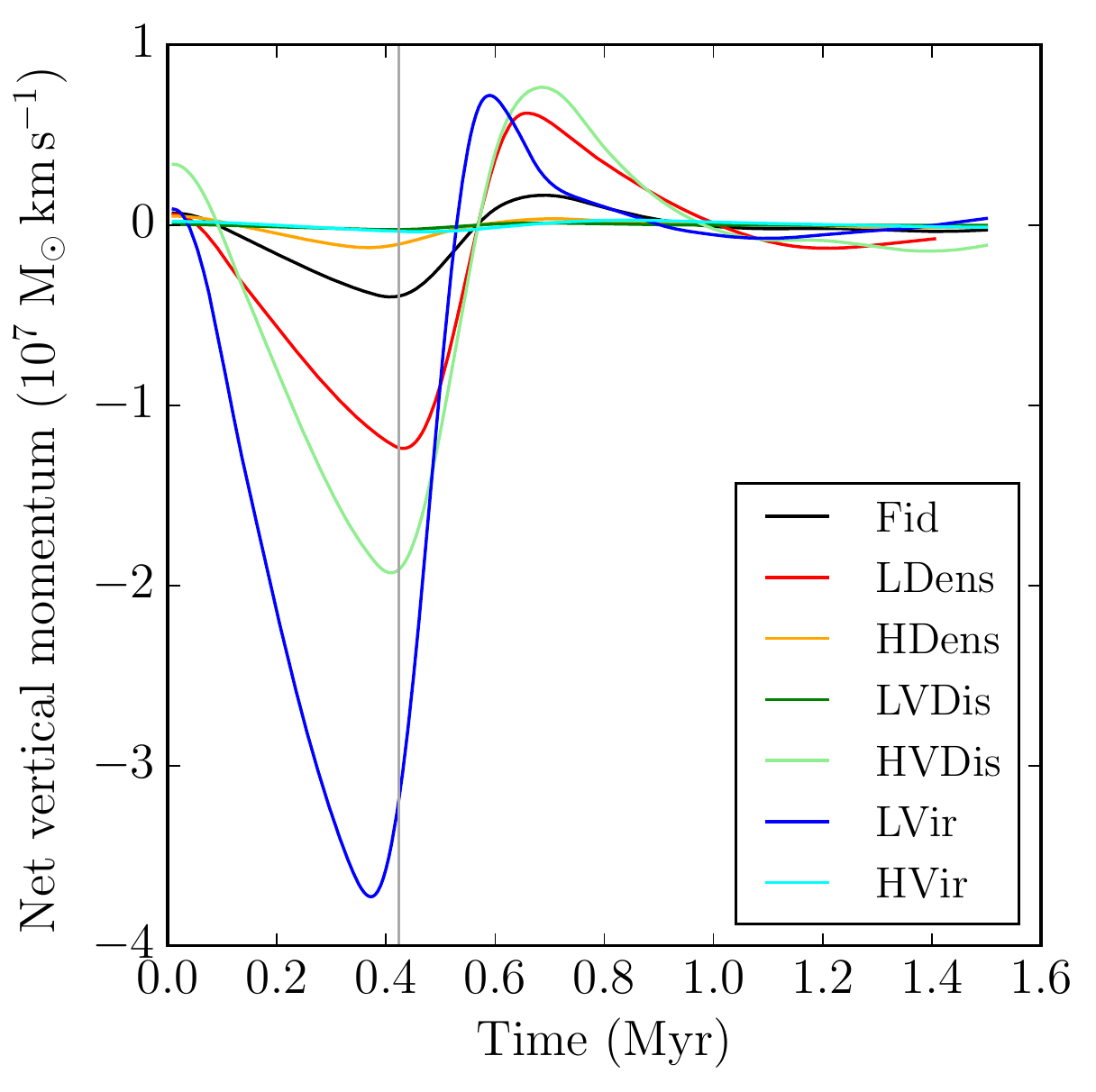}
\caption{Net momentum in the vertical direction (defined such that negative values indicate vertical contraction and positive values indicate vertical expansion) in all eccentric self--virialised simulations. The colours denote the different model clouds, as indicated by the legend in the left panel, and the vertical grey line represents pericentre passage in the eccentric orbit at 0.41\,Myr.}
\label{fig:self_mass_flux}
\end{figure}

\indent Readers are reminded that the coordinate system used in this work differs of necessity from that used in \citet{kruijssen18}. The latter paper concentrates on the observable properties of the model clouds dictated by our perspective on the Galactic Centre. This obliges the use of a Cartesian coordinate system defined by the vector joining the Sun to the Galactic Centre (equivalent to $\hat{\bf y}$ in our simulations), Galactic longitude (equivalent to $-\hat{\bf x}$) and Galactic latitude (equivalent to $\hat{\bf z}$). In this paper by contrast, we are effectively using a polar coordinate system.\\
\indent The evolution of the self--virialised clouds on eccentric and circular orbits is remarkably similar. In particular, \textit{both} sets of clouds appear to undergo an acceleration in their star formation rate on a timescale close to, but slightly longer than, that on which the eccentric clouds pass through pericentre. This is curious, since this timescale has no meaning for the clouds on the circular orbits.\\
\indent In Figure \ref{fig:sfr_ecc_over_circ}, we show the ratio as a function of time of the star formation rate in the eccentric and circular model clouds. We see that, from the beginning of the simulations, the clouds on eccentric orbits form stars slightly more rapidly than those on circular orbits. Both the clouds on eccentric and circular orbits experience very strong compression in the $z$--direction owing to their velocity fields being insufficient to resist the vertical tidal field. From the beginning of the simulations, the clouds on eccentric orbits experience stronger tidal fields, particularly in the $z$--direction, as they approach the centre of the potential. This leads to greater compression and more rapid star formation. The discrepancy continues to increase, reaching maximum values of $\approx1.5$ in most simulations, until shortly after the eccentric clouds pass through pericentre. After this point, the star formation rate in the HVir cloud remains elevated in the eccentric model for $\approx0.5$\,Myr. This is likely because this cloud is the least efficient at forming stars and therefore the most responsive to tidally--induced collapse. By contrast, the LVir calculation, which is already forming stars vigorously by the time of pericentre passage, is the least responsive to tidal compression. After $\approx0.8$\,Myr, both circular and eccentric clouds are close to gas exhaustion and the relative star--formation rates after this point are largely stochastic. The other simulations rapidly return to a state where the eccentric and circular clouds are forming stars at very similar rates.\\
\indent It is clear from comparing the isolated and orbiting clouds that the evolution of the latter models is dominated by tidal compression, which is strongest in the $z$--direction. The compression drives the $z$--component of the clouds' velocity fields up as the tidal field does work on the gas, accelerating it towards the clouds' instantaneous midplanes. The collapse is approximately homologous in this direction, so that most of the gas arrives at the midplane at the same time, forming a very thin structure (this is most easily seen in an animation of the simulation, provided in the Supporting Information).\\
\indent We estimate the timescale on which the clouds reach maximum vertical compression, and thus maximum star formation rate, by computing the ratio of the clouds' instantaneous half--sizes in the $z$--direction $\delta Z(t)$ to the time--integral of their velocity dispersions in the $z$--direction, $\sigma_{Z}$. This results in a normalised crossing time, $\tilde{\tau}_{z}(t)=\delta Z(t)/\int \sigma_{Z}{\rm d}t$\footnote{An integral is computed here, since both $\delta Z$ and $\sigma_{Z}$ change substantially during the simulations}. When this quantity becomes unity or larger, the cloud has had sufficient time to respond to the vertical tidal field and collapse into the midplane. In Figure \ref{fig:delsigma} we show this quantity as a function of time for the eccentric and circular clouds.\\
\indent For all self--virialised clouds, in both eccentric and circular simulations, $\tilde{\tau}_{z}$ reaches unity in the narrow range of times 0.3--0.4\,Myr. The similarity of this timescale in both Eccentric and Circular simulations explains the very similar behaviour of these two classes of run -- although this is the time in the eccentric simulations when the clouds reach pericentre, it is not connected to this event. Note also that this timescale is substantially shorter than the clouds' initial freefall times, indicating once again that it is not their self--gravity that governs their evolution. The similarity of these timesscales is instead likely to be driven by the external potential, which sets both the orbital time and the collapse timescale.\\
\indent All the self--virialised clouds therefore reach their flattest and densest state, and achieve their highest star formation rates after $\approx0.4$\,Myr when homologous collapse brings virtually all the gas together in the clouds' instantaneous vertical midplanes. However, because the gas is highly structured, it does not simply form a dense overpressured layer, as is found, for example, in simulations of smooth colliding clouds \citep[e.g.][]{2015MNRAS.453.2471B}. Some dense substructures approaching each other from opposite vertical directions collide and shock, producing very high densities and star formation rates, but others meet no significant resistance and pass straight through the clouds' instantaneous midplanes. This has the effect that the clouds then become more extended in the vertical direction, and the vertical velocity dispersion declines, as the velocity field instead begins to do work against the tidal and self--gravitational fields.\\
\indent This behaviour is illustrated clearly in Figure \ref{fig:self_mass_flux}. At each timestep, we find the location of the clouds' midplane in the $z$ direction, $z_{0}$ and compute for all $i$ SPH particles the quantity $p_{i}=m_{i} {\bf v_{z,i}}.[\mathrm{sign}(z_{i}-z_{0})\hat{{\bf z}}]$. This quantity is a momentum defined such that negative values indicate vertical motion towards the midplane, and positive values indicate vertical motion away from the midplane. Summing over all particles results in a total vertical momentum whose sign indicates whether the cloud is globally vertically contracting (negative values) or expanding (positive values). Figure \ref{fig:self_mass_flux} shows the result of this procedure applied to all self--virialised clouds on eccentric orbits. The vertical momenta are strongly negative (indicating contraction) up to $\approx0.4$\,Myr, before very abruptly transitioning to positive values (vertical expansion), as significant quantities of material pass straight through the clouds' midplanes.\\
\\
\subsubsection{Shortcomings of the self--virialised cloud models}
\indent Clearly the self--virialised cloud models are unable to reproduce the basic properties of the CMZ clouds. They are unable to support themselves against the tidal forces they experience, and they undergo very strong compression, leading to very high star formation rates and efficiencies, often resulting in gas exhaustion in a fraction of an orbital time and less than the clouds' initial freefall times. In particular, most of the self--virialised clouds are forming stars vigorously by the time they  reach the position of the Brick at $\approx0.4$\,Myr.\\
\indent These simulations lack two important physical processes which would be expected to reduce the star formation rates of the clouds. Magnetic fields provide support to molecular clouds at intermediate scales by exerting an additional (anisotropic) pressure. However, on $>$~pc scales, it is very unlikely that the magnetic pressure is more than comparable to the dynamic pressure, and the magnetic fields present in the Galactic Centre clouds certainly do not appear strong enough to support the clouds against collapse \citep[e.g.][]{2014MNRAS.440.3370K}.\\
\indent Alternatively, stellar feedback, particularly from massive stars in the form of ionising radiation and winds, could in principle prevent the clouds collapsing, or even disrupt them. However, as shown by \cite{2012ApJ...758L..28B,2014MNRAS.442..694D,2016MNRAS.461.2953H}, photoionisation is ineffective on cloud scales if the escape velocity of the cloud is greater than the canonical 10\,km\,s$^{-1}$ sound speed in ionised gas, which is true of all the clouds modelled here. Stellar wind bubbles do not suffer from the limitation of a maximum expansion velocity, but the inhomogeneous nature of molecular clouds likely permits most of the injected energy to escape by advection of hot gas \citep[e.g.][]{2013MNRAS.431.1337R}.\\
\indent In molecular clouds which are dense enough to efficiently trap photons, radiation pressure may be an important feedback mechanism \citep[e.g.][]{2015ApJ...809..187S}. The CMZ clouds certainly possess high enough mean column densities that radiation pressure may be important \citep[e.g.][]{2010ApJ...710L.142F}, but it strains credulity to invoke stellar feedback to explain the low--star formation rates in the CMZ clouds, since many of the clouds appear not to have even started forming stars (supernovae can be immediately dismissed for the same reason).\\
\indent We instead investigate the possibility that the star formation rates in the CMZ clouds are slowed down by the much higher turbulent velocities required to support them against the external tidal field they inhabit.\\
\subsection{Tidally--virialised clouds}
\subsubsection{Star formation efficiencies and rates}
\begin{figure*}
\captionsetup[subfigure]{labelformat=empty}
\centering
\subfloat[]{\includegraphics[width=0.32\textwidth]{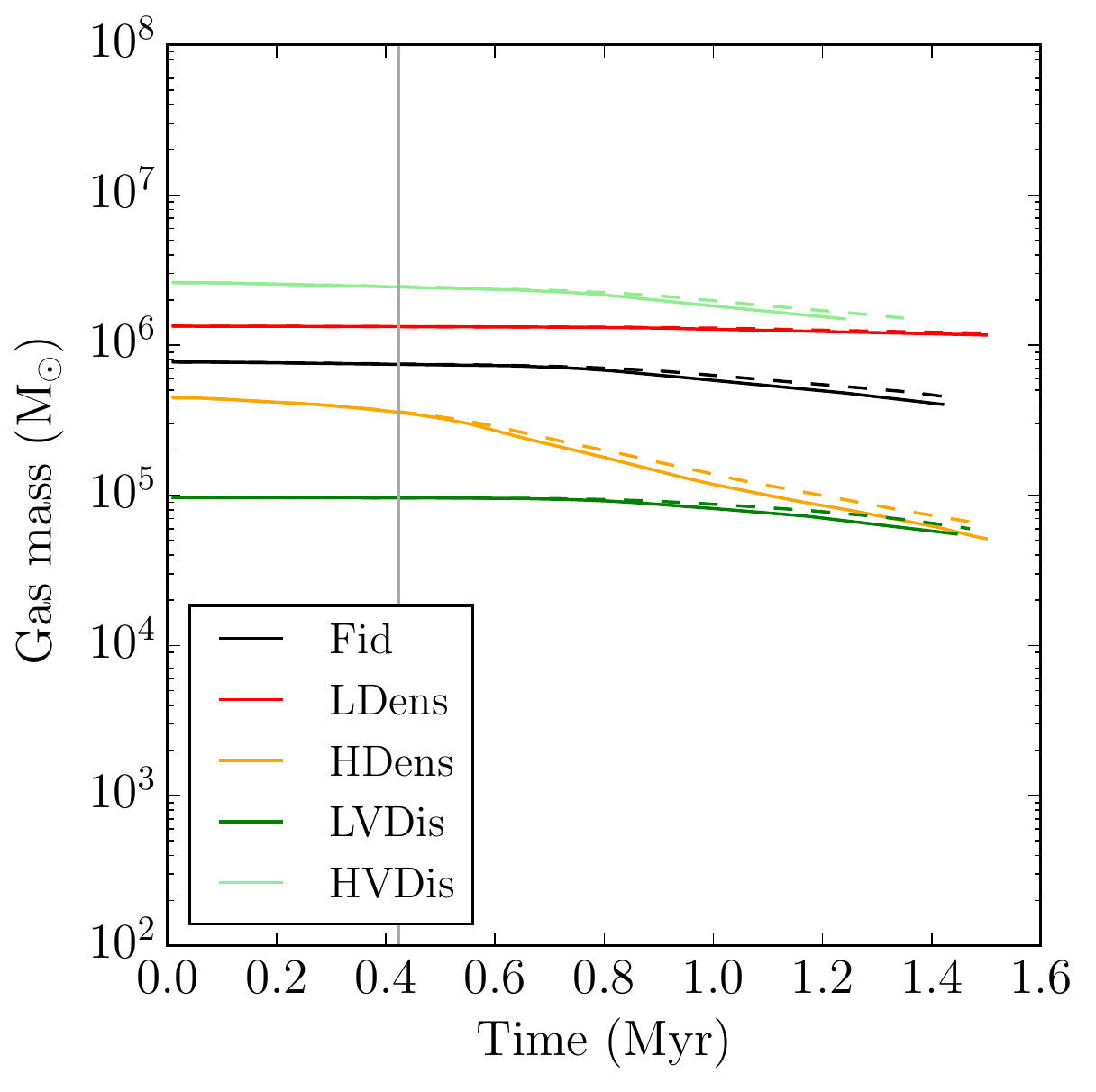}}     
     \hspace{.1in}
\subfloat[]{\includegraphics[width=0.32\textwidth]{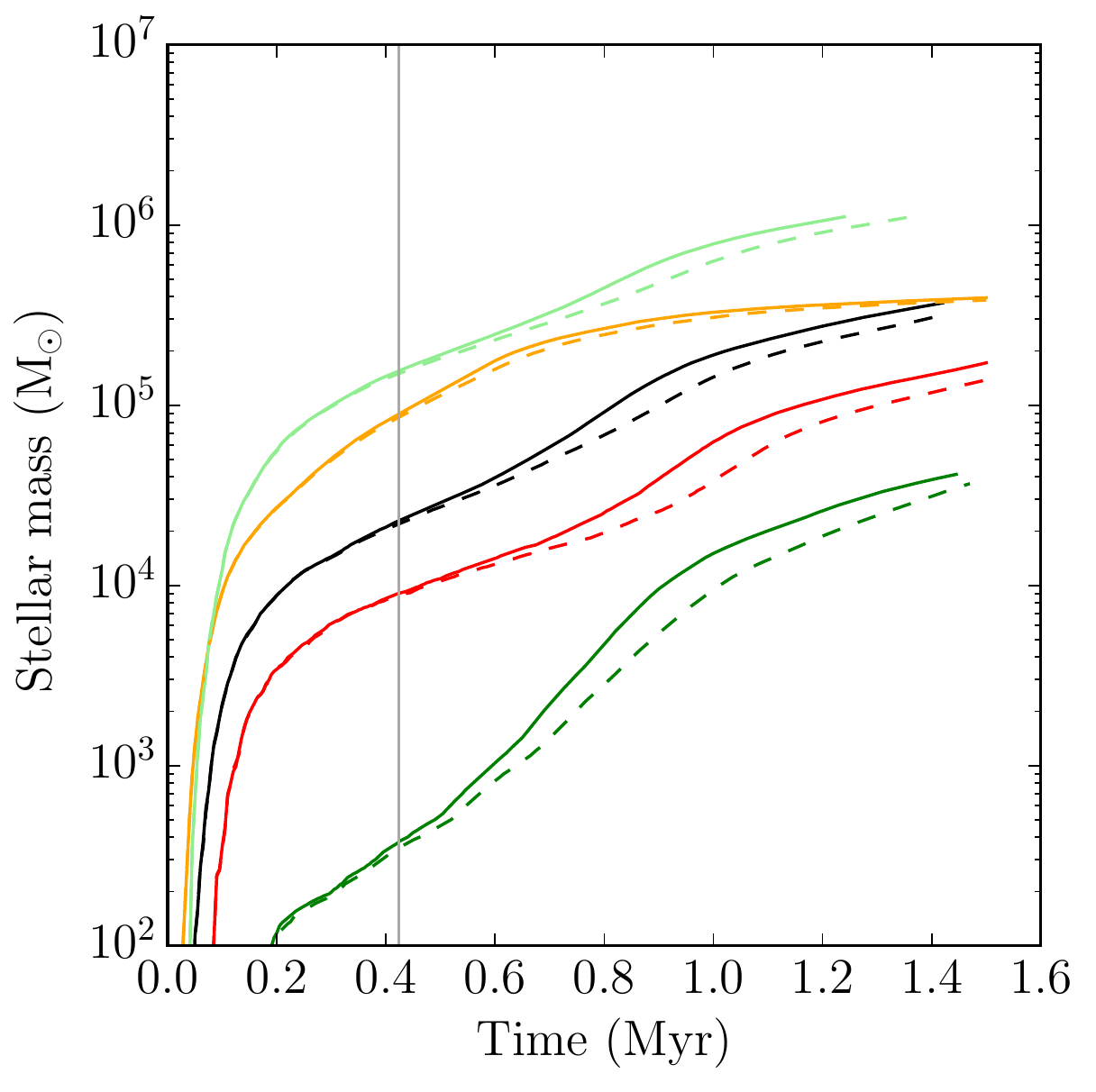}}
     \hspace{.1in}   
\subfloat[]{\includegraphics[width=0.32\textwidth]{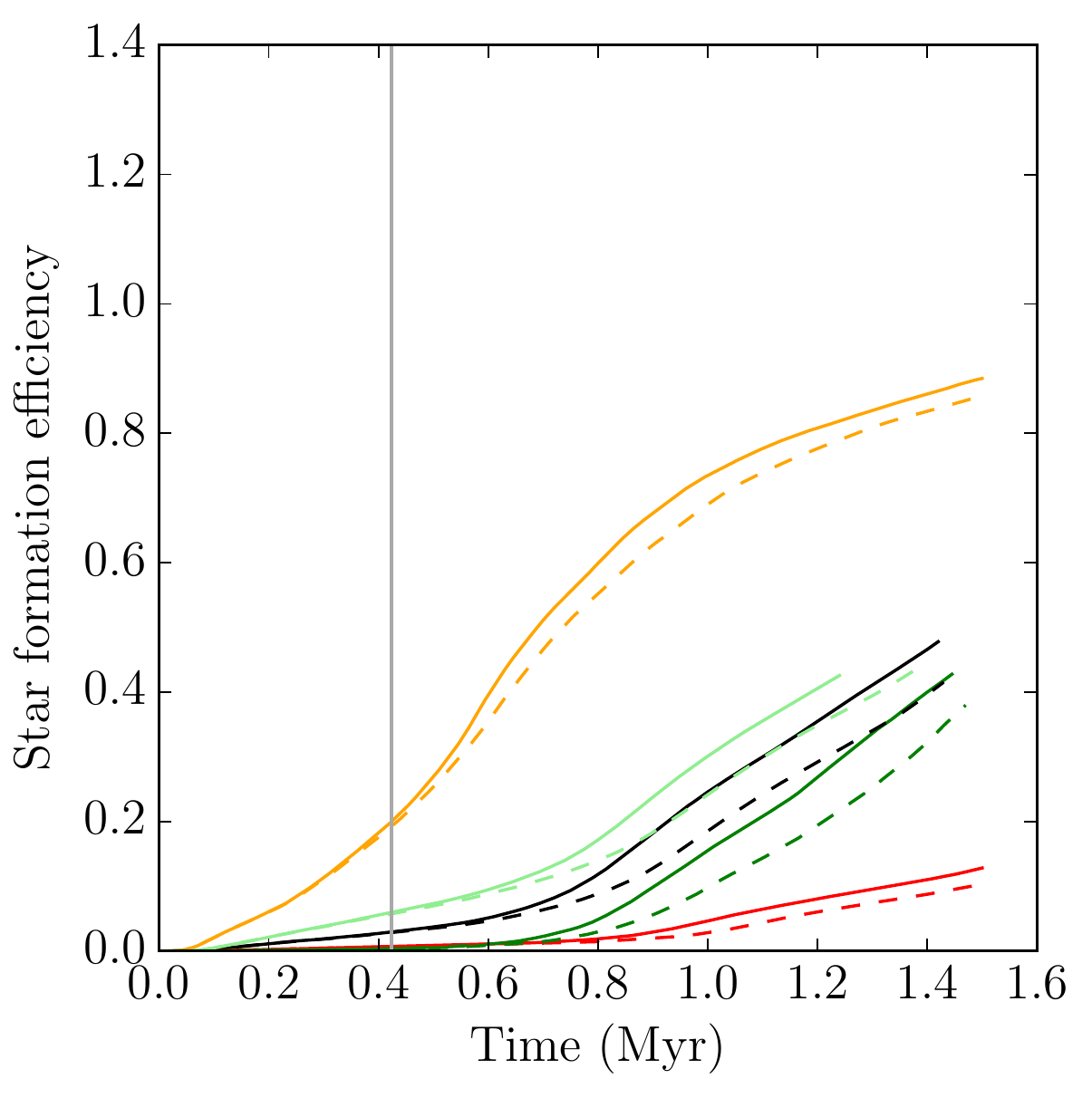}}
     \vspace{-.3in}   
\subfloat[]{\includegraphics[width=0.32\textwidth]{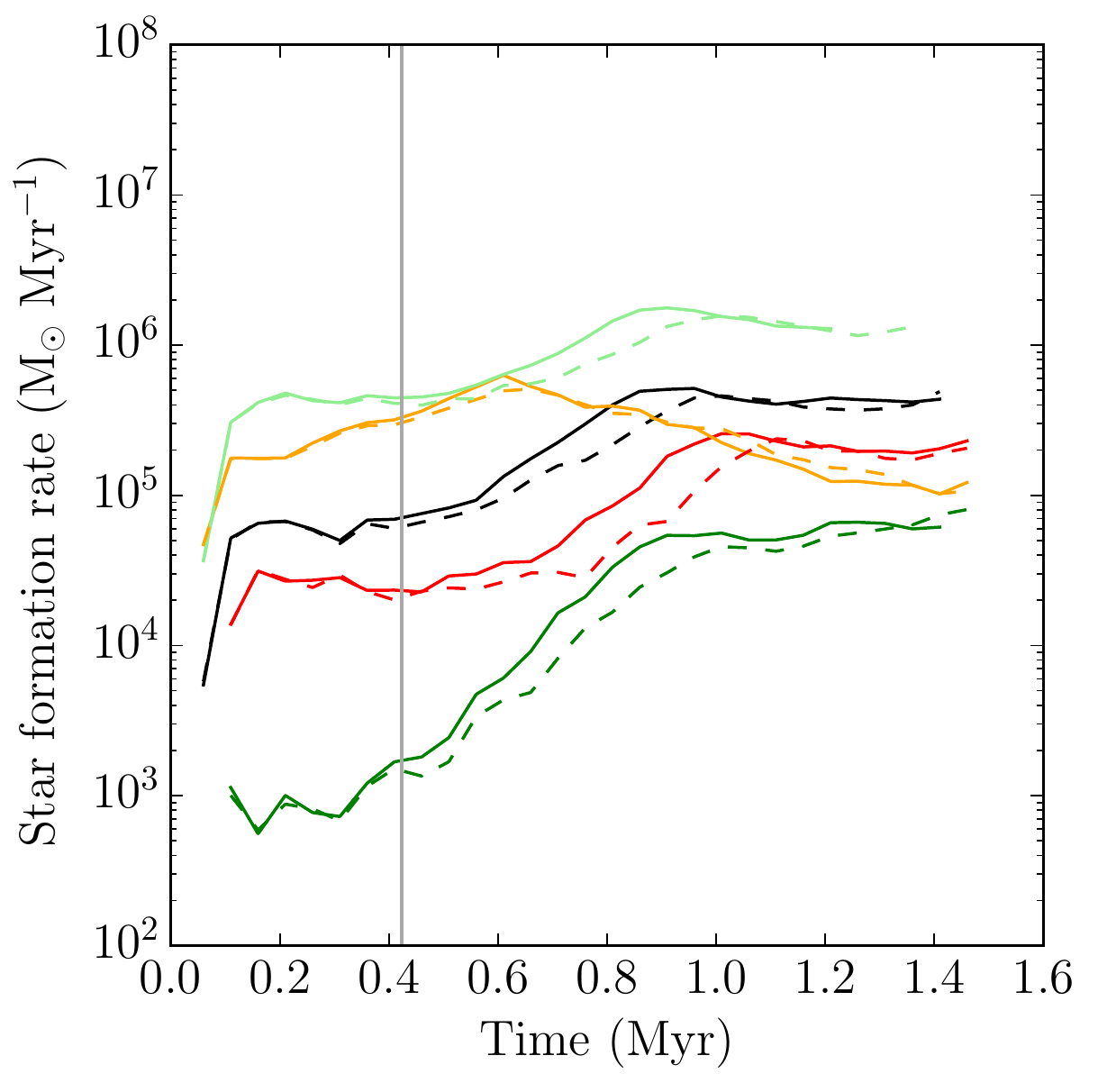}}     
     \hspace{.1in}
\subfloat[]{\includegraphics[width=0.32\textwidth]{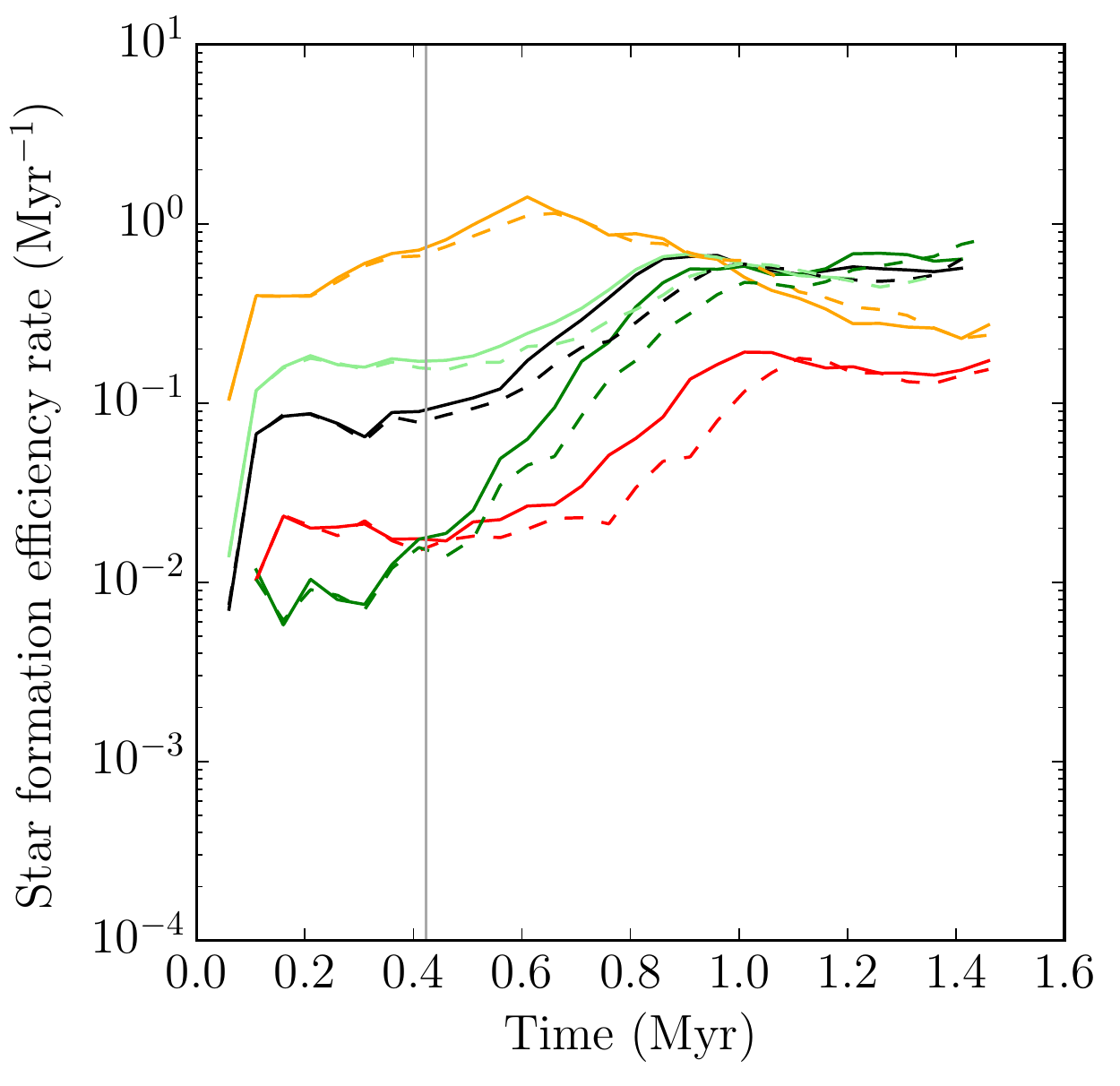}}
     \hspace{.1in}   
\subfloat[]{\includegraphics[width=0.32\textwidth]{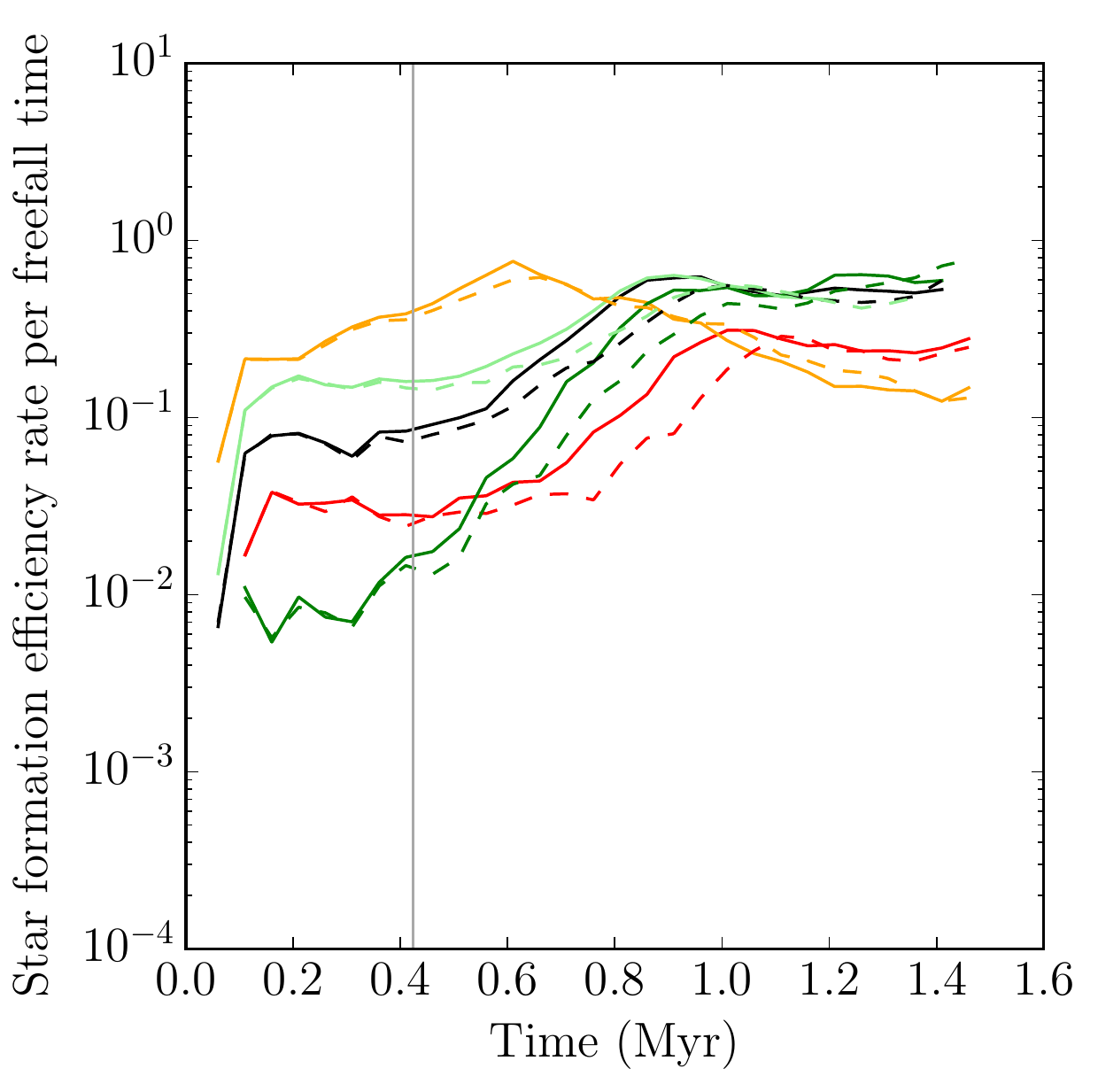}}
\caption{Time evolution of several quantities describing all of the tidally-virialised simulations. The panels show the gas mass (top left), stellar mass (top centre), star formation efficiency (top right), absolute star formation rate (bottom left), star formation efficiency rate (bottom centre), star formation efficiency rate per free-fall time (bottom right). This figure shows that star formation rates in the tidally--virialised clouds are much lower than in the self--virialised models. In all panels, solid lines are eccentric simulations, dashed lines are circular simulations, colours denote the different model clouds, as indicated by the legend in the first panel, and the vertical grey line represents pericentre passage in the eccentric orbit at 0.41\,Myr.}
\label{fig:sf_rates}
\end{figure*}

\begin{figure}
\includegraphics[width=0.45\textwidth]{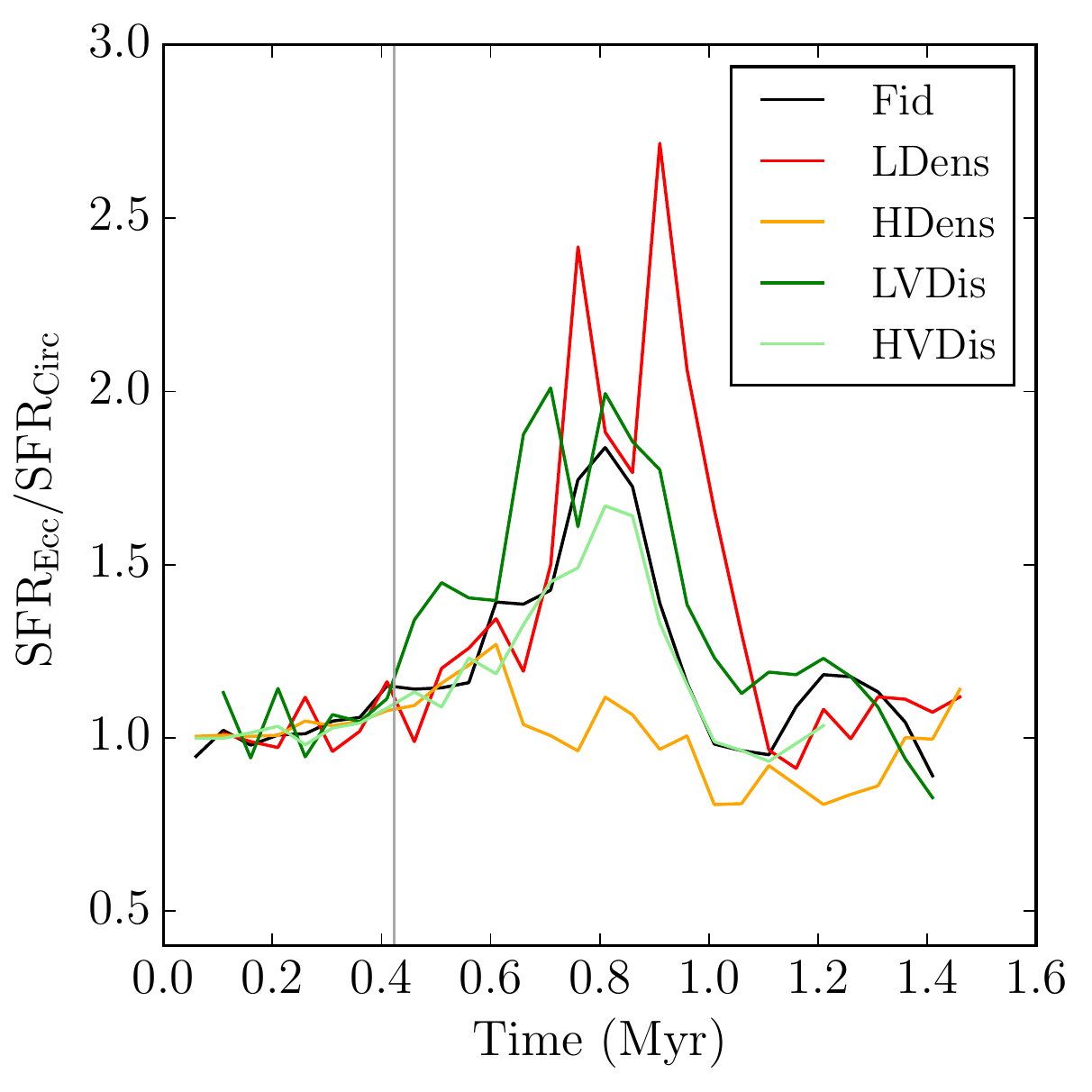}
\caption{Time evolution of the star formation rates of the tidally-virialised clouds on eccentric orbitts relative to those on circular orbits, demonstrating a modest difference. The colours denote the different model clouds, as indicated by the legend, and the vertical grey line represents pericentre passage in the eccentric orbit at 0.41\,Myr.}
\label{fig:talpha_sfr_ecc_over_circ}
\end{figure}

\indent As we did for the self--virialised clouds, we first examine the rates of star formation in the tidally--virialised objects. In Figure \ref{fig:sf_rates}, we show on the top row the remaining gas mass (left panel), the stellar mass (centre panel) and the star formation efficiency (right panel) in the tidally--virialised clouds. In the lower panels of Figure \ref{fig:sf_rates}, we show the star formation rates and efficiencies of the simulations.\\
\indent Models of isolated turbulent clouds without feedback generally produce star formation efficiency rates per freefall time of several tens of percent. This is precisely what we observed the Isolated clouds described in the previous section, with the exception of the unbound HVir model. Placing the self--virialised clouds on the galactic centre orbit increased the star formation rates, sometimes by more than an order of magnitude.\\
\indent In stark contrast, the star formation rates and rate of growth in star formation efficiency in the tidally--virialised clouds are much slower. For all except the HDens model, SFER$_{\rm ff}$ is less than 0.15 until approximately 0.6\,Myr, substantially after pericentre passage. Star formation efficiencies after one cloud freefall time are in the range 0.05--0.25, instead of 0.25--1.0. The HDens model cloud reaches a star--formation efficiency of $\approx0.5$ at 0.7 Myr, some 0.3 Myr after pericentre (matching the position of the Brick, see Figure 1). Likewise, the high velocity dispersion cloud achieves $\approx0.2$, the fiducial cloud $\approx0.15$, and the two other clouds $\approx0.05$. These values are still higher than the star formation efficiencies seen in the CMZ clouds, but they are much lower than those achieved in the self--virialised clouds and are in fact similar to or smaller than the efficiencies observed in the isolated clouds, even if such a comparison should be made carefully, since the velocity dispersions in the tidally--virialised clouds are so much higher. Readers are reminded that these are strict upper limits to the star formation rates/efficiencies, due to our inability to resolve individual stars, the use of sink particles, and our neglect of feedback and magnetic fields.\\
\indent If we compare the rate of star formation in the clouds on eccentric and circular orbits, as depicted in Figure \ref{fig:talpha_sfr_ecc_over_circ}, we see that, apart from the HDens cloud which is the cloud least affected by (i.e. most decoupled from) the tidal field, the eccentric clouds experience a substantially larger relative jump in star formation rate than was observed in the self--virialised models, greater than a factor of two in some cases. In addition, there is a substantial delay of $\approx0.4$\,Myr between passage through pericentre and the peak in the relative difference between star formation rates in the eccentric and circular clouds. This shows that the compression induced by the pericentre passage requires time to influence the star formation rate. Fig.~4 of \citet{kruijssen18} demonstrates that this is partly caused by the delayed conversion of tidal--shear--driven rotational energy into internal (possibly turbulent) kinetic energy that can be dissipated. We show below that tidal compression, in the form of delayed non--homologous collapse in the vertical direction is also a contributing factor (see Table 3 of \citealt{kruijssen18} for a summary of the effects of various physical mechanisms on the observable properties of the clouds). As a result, star formation may remain inhibited until the position of `cloud~b' on the Galactic centre dust ridge (see Figure 1).\\ 
\subsubsection{Response of the tidally--virialised clouds to the tidal field}
\begin{figure}
\includegraphics[width=0.48\textwidth]{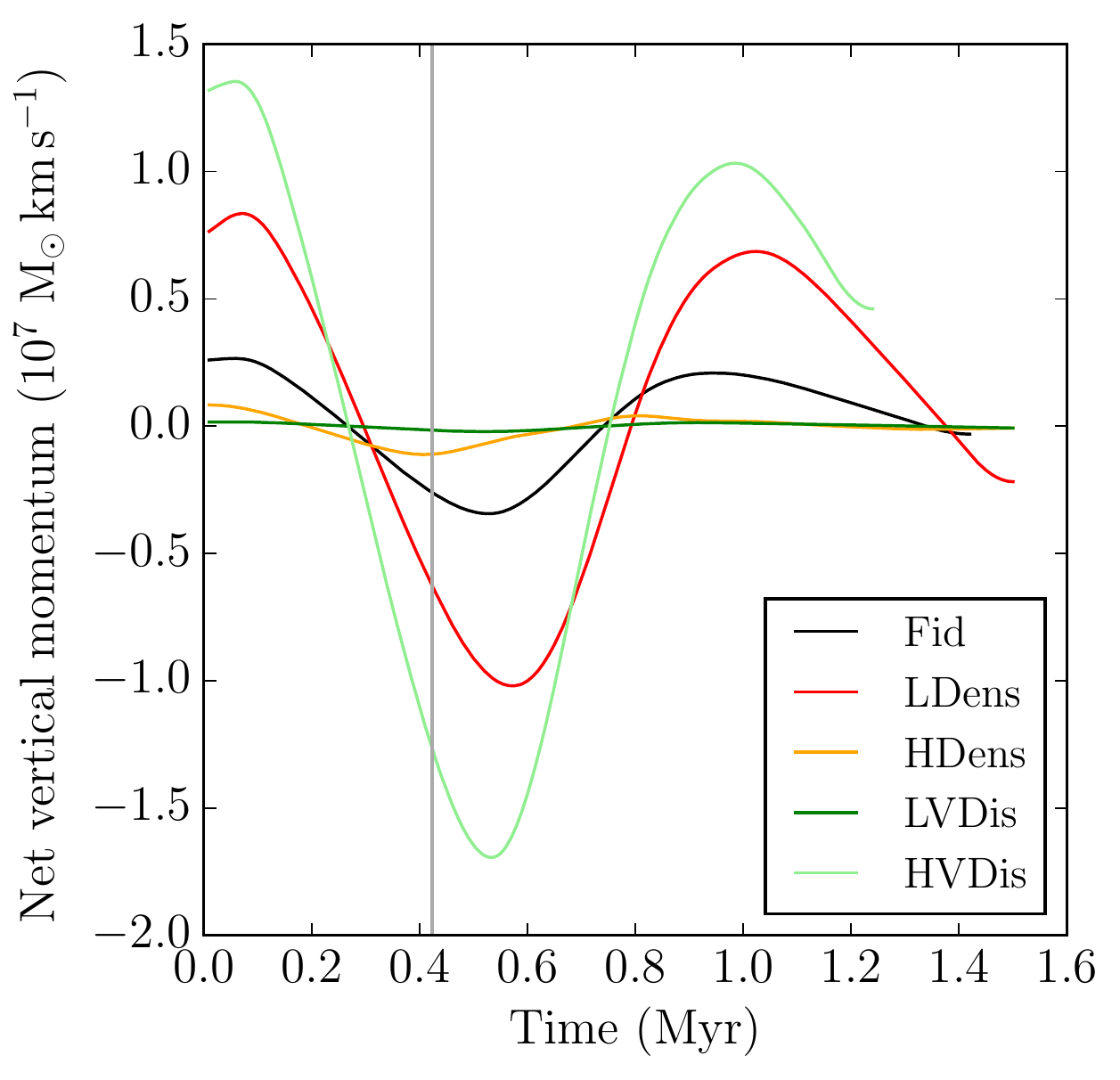}
\caption{Net momentum in the vertical direction (defined such that negative values indicate vertical contraction and positive values indicate vertical expansion) in all eccentric tidally--virialised simulations. The colours denote the different model clouds, as indicated by the legend, and the vertical grey line represents pericentre passage in the eccentric orbit at 0.41\,Myr.}
\label{fig:talpha_mass_flux}
\end{figure}

\begin{figure}
\centering
\subfloat[]{\includegraphics[width=0.48\textwidth]{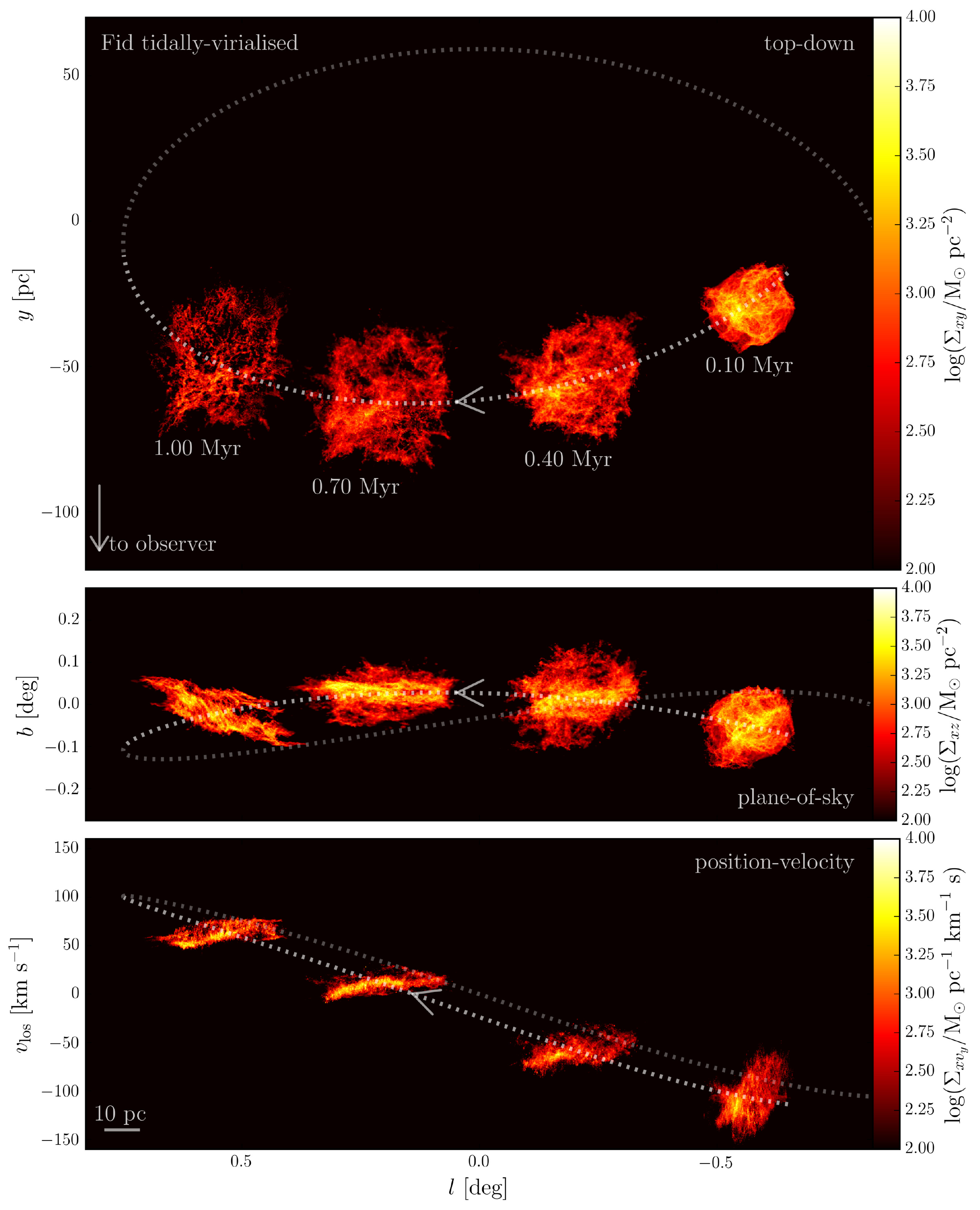}}
\caption{Position--position and position--velocity renders of the tidally--virialised Fid simulation at four timesteps. All images are logarithmically--scaled. Dashed lines represent the orbit fitted by . Top panel: Top--down position--position view (along the $z$--axis) with an arrow indicating the direction of the Sun. Middle panels: Line--of--sight position--position view (along the $y$--axis). Bottom panel: Line--of--sight position--velocity view.}
\label{fig:sim_PP_PV_Fid}
\end{figure}

\begin{figure*}
\captionsetup[subfigure]{labelformat=empty}
\subfloat[]{\includegraphics[width=0.48\textwidth]{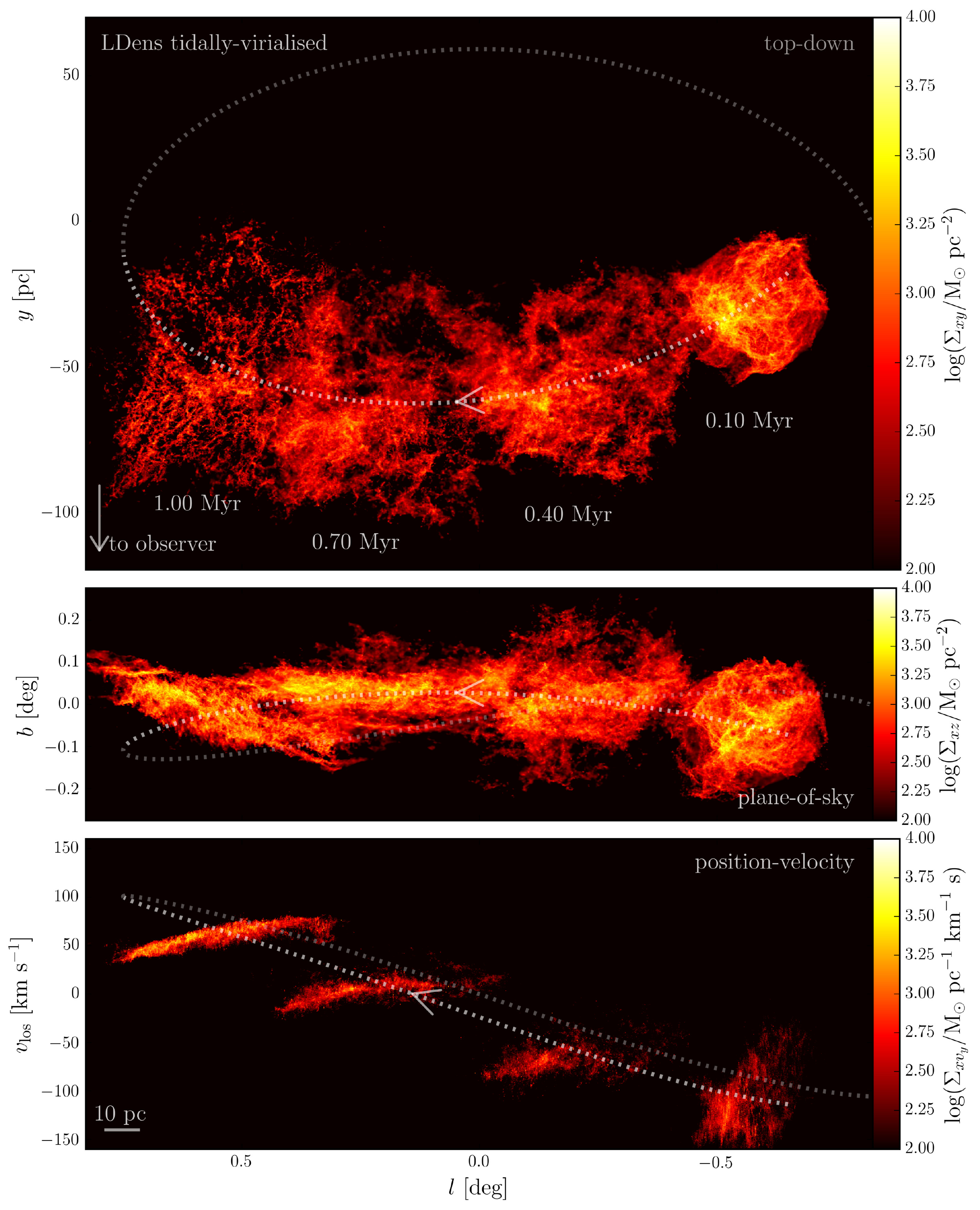}}
\hspace{-0.05in}
\subfloat[]{\includegraphics[width=0.48\textwidth]{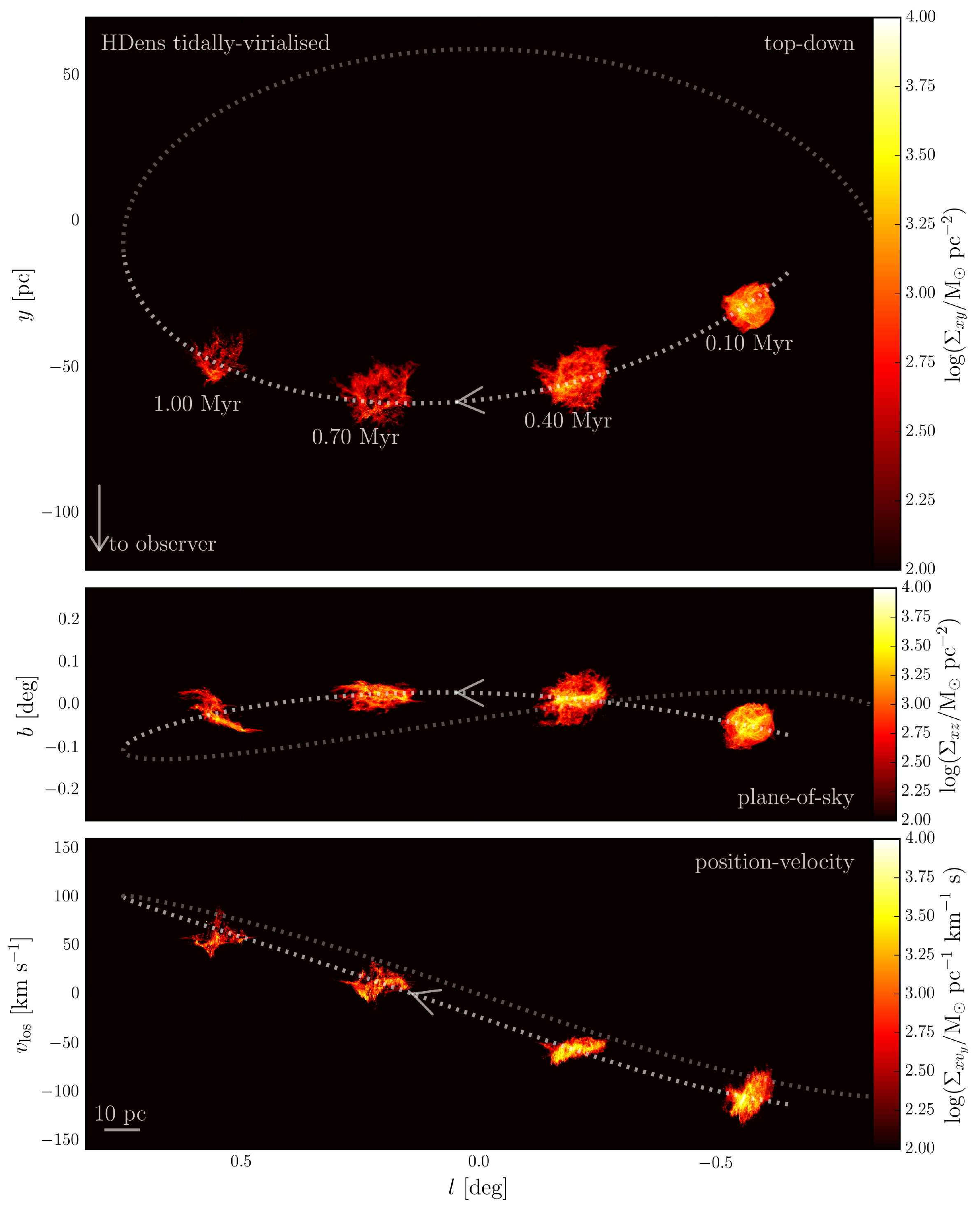}}
    \vspace{-.2in}
 \subfloat[]{\includegraphics[width=0.48\textwidth]{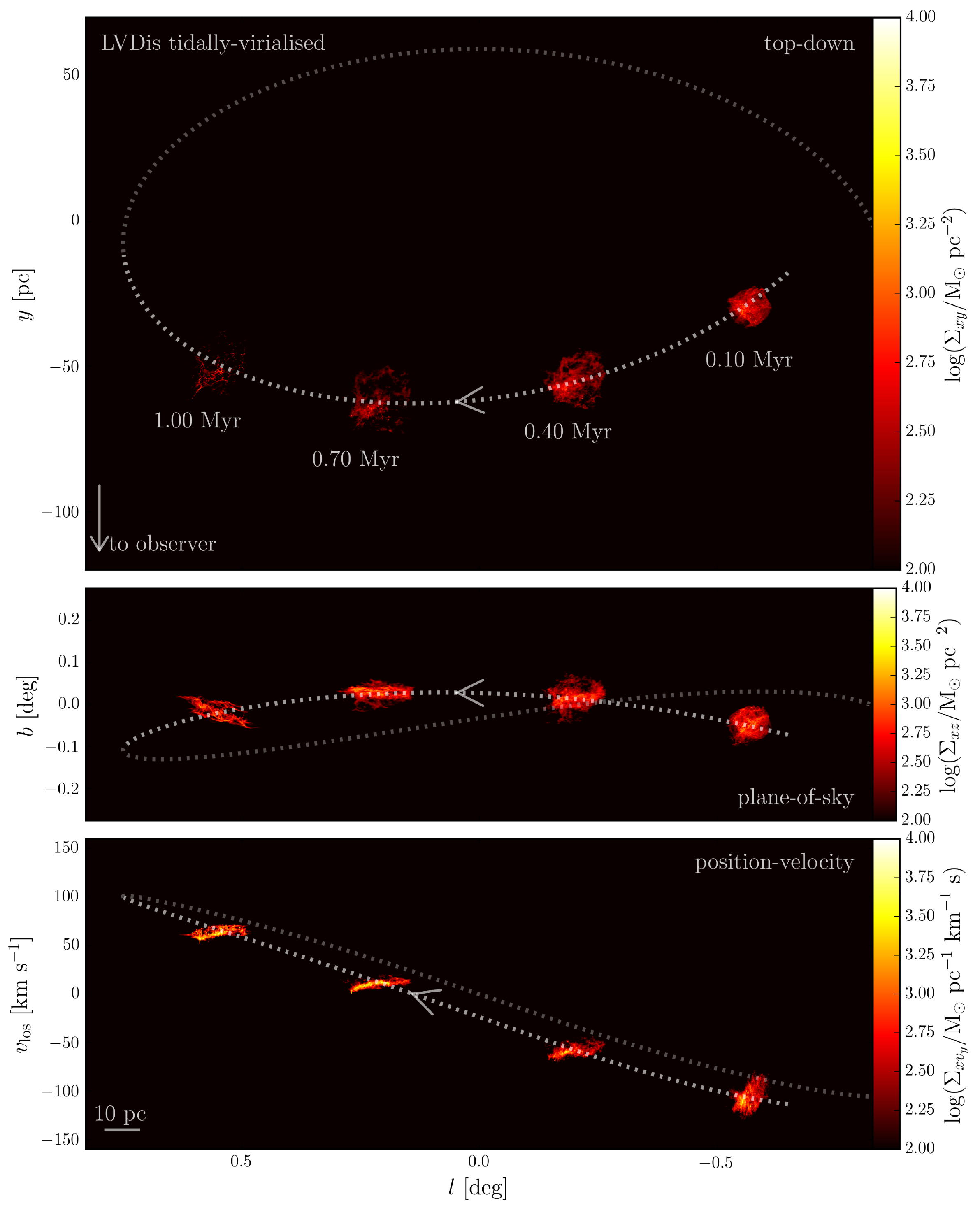}}
     \hspace{-.05in}
\subfloat[]{\includegraphics[width=0.48\textwidth]{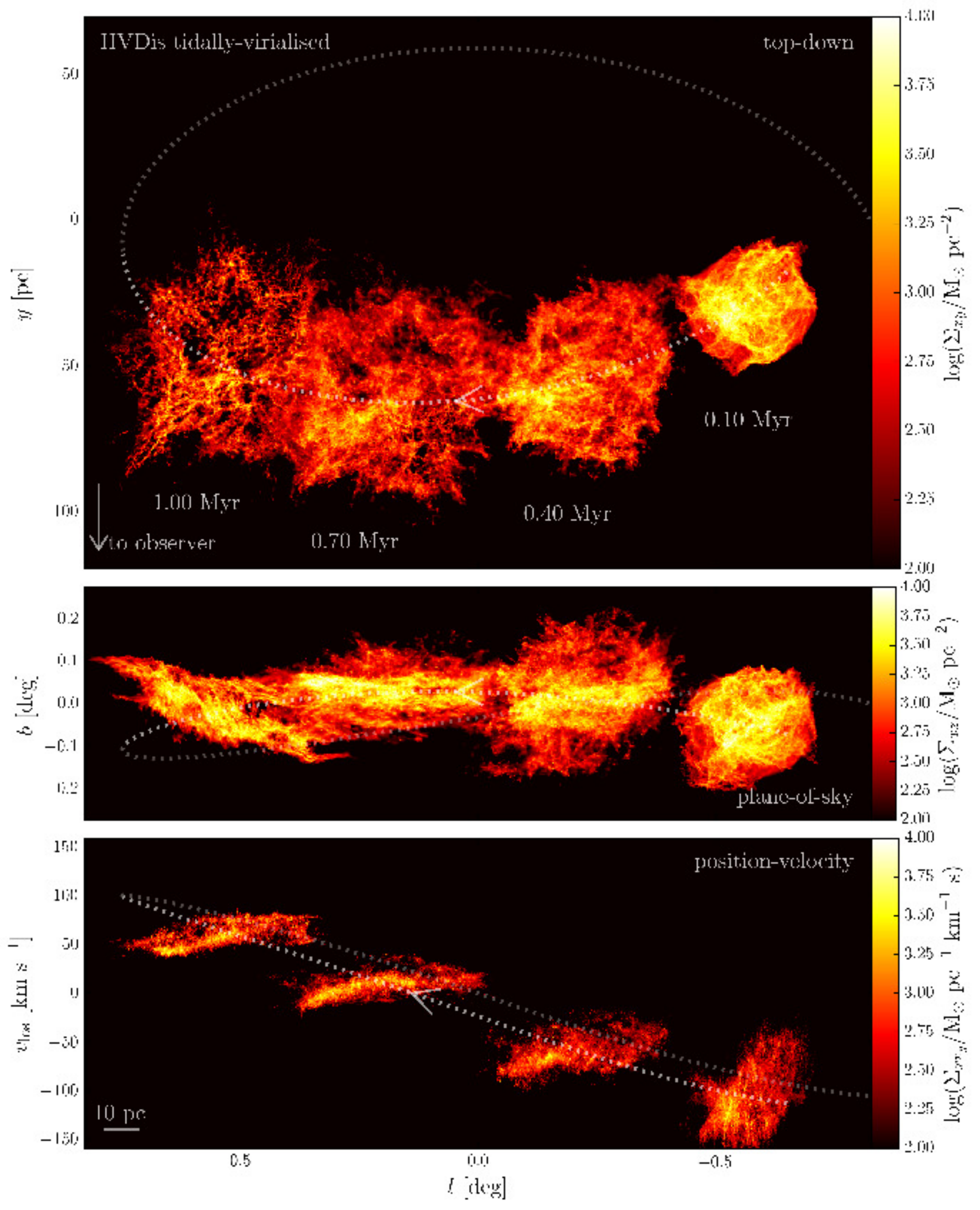}}
\caption{Position--position and position--velocity renders of the tidally--virialised LDens (top--left), HDens (top--right), LVDis (bottom--left) and HVDis (bottom--right) simulations at four timesteps. All images are logarithmically--scaled. Dashed lines represent the orbit fitted by Kruijssen et al., (2018). Top panels: Top--down position--position views (along the $z$--axis) with an arrow indicating the direction of the Sun. Middle panels: Line--of--sight position--position views (along the $y$--axis). Bottom panel: Line--of--sight position--velocity views.}
\label{fig:sim_PP_PV}
\end{figure*}

\indent In Figures \ref{fig:sim_PP_PV_Fid} and \ref{fig:sim_PP_PV} we show position--position and position--velocity diagrams for the eccentric tidally--virialised simulations, for comparison with the self--virialised cloud shown in Figure \ref{fig:sim_PP_PV_trad}. The tidally--virialised cloud is plainly larger in physical space and more spread in velocity space than the self--virialised model. The tidally--supported cloud does become flattened, but not to the degree of its self--virialised counterpart, and its larger extent in the direction tangential to the orbit gives it a larger aspect ratio than the self--virialised clouds. There is very little of the fading observed in the self--virialised model, since the rate of gas consumption is much lower.\\
\indent It is instructive to examine the other tidally--virialised cloud models in this fashion. A detailed discussion is given in \cite{kruijssen18} and we only give a brief exposition here. The spreading of the clouds in the $x$ (i.e. in Galactic longitude) and $y$ (i.e. line--of--sight) directions is a feature of all the clouds save the HDens model, although the interpretation of this cloud is complicated by the fact that it approaches gas exhaustion by $\sim0.8$\,Myr, leading to the gas column--density declining strongly by this time. The spreading is strongest along the line--of--sight, since this direction is closest to the radial direction for most times during the simulations, and it is in this coordinate that the clouds are oversupported against the tidal field. As the clouds approach apocentre, however, this spreading produces a significant extension projected along the orbit, which is further contributed to by tidally--driven shear. As discussed in \cite{kruijssen18}, this leads to an increase in observed velocity dispersion as the clouds approach the position of Sgr B2. The spreading is, to a large degree, due to disturbances in the outer layers of the clouds, which are least strongly bound, and also most subject to tidal shear.\\
\indent The $(x,z)$ morphology of the clouds show two notable features. Firstly, the clouds are strongly compressed in the vertical direction -- that in which they are undersupported -- but not as strongly as the self--virialised models due to their elevated velocity dispersions. The denser inner regions of the clouds are initially compacted more strongly, leaving low--density haloes above and below the clouds as they pass through pericentre. However, by the time the clouds reach $x\approx -50$~pc (corresponding to the position of the Brick in Figure 1), they are globally compacted into pancake structures whose projected long axes are nearly parallel to the orbit projected into the $(x,z)$ plane. This leads to a strong drop in the clouds' aspect ratios \citep[][]{kruijssen18}.\\
\indent As the clouds pass from $x\approx -50$~pc to $x\approx -100$~pc (corresponding to the longitude range of the Galactic centre `dust ridge', see Figure 1), the clouds are tilted with respect to the orbit. This clockwise rotation about the line of sight is caused by the torques accompanying pericentre passage and increases the aspect ratios once more.\\
\indent The position--velocity images reveal, as in the case of the self--virialised clouds, that the imprint of the clouds' initial turbulent velocity fields is largely erased by the time of pericentre passage, with the clouds acquiring a pronounced velocity gradient in the opposite sense to that of the orbit. This counter--rotation is induced by tidal shear \citep{kruijssen18}. The exception is the HDens cloud, where localised collapse again dominates the position--velocity plots by $0.7$\,Myr and the fading due to gas exhaustion observed in the self--virialised clouds is also present. Overall, however, the structure and kinematics of the tidally--virialised simulations resemble much more closely what is observed in the  than do those of the self--virialised models (see Figure 2 of \cite{kruijssen18}).\\
\indent Examination of the vertical mass fluxes, depicted in Figure \ref{fig:talpha_mass_flux} reveals crucial differences between these clouds and their self--virialised counterparts.
The net vertical momenta of all the tidally--virialised clouds are initially away from the clouds' midplanes. This reverses quite rapidly after $\approx0.2$\,Myr but, as can be seen in the position--position images and the animations in the Supporting Information, the initial expansion creates a low--density halo above and below the clouds and, when they do begin to contract vertically, they do so non--homologously, with the more weakly--bound halo taking longer to begin contracting than the dense regions closer to the midplane. The contraction occurs over a more extended period of time and, because the gas does not all arrive at the midplane at the same time, as in the self--virialised clouds, the increase in star formation activity is much slower. For the same reason, the tidally--virialised clouds are never as thin in the vertical direction or as dense. Substructure which passes through the midplanes without being shocked continues to meet still infalling material from the opposite vertical side of the cloud until $\sim0.8$\,Myr. After this time, essentially all the material in the clouds has been shocked by vertical tidal compression.\\
\section{Discussion}
\subsection{Effect of tidal forces on cloud dimensions and velocity dispersions}
\begin{figure*}
\centering
\subfloat[]{\includegraphics[width=0.32\textwidth]{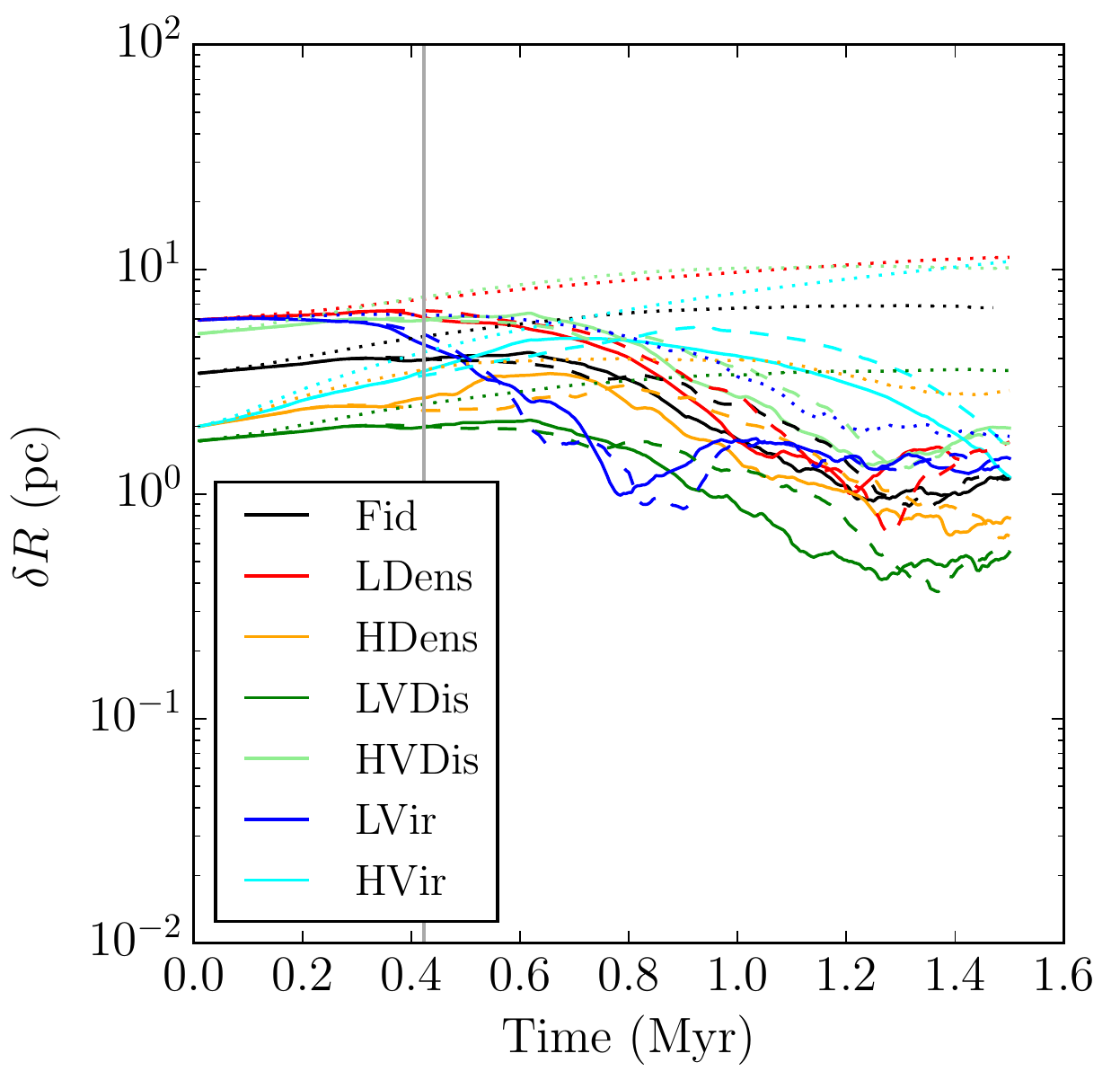}}   
    \hspace{.1in}
\subfloat[]{\includegraphics[width=0.32\textwidth]{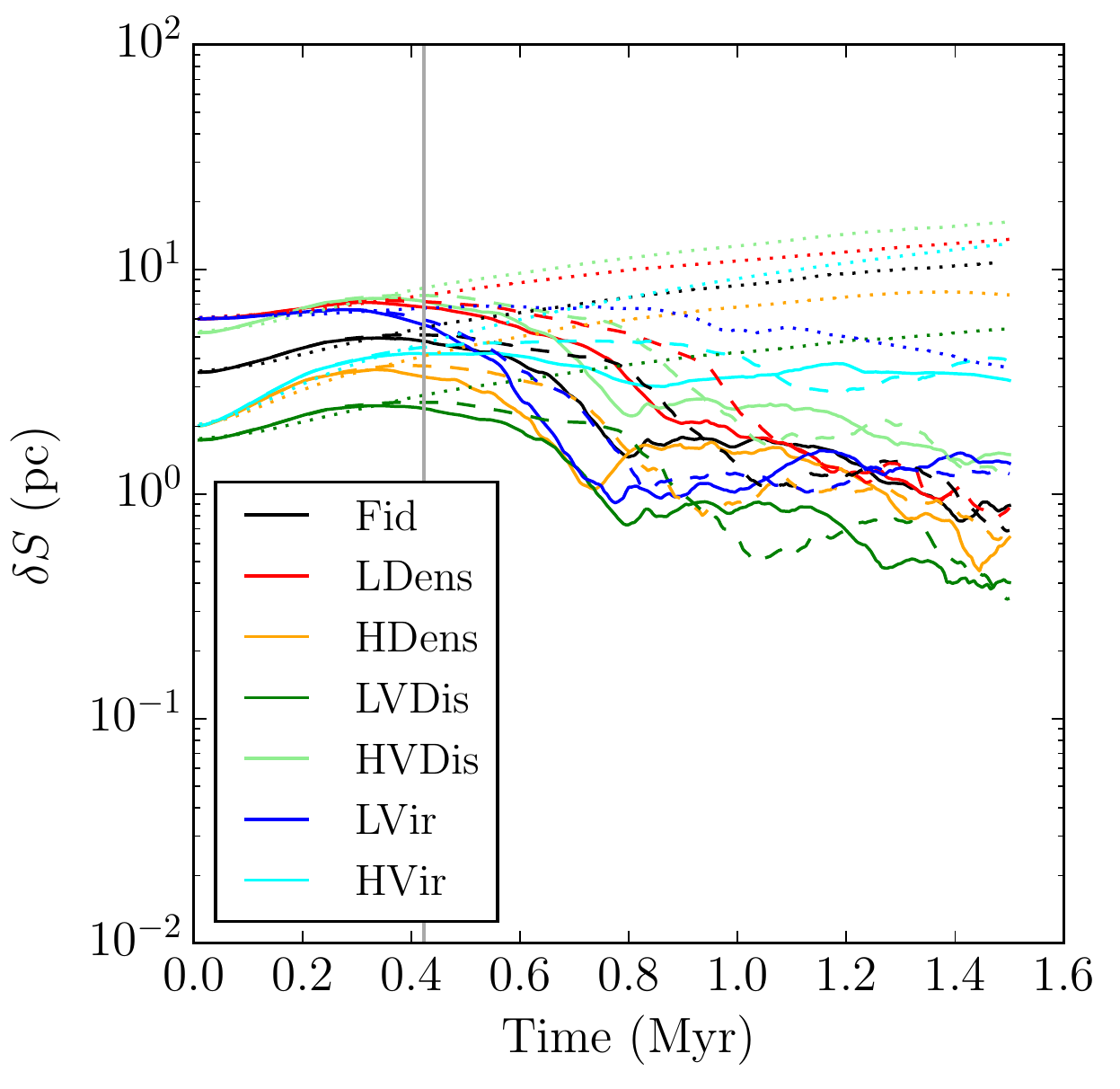}}     
    \hspace{.1in}
\subfloat[]{\includegraphics[width=0.32\textwidth]{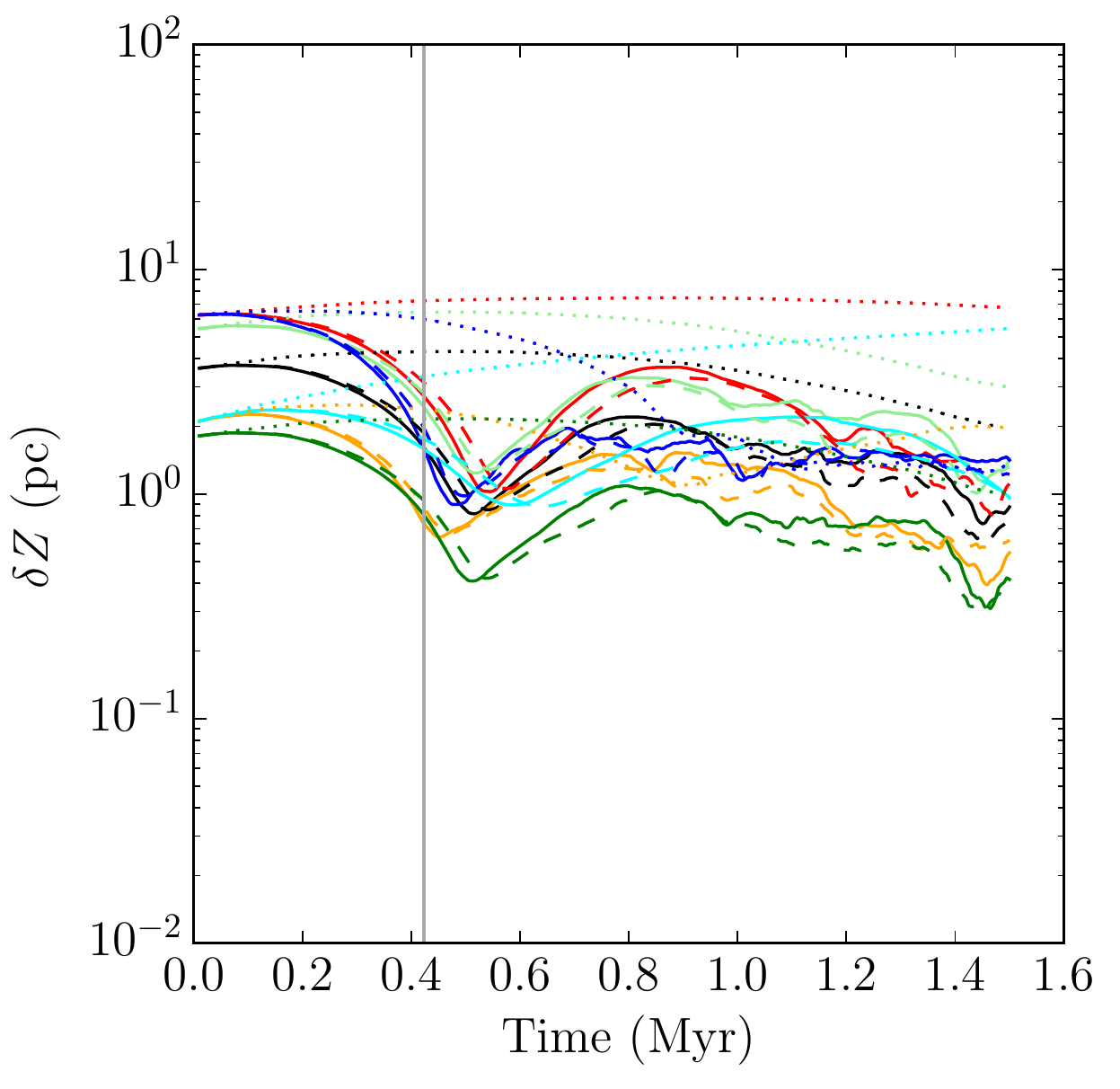}}
\vspace{-.1in}
\subfloat[]{\includegraphics[width=0.32\textwidth]{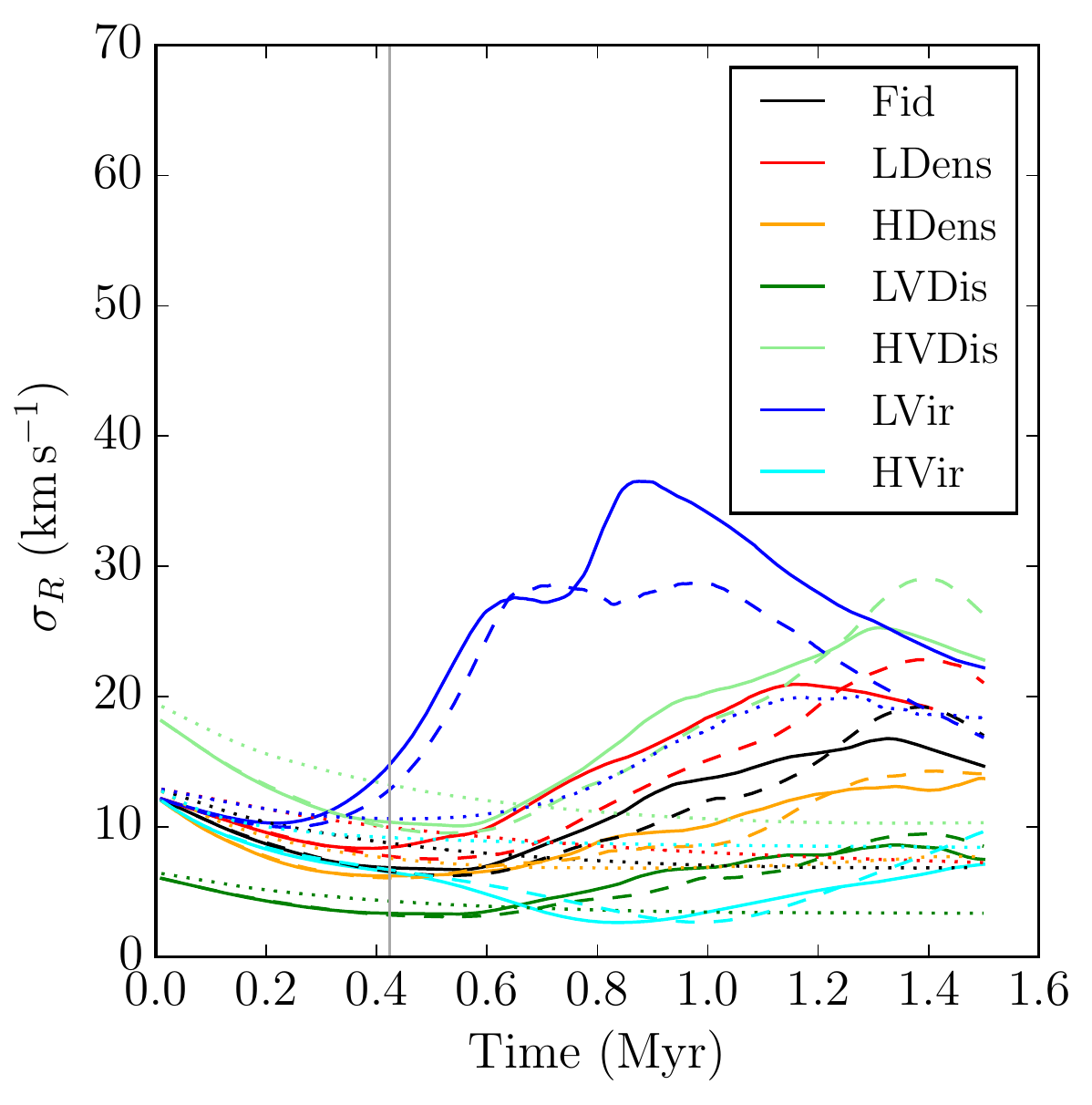}}   
    \hspace{.1in}
\subfloat[]{\includegraphics[width=0.32\textwidth]{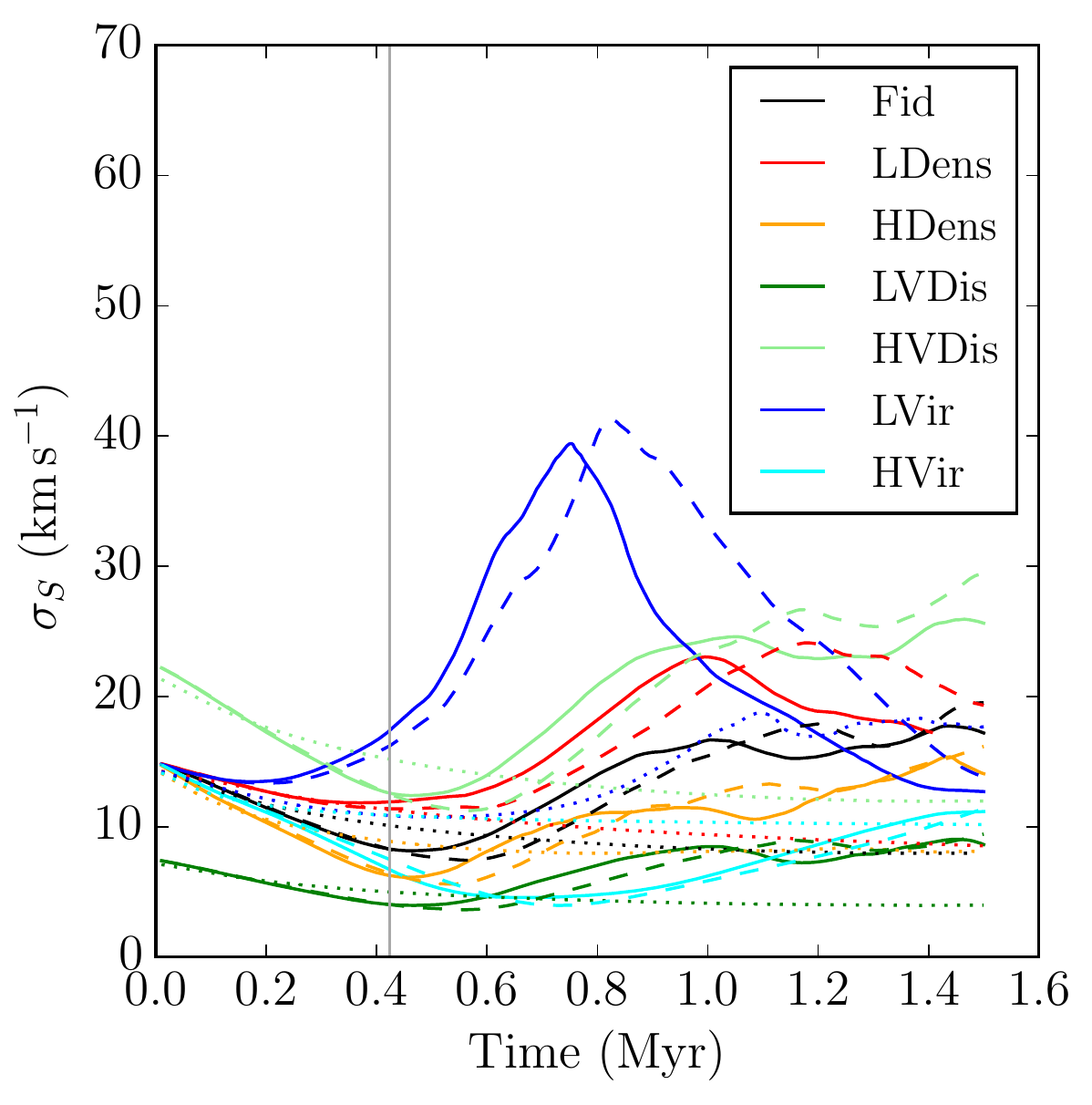}}     
    \hspace{.1in}
\subfloat[]{\includegraphics[width=0.32\textwidth]{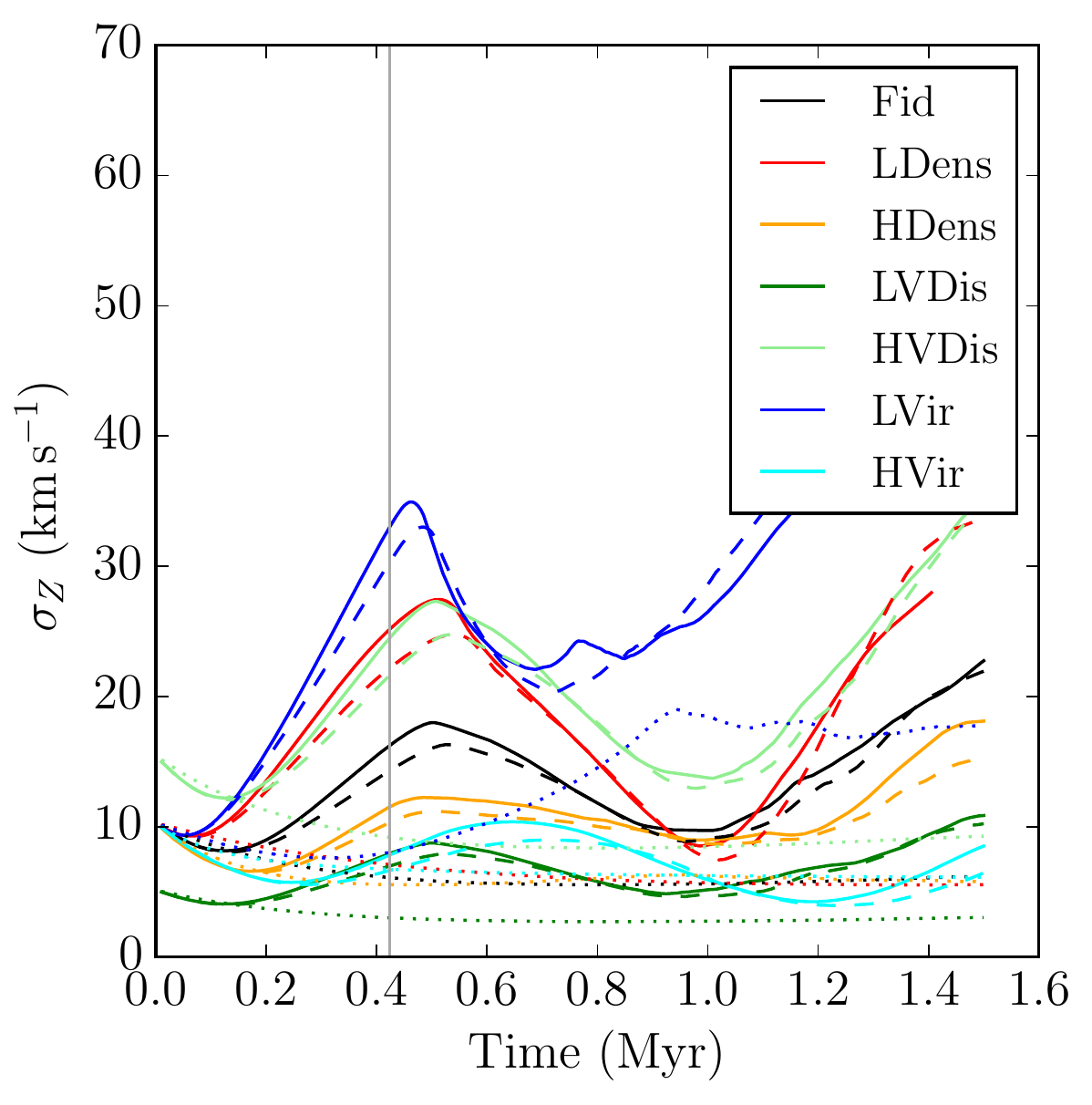}}
\caption{Top row: Sizes of the self--virialised clouds plotted against time in the radial (left panel), tangential (centre panel) and vertical (right panel) directions, showing tidal compression along all axes. Bottom row: velocity dispersions in the self--virialised clouds plotted against time in the radial (left panel), tangential (centre panel) and vertical (right panel) directions. In all panels, solid lines are eccentric simulations, dashed lines are circular simulations, dotted lines are isolated simulations, colours denote the different model clouds, as indicated by the legend in the left panels, and the vertical grey line represents pericentre passage in the eccentric orbit at 0.41\,Myr.}
\label{fig:delta_sigma_self}
\end{figure*}

\begin{figure*}
\centering
\subfloat[]{\includegraphics[width=0.32\textwidth]{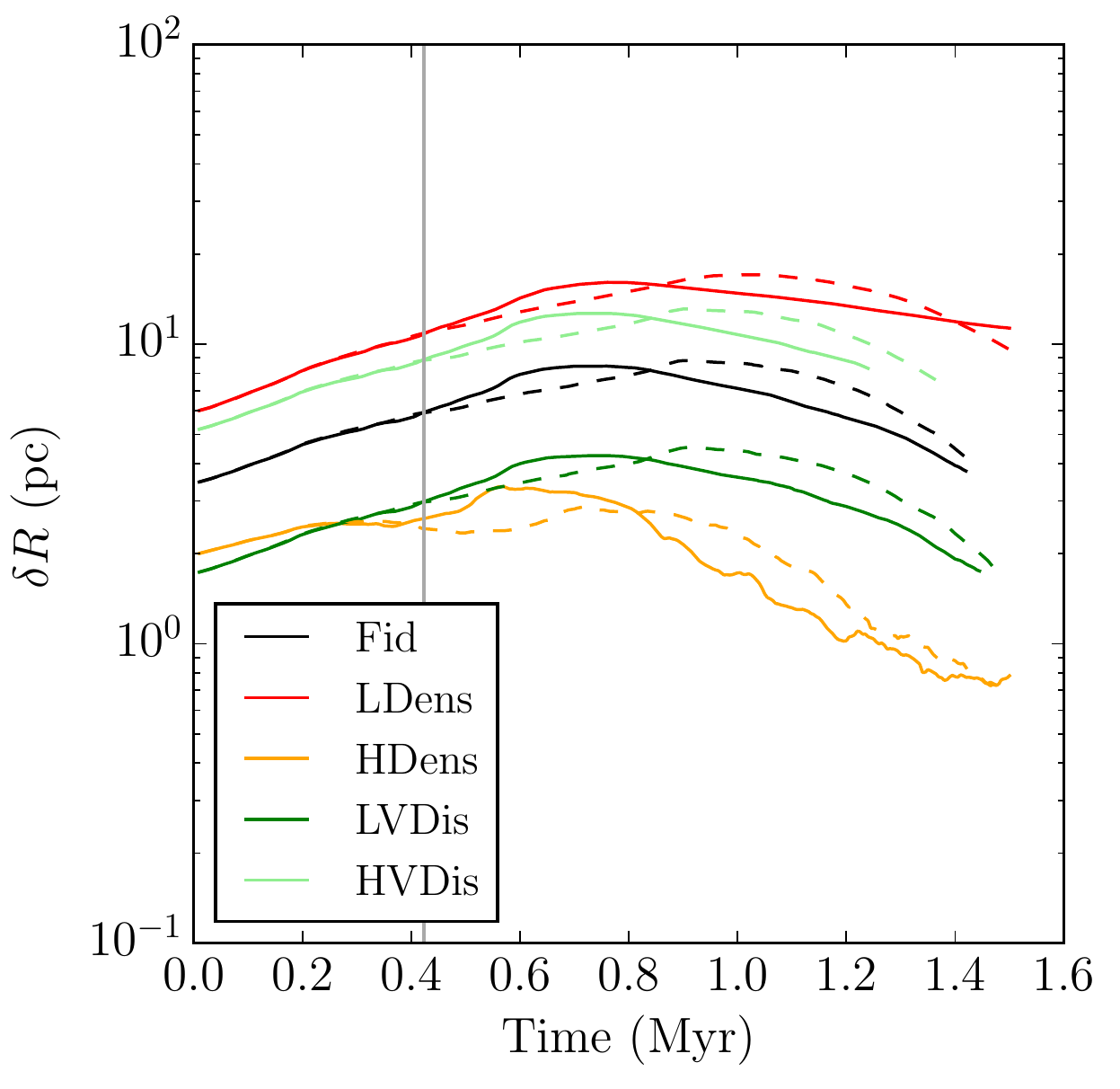}}   
    \hspace{.1in}
\subfloat[]{\includegraphics[width=0.32\textwidth]{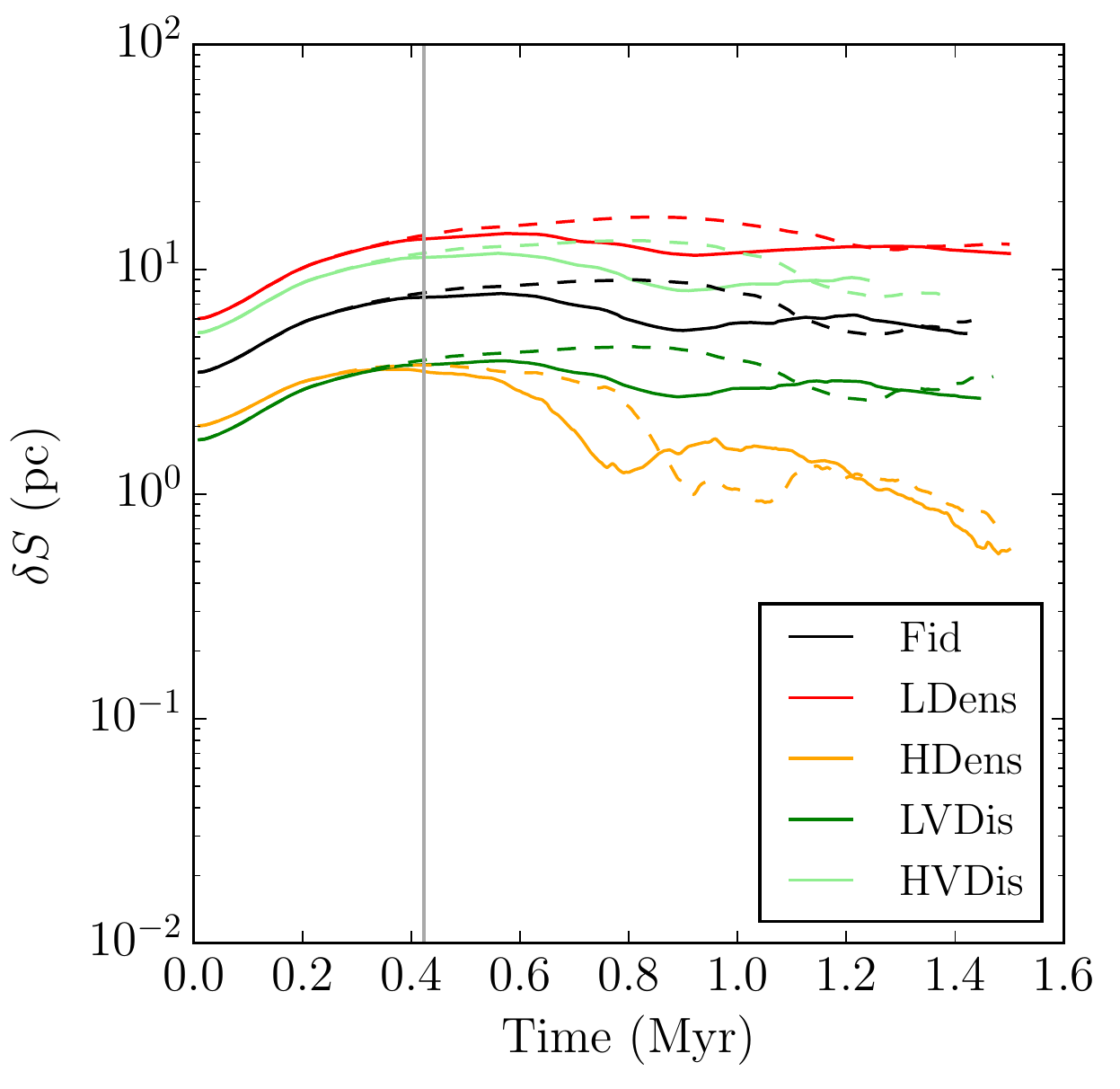}}     
    \hspace{.1in}
\subfloat[]{\includegraphics[width=0.32\textwidth]{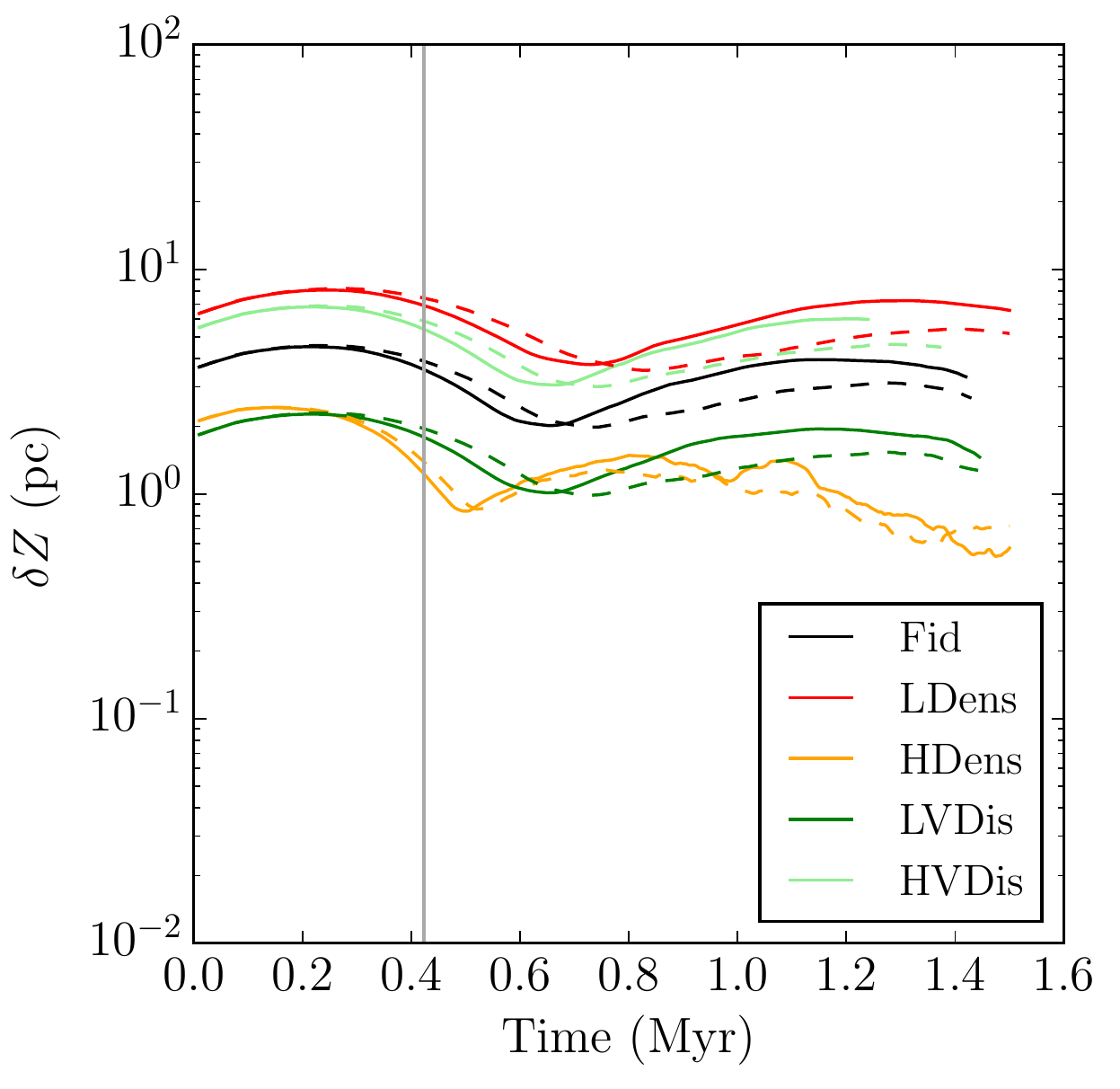}}
\vspace{-.1in}
\subfloat[]{\includegraphics[width=0.32\textwidth]{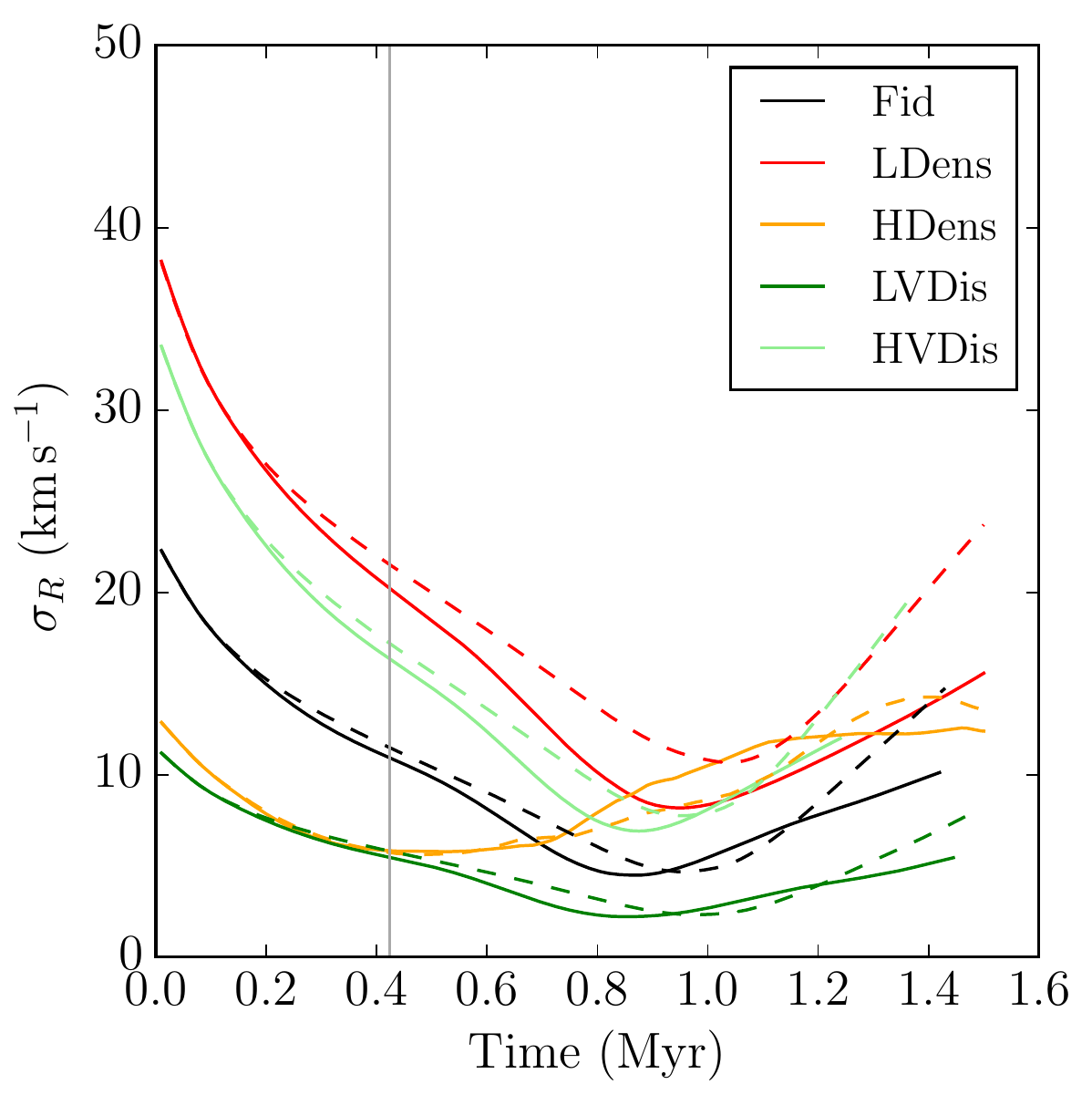}}   
    \hspace{.1in}
\subfloat[]{\includegraphics[width=0.32\textwidth]{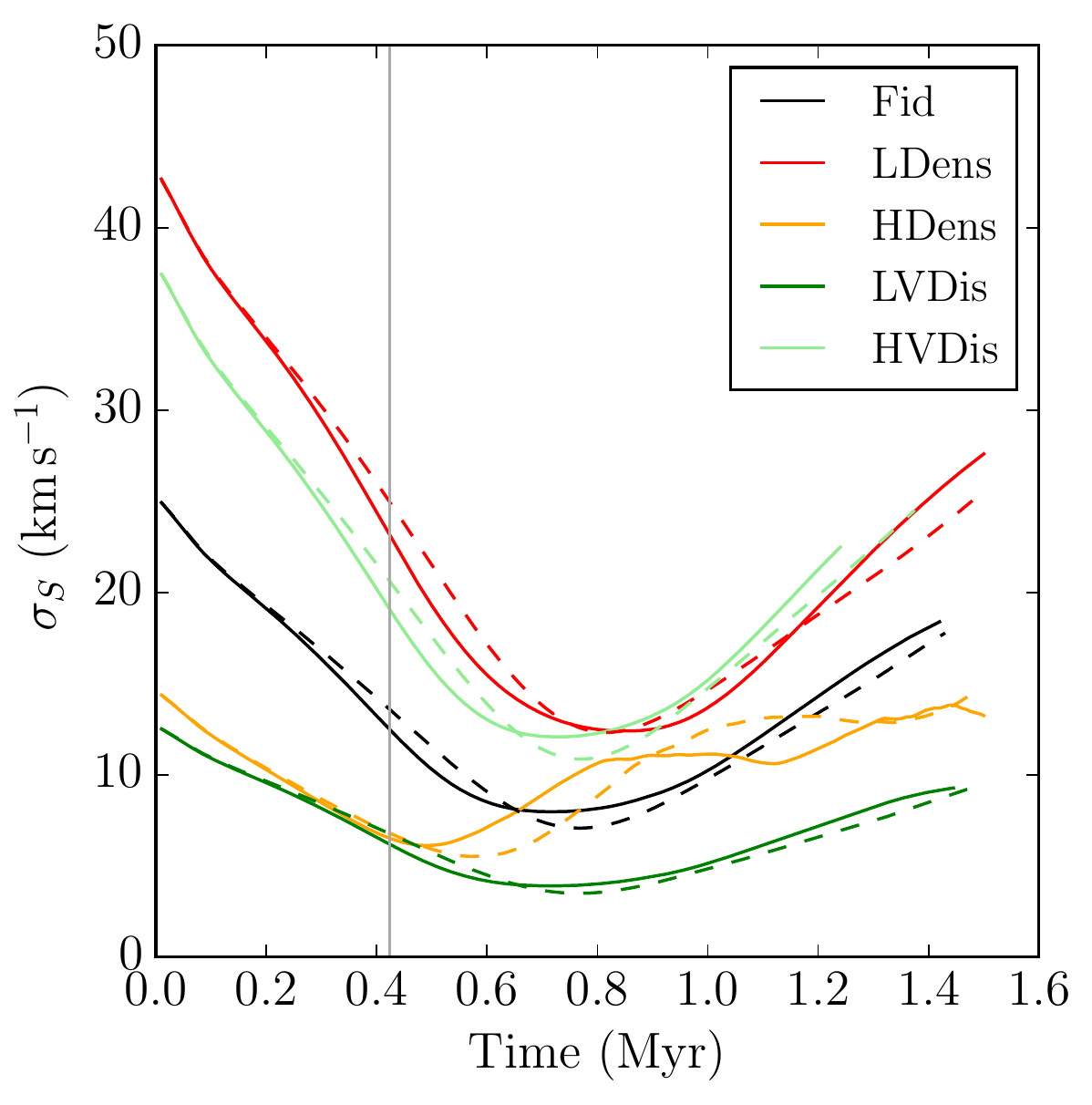}}     
    \hspace{.1in}
\subfloat[]{\includegraphics[width=0.32\textwidth]{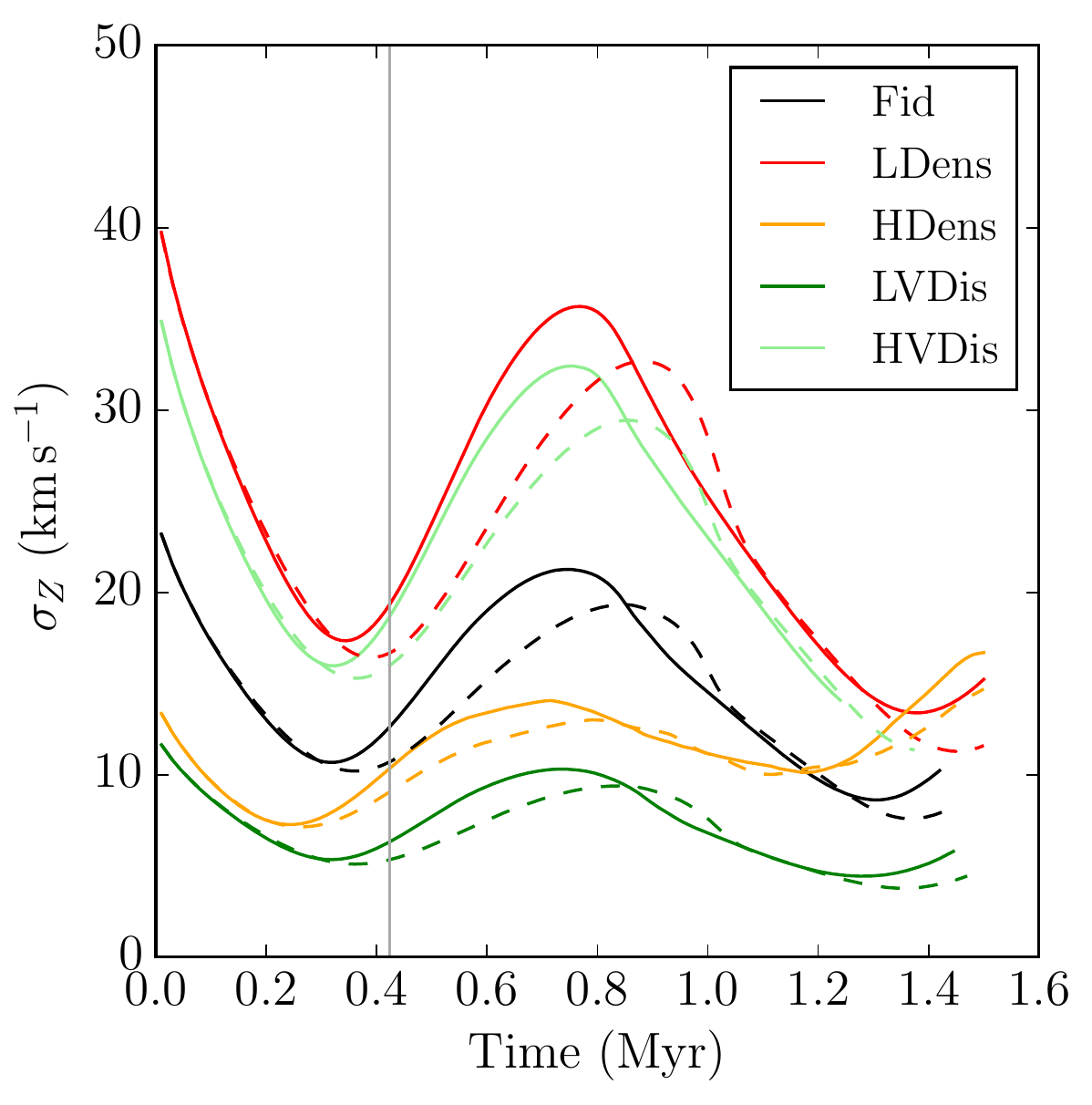}}
\caption{Top row: Sizes of the tidally--virialised clouds plotted against time in the radial (left panel), tangential (centre panel) and vertical (right panel) directions, revealing that these clouds are much more ale to resist tidal compression. Bottom row: velocity dispersions in the tidally--virialised clouds plotted against time in the radial (left panel), tangential (centre panel) and vertical (right panel) directions, showing an initial decline, followed increases driven by tidally--driven shear in the radial and azimuthal directions. In all panels, solid lines are eccentric simulations, dashed lines are circular simulations, dotted lines are isolated simulations, colours denote the different model clouds, as indicated by the legend in the left panel, and the vertical grey line represents pericentre passage in the eccentric orbit at 0.41\,Myr.}
\label{fig:delta_sigma_tidal}
\end{figure*}

To understand the influence of the tidal field on the clouds, we first compare the evolution of the clouds' dimensions and velocity dispersions in the three principal directions -- radial, tangential and vertical -- in the self--virialised (Figure \ref{fig:delta_sigma_self}) and tidally--virialised (Figure \ref{fig:delta_sigma_tidal}) cases. We locate the clouds' instantaneous centres of mass, determine the interval centred on this point in each direction enclosing half--mass, and halve these distances to generate half--mass half--extents $\delta R$, $\delta S$ and $\delta Z$. We compute velocity dispersions $\sigma_{R}$, $\sigma_{S}$ and $\sigma_{Z}$ of all the gas within the cuboid defined by the volume $2\delta R\times2\delta S\times2\delta Z$, centred on the centre of mass.\\
\indent Because of the nature of the mass distribution generating the tidal field, all components of the tidal tensor are compressive. Since the self--virialised clouds, with the exception of the HVir model, have kinetic energy densities too small to resist the tidal forces in any direction, Figure  \ref{fig:delta_sigma_self} shows that they are all contracting in all three principal directions by the time they reach pericentre, with the exception of the HVir  and HDens clouds, which are expanding mildly in the radial direction, but contracting in the other two. These two clouds are respectively the most unbound and the least susceptible to tidal compression of this suite of simulations, and the radial direction is that in which tidal compression is weakest, so this behaviour is not unexpected. In general, however, it is clear that the clouds are being crushed by the tidal field. The compression is fastest and strongest in the vertical direction, that in which the tidal compression is strongest. Additionally, tidal shear--induced spreading in the tangential direction is suppressed by tidal compression along this coordinate.\\
\indent The LVir cloud is the most weakly--supported of all the models and compression drives up the velocity dispersion in all three principal directions in this cloud, almost from the beginning of the simulations. In the other clouds, by contrast, the velocity dispersions in the radial and tangential directions decline for approximately $0.4$\,Myr before rising modestly at later times. The initial decline is due to the dissipation of the (non--artificially--driven) turbulence by shocks, whereas the rise at later times is due to tidal shear in the tangential direction, as well as local fragmentation and collapse -- the clouds have mostly reached very high star--formation efficiencies by this point. The velocity dispersion in the vertical direction, however, increases almost immediately in all simulations, due to the rapid collapse of the clouds on this axis. After approximately $0.5$\,Myr, the clouds rebound in the $z$--direction, as material passes though the instantaneous vertical midplane. Their thicknesses increase, and the velocity dispersion in the $z$--direction drops as the velocity field does work against the tidal and self--gravitational fields.\\
\indent The differences between clouds on eccentric and circular orbits are of a quantitative nature only and are modest. The simulations depart from each other as they approach and pass through pericentre, with compression in the tangential and vertical directions being slightly greater in the eccentric simulations. There is no pattern in the radial compression experienced by the clouds, but the differences between the two classes of simulation are small. The velocity dispersions in the eccentric clouds tend to be slightly larger in all three principal directions for most of the simulations but the differences are generally $\lesssim10\%$. Pericentre passage therefore has little additional influence on the eccentric clouds, probably because all the clouds are so strongly affected by the tidal forces.\\  
\indent The behaviour of the tidally--virialised clouds, depicted in Figure \ref{fig:delta_sigma_tidal}, is markedly different. The isotropic velocity field given to these clouds leaves them somewhat radially oversupported, vertically undersupported, and in approximate tangential equilibrium, although still prone to spreading by tidal shear. The clouds respond as expected. At pericentre, they are growing radially, shrinking vertically and roughly stable in tangential extent. At later times, dissipation of internal kinetic energy allows them to shrink in radial extent, but tidal shear--induced spreading stabilises them in the tangential direction.\\
\indent The velocity dispersion of the tidally-virialised clouds are characterised by steady falls in the radial and tangential components until well after pericentre, due to a combination of dissipation in shocks and doing work against the tidal field. The tangential velocity dispersion remains higher than the radial because tidal shearing motions also contribute to the former component. Contraction in the vertical direction leads to an initial increase in the velocity dispersion in this axis as, conversely, the tidal forces do work on the velocity field. This persists until material crosses the vertical midplane and the vertical velocity dispersion declines after $\approx0.8$\,Myr. The differences induced by pericentre passage are again very modest.\\
\subsection{Effect of tidal forces on cloud energies and virial ratios}
\begin{figure*}
\captionsetup[subfigure]{labelformat=empty}
\centering
\subfloat[]{\includegraphics[width=0.48\textwidth]{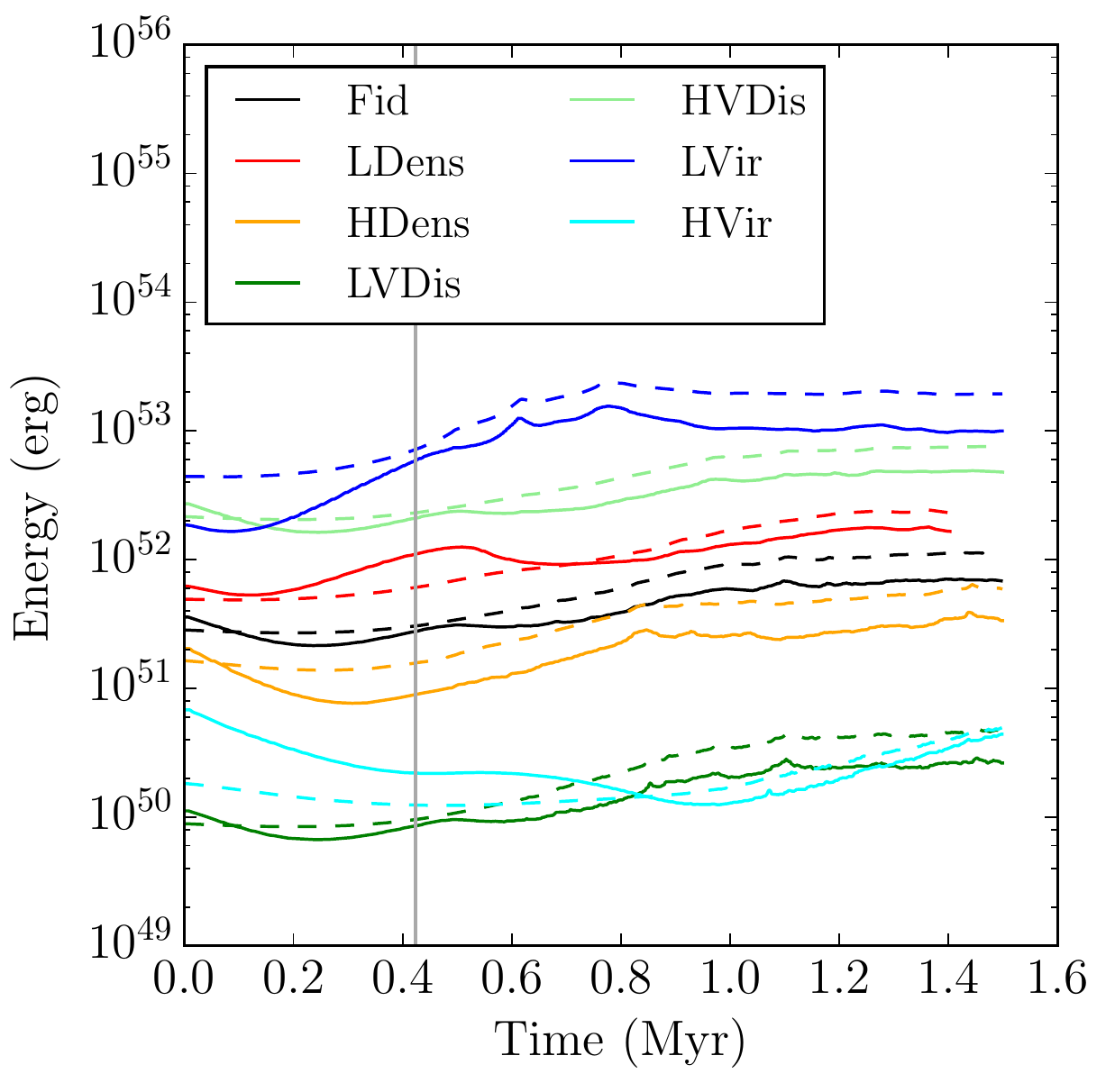}}
\hspace{-.1in}
\subfloat[]{\includegraphics[width=0.48\textwidth]{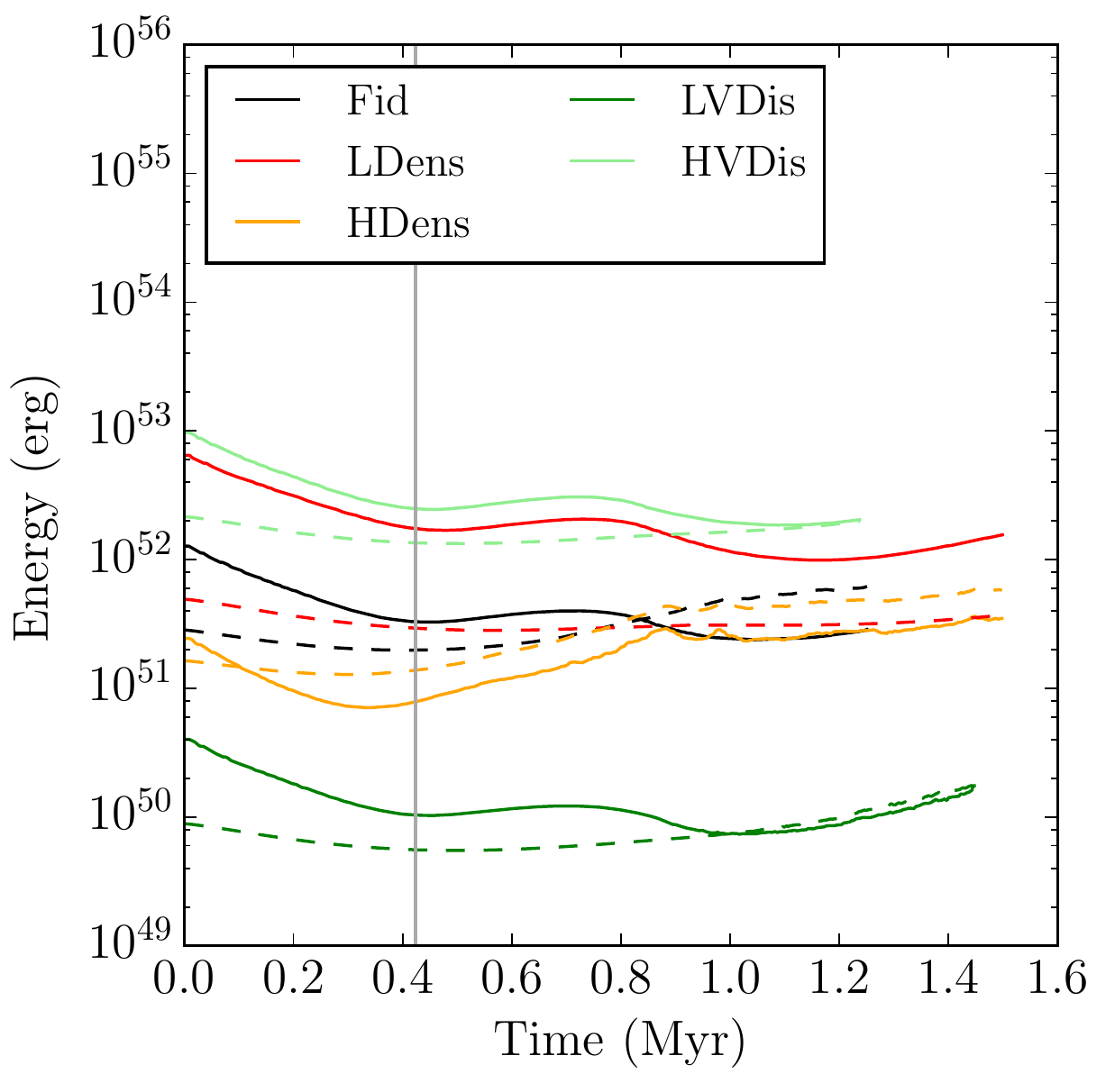}}
\caption{Kinetic energy in the centre--of--mass frame $T$ (solid lines) and the magnitude of the self--gravitational energy $V$ (dashed lines) in all eccentric self--virialised (left panel) and tidally--virialised (right--panel) calculations. The colours denote the different model clouds, as indicated by the legends, and the vertical grey lines represent pericentre passage in the eccentric orbit at 0.41\,Myr.}
\label{fig:energies}
\end{figure*}

\begin{figure*}
\captionsetup[subfigure]{labelformat=empty}
\centering
\subfloat[]{\includegraphics[width=0.48\textwidth]{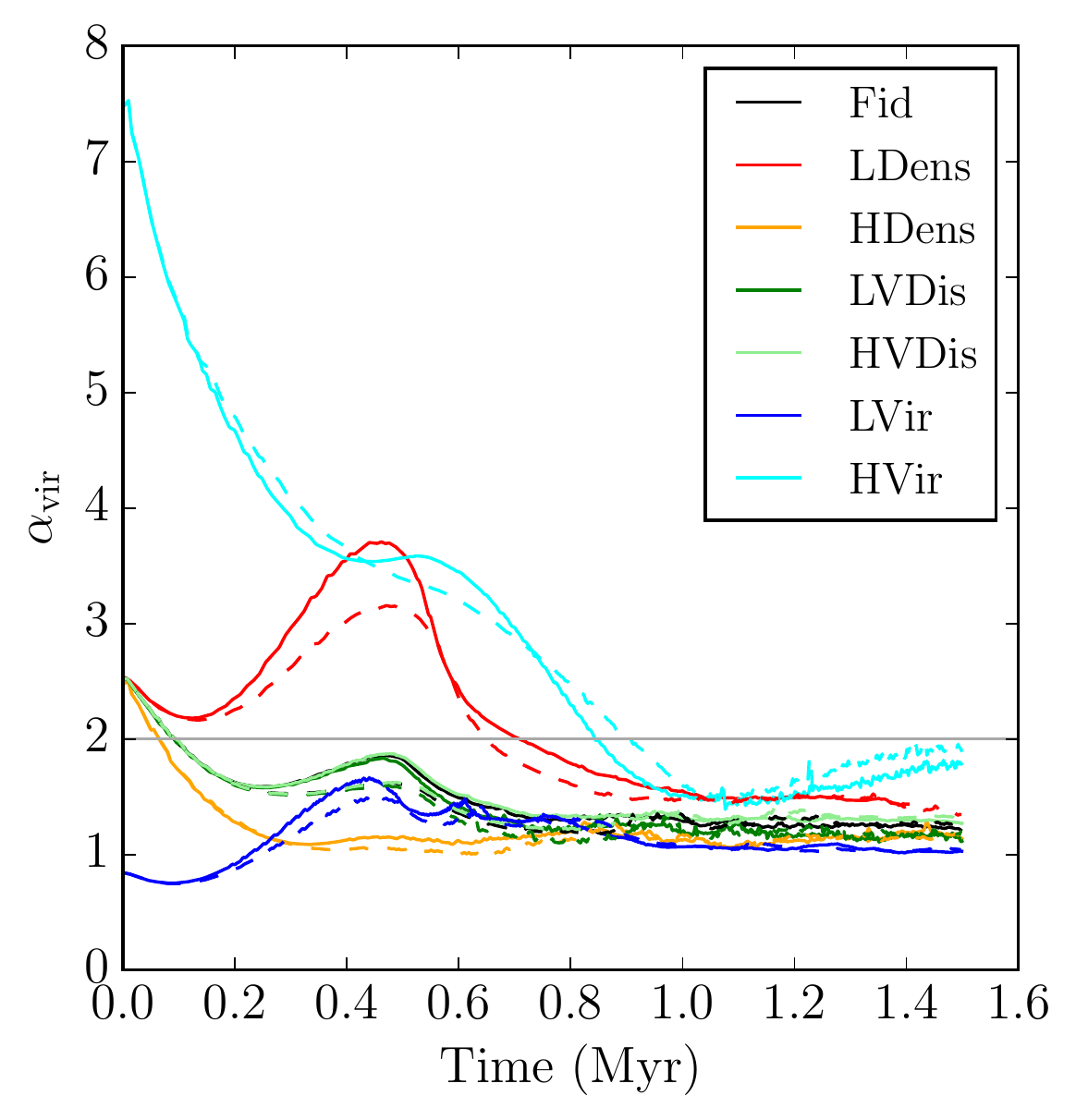}}
\hspace{-.1in}
\subfloat[]{\includegraphics[width=0.48\textwidth]{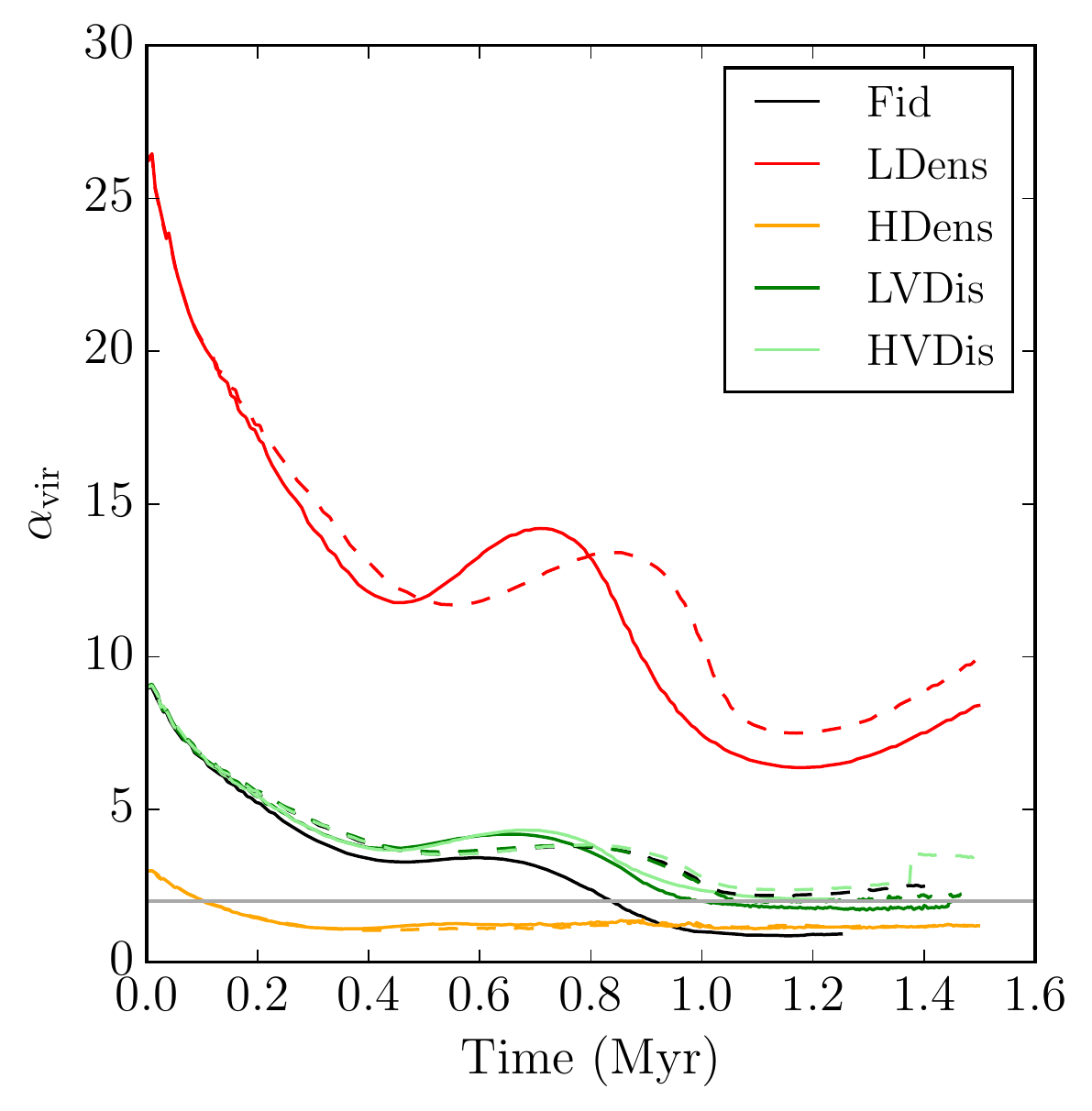}}
\caption{Virial ratios (defined as $\alpha_{\rm vir}=2T/|V|$, with $T$ the total kinetic energy in the centre of mass frame and $V$ the self--gravitational potential energy) in the self--virialised (left panel) and tidally--virialised (right panel) eccentric calculations. Clouds on Eccentric orbits are denoted by solid lines, and those on circular orbits by dashed lines. The colours denote the different model clouds, as indicated by the legends, and the vertical grey lines represent pericentre passage in the eccentric orbit at 0.41\,Myr. The grey horizontal lines denotes virial equilbrium, defined as $\alpha=2$.}
\label{fig:alphas}
\end{figure*}

\indent We now examine the evolution of the kinetic and gravitational potential energies, shown in Figure \ref{fig:energies}, and the virial ratio, defined as twice the modulus of the ratio of these energies, shown in Figure \ref{fig:alphas}.\\
\indent As expected, all of the self-virialised clouds save the LVir simulation initially have kinetic energies in excess of their gravitational potential energies. Since the clouds are taken to have fixed masses, the evolution of their self--energy content depends only on their internal velocity dispersions and their sizes. Since, by the time the clouds reach pericentre,  they are all (except for the HVir cloud) contracting, the gravitational potential energy increases in magnitude after this point, causing the kinetic energy to drop below the absolute gravitational potential energy. As a consequence of the contraction, the tidal and self--gravitational forces do work on the velocity field (in all but the HVir cloud) and the kinetic energy, which initially declines as shocks dissipate the turbulence, rises.\\
\indent With the exception of the HDens cloud, the tidally--virialised clouds behave in a radically different fashion. They all initially have kinetic energies larger than their gravitational energies, so they begin in a slightly over--stable state. As with the self--virialised clouds, their turbulent kinetic energies decline due to the action of shocks. However, since they enjoy adequate support against the tidal field in the radial and tangential directions, the tidally--virialised clouds do not contract overall, and in fact expand, causing their gravitational potential energies to drop. Their velocity fields are therefore doing work against the tidal and self--gravitational forces, and their kinetic energies also decline. Eventually, this results in kinetic energies and gravitational potential energies that are similar in magnitude.\\
\indent However, in all tidally--virialised clouds, at $\approx0.4$\,Myr, the decline in kinetic energy is stopped or reversed due to an increase in $\sigma_{z}$. At this point in time, the clouds begin to  contract in the vertical direction, so gravitational energy is converted to energy stored in this velocity component. After $\approx0.8$\,Myr, the extra energy injected into the vertical velocity component is inevitably lost as the non--homologous contraction in the $z$--direction at last overruns all the gas. The total kinetic energies eventually begin to decline once more at this point, although this is partly offset by the increase in the tangential velocity dispersion induced by tidal shear. The tidal forces acting on the clouds effectively increase the internal kinetic energies of the gas in a strongly anisotropic fashion, with different components gaining and losing dominance at different stages of the orbit.\\
\indent The virial ratios of the clouds, shown in Figure \ref{fig:alphas}, determine when the clouds become unstable and star formation commences. Most of the self--virialised clouds rapidly achieve a virial ratio less than unity and star formation therefore begins quickly. The LDens and particularly the HVir clouds delay this occurrence owing to, respectively, a longer initial freefall time and an early expansion.\\
\indent Since the tidally--virialised clouds are contracting only in the $z$--direction while they expand or are stable in the other two, their gravitational potentials remain relatively flat, and it is largely the change in the kinetic energy which determines if and when the two energies become equal and the virial ratio drops below unity. In the case of the HDens cloud, the decline in velocity dispersion due to both shocks and expansion is the fastest and the gap between the turbulent and gravitational energies is narrow, so this rapidly drives the cloud to an unstable state. These results show that the self-virialised and tidally-virialised clouds respond in radically-different, yet explicable ways to the external tidal field.\\
\indent A corollary of the energetic ratios of the tidally--virialised clouds is that, particularly in the early stages of their evolution \textit{the clouds appear strongly unbound even though they are not}. This is an obvious consequence of the action of the tidal field, which is what is preventing these clouds from rapidly dispersing. However, this makes inferring the future evolution and star--formation activity in thse clouds observationally very difficult, since their dynamics and evolution cannot be understood from observations of the clouds alone. Instead, the external gravitational potential must be accounted for.\\
\subsection{Effect of tidal forces on star formation rates}
\indent As noted above, the star formation rates and efficiencies in the tidally--virialised clouds are much lower than in the self--virialised models. Comparison of Figures \ref{fig:traditional_sf_rates} and \ref{fig:sf_rates} shows that, in the former case, star formation accelerates smoothly for $\approx0.5$\,Myr in most simulations, driven by tidally--induced contraction, while in the latter, in all but the HDens run, star formation remains at a very low level until $\approx0.8$\,Myr ($\approx0.4$\,Myr after pericentre), when it accelerates (although only weakly in the LDens simulation). This is a particularly interesting observation, because \cite{barnes17} recently reported that clouds in the CMZ dust ridge remain quiescent for 0.3--0.5\,Myr after passing through pericentre, before rapidly transitioning into a star--forming mode. As discussed above, we find in \citet{kruijssen18} that this is likely caused by the delayed conversion of tidal shear-driven rotational energy into internal, possibly turbulent, kinetic energy that can be dissipated, thus enabling collapse. Additionally, because of the non--homologous collapse of the tidally--virialised clouds, there is a considerable delay in vertical compression shocking all of the gas.\\
\indent In both classes of simulation, the onset of star formation is well correlated with the time when the clouds' virial ratios drop below 2. Most of the self--virialised clouds achieve this rapidly due to their smaller initial virial ratios and rapid contraction, and therefore begin forming stars rapidly, and before pericentre. The HVir cloud takes $\approx0.8$\,Myr to achieve this transition, resulting in a delay in the initiation of star formation relative to the other clouds.\\
\indent The tidally--virialised clouds again behave differently. The LDens cloud never achieves a virial ratio less than unity, which explains the persistent low star formation rate in this run. In contrast, the HDens cloud, which has the lowest initial virial ratio of these models, rapidly attains an unstable state as shocks drain energy from its velocity field. This results in prompt and continuing rapid star formation in this run. In the three remaining clouds, the crossover of turbulent kinetic and gravitational potential energies is delayed by tidal shear, by non--homologous contraction in the $z$--direction, and by the cloud expansion in the radial and tangential directions (also induced by tidal shear), which keeps the gravitational potential energy roughly constant. These effects combine to stabilise the clouds until $\approx0.8$--1.0\,Myr. After this point,  star formation also initiates in these models.\\
\section{Conclusions}
We have performed the first hydrodynamic simulations of isolated molecular clouds in the Galactic Centre in which the clouds are placed on accurately--determined orbits in a realistic external potential and self--consistently experience appropriate tidal forces. We find that these forces profoundly influence the evolution of the clouds. Our principal conclusions are summarised as follows:\\
\indent (i) The flattened potential of the Galactic stellar mass distribution dominates the dynamics of the gas in the $\approx$100\,pc stream of molecular gas orbiting the Galactic Centre. Simulations of isolated clouds cannot reproduce the evolution of the GMCs in this region.\\
\indent (ii) Adopting the traditional approach of scaling the clouds' turbulent velocity dispersions to put them close to equilibrium with the clouds' self--gravity results in objects which are completely unable to resist the tidal forces generated by the external potential. These objects rapidly collapse in all three principal directions, leading to rapid star formation and, in most cases, to gas exhaustion in a small fraction of an orbital time. Such models are, then, also incapable of reproducing the basic characteristics of the CMZ clouds.\\
\indent (iii) Scaling the turbulent velocity dispersions in the clouds to instead balance the tidal forces acting upon them results in objects which are much more stable and whose star formation rates and efficiencies, and overall morphology are in much better agreement with observed CMZ cloud properties. These simulations support the conclusion by \cite{2013MNRAS.433L..15L,2014MNRAS.440.3370K} that the low star formation rates per surface or volume density seen in clouds in the Central Molecular Zone are likely to be largely due to their high turbulent velocity dispersions. These elevated turbulent velocity dispersions are expected if the clouds initially form out of the shearing medium as gas moves radially inwards and enters the CMZ (see the discussion in \citealt{kruijssen18}).\\
\indent (iv) In such models, the high velocity dispersion regulates the star formation rate, while the compressive tidal forces prevent the gas motions dispersing the clouds, and delay the decay of the turbulence. The success of these models depends therefore on both the high initial velocities \textit{and} the tidal field.\\
\indent (v) The formal virial ratios of four of the five tidally--virialised clouds, as could be inferred by observing their linewidths or measured by comparing the energies stored in their velocity fields with their self--gravitational potential energies, are in the range 9.4--27.4. This is at and over the higher end of the observed range in typical Galactic--disk clouds 0.1--10 \citep[e.g.][]{2011MNRAS.413.2935D}, indicating that these objects are far from typical molecular clouds. Additionally, since their evolution is dominated by the tidal field, their dynamical states cannot be inferred purely from measurements of their intrinsic properties.\\
\indent (vi) The eccentric shape of the orbit inferred by \cite{2015MNRAS.447.1059K} has a moderate effect on the evolution of the clouds. The stronger tidal forces experienced by the clouds on the eccentric orbits compared to those on circular orbits do accelerate collapse and star formation by a factor of $\sim2$ over timescales of $\approx0.5$\,Myr, or about one sixth of an orbital time.\\
\indent Although these simulations are better able to reproduce the properties of clouds (star formation deficiency, morphology, kinematics) in the Galactic Centre environment (also see \citealt{kruijssen18}), the star formation rates and efficiencies they produce are still factors of a few to an order of magnitude too large. However, we stress that the rates and efficiencies we report are strict upper limits. There are several physical processes, such as magnetic fields and stellar feedback, which we have not modelled and which likely also play a part in regulating star formation. We defer a quantitative exploration of these issues to later work.\\
\section*{Acknowledgments}
We thank an anonymous referee for provided a detailed and careful report which substantially improved the paper.\\
JMDK gratefully acknowledges funding from the German Research Foundation (DFG) in the form of an Emmy Noether Research Group (grant number KR4801/1-1) and from the European Research Council (ERC) under the European Union's Horizon 2020 research and innovation programme via the ERC Starting Grant MUSTANG (grant agreement number 714907).

\bibliography{myrefs}

\section*{Supporting information}

Supplementary data are available at MNRAS online.\\[1ex]
\noindent \textbf{Movies.} Evolution of the simulated clouds: animated versions of the centre panels of Figures \ref{fig:sim_PP_PV_trad} and \ref{fig:sim_PP_PV_Fid}, showing the $x$--$z$ column--density projections of the Fiducial simulation showing the full evolutionary time sequence.\\[1ex]
\noindent Please note: Oxford University Press is not responsible for the content or functionality of any supporting materials supplied by the authors. Any queries (other than missing material) should be directed to the corresponding author for the article.

\appendix
\section{Anisotropy of the initial velocity field}
\begin{figure}
\centering
\includegraphics[width=0.48\textwidth]{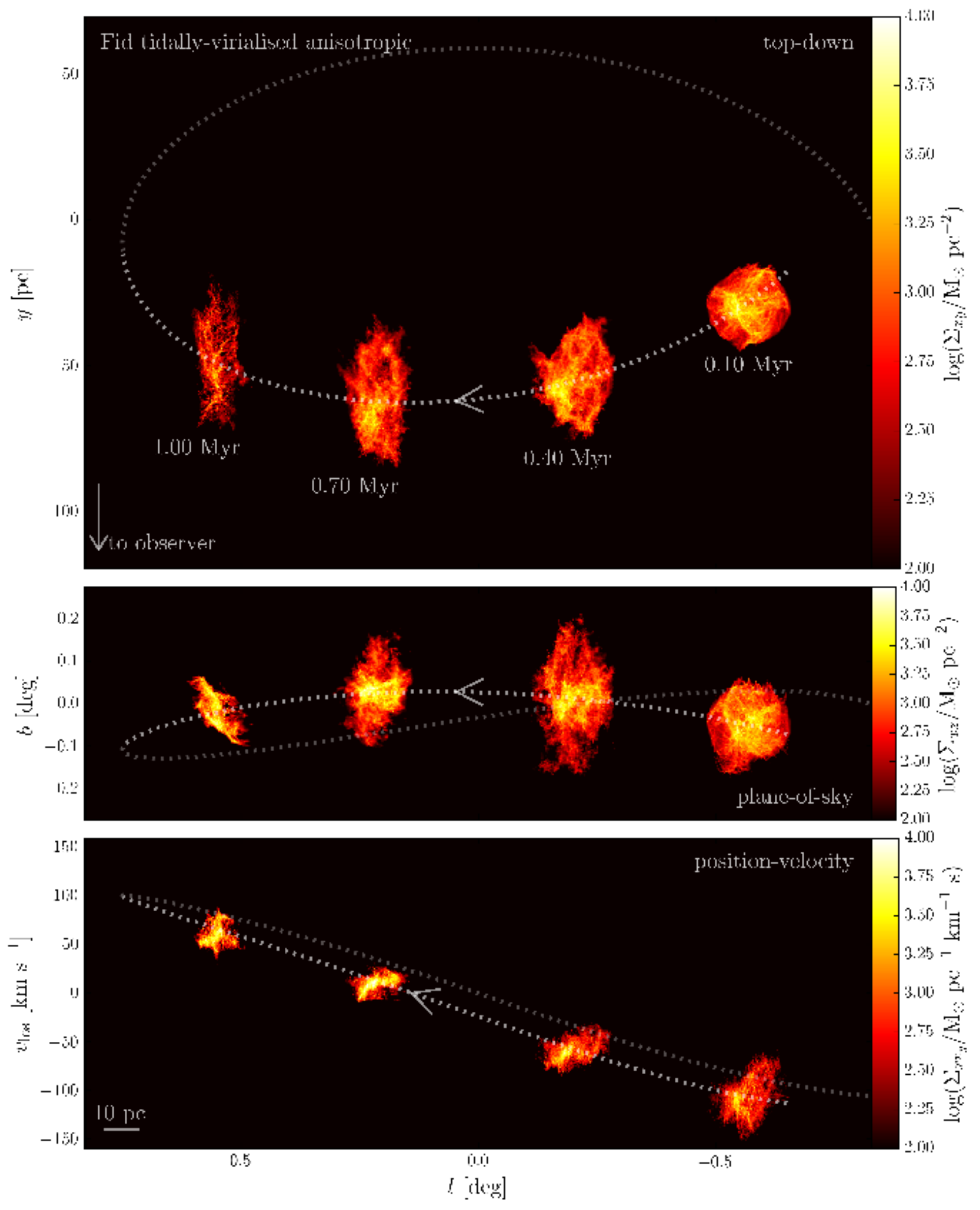}
\caption{PP and PV diagrams of the Fiducial eccentric tidally--supported cloud simulation with an initially anisotropic velocity field.}
\label{fig:sim_PP_PV_aniso}
\end{figure}

\indent To construct the tidally--supported clouds in this work, the one--dimensional velocity dispersions required to resist the three components of the tidal tensor have been computed and then geometrically averaged, so that the initial velocity field is isotropic. Since the radial tidal forces are the weakest, and the vertical forces are strongest, this leads to over--support of the clouds in the radial direction and under--support in the vertical direction.\\
\indent To evaluate the consequences of this assumption, we also ran additional realisations of the Fiducial cloud in which the velocity dispersions in the initially radial, tangential and vertical directions is set to those computed from the tidal tensor, resulting in a strongly anisotropic velocity field where the dispersion in the vertical direction is about four times that in the radial. The differences between the simulations with the isotropic and anisotropic velocity fields are modest, and are strongest in terms of morphology, as shown in Figure \ref{fig:sim_PP_PV_aniso}. Unsurprisingly, this cloud in particular shows little if any flattening in the vertical direction, and very little tendency to spread in the orbital direction. Overall, its morphological properties do not provide as good a match to the observed CMZ clouds as the clouds with isotropic velocity fields. We therefore suggest that the initial velocity field of the observed CMZ clouds were likely isotropic.\\

\label{lastpage}

\end{document}